\documentclass[modern]{aastex61}

\begin{document}
\title{Abundances and Depletions of Neutron-Capture Elements in the Interstellar Medium\footnote{Based on observations made with the NASA/ESA \emph{Hubble Space Telescope}, obtained from the Mikulski Archive for Space Telescopes (MAST). STScI is operated by the Association of Universities for Research in Astronomy, Inc., under NASA contract NAS5-26555.}}

\correspondingauthor{A.~M.~Ritchey}

\author{A.~M.~Ritchey}
\affiliation{Department of Astronomy, University of Washington, Seattle, WA 98195, USA}
\email{aritchey@astro.washington.edu}

\author{S.~R.~Federman}
\affiliation{Department of Physics and Astronomy, University of Toledo, Toledo, OH 43606, USA}

\author{D.~L.~Lambert}
\affiliation{W. J. McDonald Observatory and Department of Astronomy, University of Texas at Austin, Austin, TX 78712, USA}

\begin{abstract}
We present an extensive analysis of the gas-phase abundances and depletion behaviors of neutron-capture elements in the interstellar medium (ISM). Column densities (or upper limits to the column densities) of Ga~{\sc ii}, Ge~{\sc ii}, As~{\sc ii}, Kr~{\sc i}, Cd~{\sc ii}, Sn~{\sc ii}, and Pb~{\sc ii} are determined for a sample of 69 sight lines with high- and/or medium-resolution archival spectra obtained with the Space Telescope Imaging Spectrograph onboard the \emph{Hubble Space Telescope}. An additional 59 sight lines with column density measurements reported in the literature are included in our analysis. Parameters that characterize the depletion trends of the elements are derived according to the methodology developed by Jenkins. (In an appendix, we present similar depletion results for the light element B.) The depletion patterns exhibited by Ga and Ge comport with expectations based on the depletion results obtained for many other elements. Arsenic exhibits much less depletion than expected, and its abundance in low-depletion sight lines may even be supersolar. We confirm a previous finding by Jenkins that the depletion of Kr increases as the overall depletion level increases from one sight line to another. Cadmium shows no such evidence of increasing depletion. We find a significant amount of scatter in the gas-phase abundances of Sn and Pb. For Sn, at least, the scatter may be evidence of real intrinsic abundance variations due to $s$-process enrichment combined with inefficient mixing in the ISM.
\end{abstract}


\section{INTRODUCTION}
The production of elements heavier than those of the Fe peak occurs via neutron-capture ($n$-capture) processes that are either slow ($s$) or rapid ($r$) compared to the timescale for $\beta$ decay (e.g., Sneden et al.~2008). The main $s$-process operates in low- and intermediate-mass asymptotic giant branch (AGB) stars (e.g., Gallino et al.~1998; Lugaro et al.~2003; Karakas et al.~2009), whose stellar winds enrich the interstellar medium (ISM) with the products of $s$-process nucleosynthesis. A weak $s$-process occurs in the He- and C-burning shells of massive stars (e.g., Raiteri et al.~1993; The et al.~2007), which distribute their products into the ISM via supernova explosions. The astrophysical site of the $r$-process is still debated, but the most likely sites include core collapse supernovae and binary neutron star mergers (e.g., Argast et al.~2004; Beniamini et al.~2016). The latter production sites have received considerable attention recently due to the detection of the electromagnetic counterpart to the gravitational wave source GW170817 (e.g., Abbott et al.~2017; Tanvir et al.~2017). When considering the effects of $n$-capture nucleosynthesis on the chemical evolution of the Galaxy, the important distinction is between the processes associated with long-lived low- and intermediate-mass stars (i.e., the main $s$-process) and those attributed to massive stars with short evolutionary lifetimes (i.e., the weak $s$-process and the $r$-process).

Detailed studies of the interstellar abundances of $n$-capture ($Z>30$) elements provide a means to probe the enrichment of the ISM in the products of $s$- and $r$-process nucleosynthesis. Unfortunately, the low cosmic abundances of most heavy elements preclude their detection via absorption from resonance lines in the UV and visible, even those arising from the dominant ionization stages of the elements in interstellar clouds. A further complication in using the interstellar abundances of heavy elements to study nucleosynthetic processes is the unknown amount of depletion onto grains exhibited by a given element in a given line of sight. The most well-studied of the $n$-capture elements thus far detected in the ISM is krypton ($Z=36$), which, as a noble gas, would generally not be expected to be depleted along any lines of sight. Cartledge et al.~(2001, 2003, 2008) examined gas-phase Kr abundances in approximately 50 sight lines, finding that the average Kr abundance is independent of such external sight line properties as the average line-of-sight hydrogen density and the molecular hydrogen fraction. In absolute terms, however, the average interstellar abundance of Kr, which has changed little from one determination to another (e.g., Cardelli \& Meyer 1997; Cartledge et al.~2008), is only $\sim$50\% of the theoretical solar abundance. In a comprehensive examination of interstellar depletions, Jenkins (2009, hereafter J09) found that the depletion of Kr may actually increase (slightly) as the overall strength of depletions increases from one sight line to another. Still, the deficit in the interstellar abundance of Kr relative to the solar system, which is significant even in low depletion sight lines, remains unexplained.

In a study of rubidium ($Z=37$) isotope ratios in diffuse interstellar clouds, Walker et al.~(2009) found that the gas-phase Rb/K ratio is (on average) $\sim$40\% of the solar value. The actual interstellar abundance of Rb (relative to H) is uncertain because the observed line of Rb~{\sc i} at 7800.3~\AA{} arises from a trace neutral species (most of the interstellar Rb being singly-ionized). Thus, the observed abundance of Rb~{\sc i} is compared to the abundance of K~{\sc i}, and the ratio is converted to the elemental ratio by making appropriate corrections for the differences in the photoionization and radiative recombination rates (see Federman et al.~2004; Walker et al.~2009). While such corrections inevitably introduce uncertainties, the low Rb/K ratios are puzzling; they do not seem to be related to enhanced Rb depletion since the condensation temperature of K is higher than that of Rb (Lodders 2003). Given that the interstellar Na/K and Li/K ratios are generally found to be consistent with the solar values (Welty \& Hobbs 2001; Knauth et al.~2003), the low Rb/K ratios suggest a deficit in the abundance of Rb in interstellar clouds. Both Kr and Rb are produced primarily by massive stars, through a combination of the weak $s$-process and the $r$-process (Heil et al.~2008). Thus, the unexpectedly low interstellar abundances of Kr and Rb could potentially indicate a deficiency in the contribution from massive stars to the synthesis of heavy elements in the current epoch (Walker et al.~2009).

Existing studies of the interstellar abundances of other $n$-capture elements seem to support this conjecture. Sofia et al.~(1999) examined the gas-phase abundances of cadmium ($Z=48$) and tin ($Z=50$) in 5 and 14 sight lines, respectively, using data obtained with the Goddard High-Resolution Spectrograph (GHRS) onboard the \emph{Hubble Space Telescope} (\emph{HST}). They found that Cd shows no evidence of depletion onto grains. The Cd abundances in their small sample appear to be indistinguishable from the solar abundance regardless of the amount of molecular material in the line of sight. In contrast, the gas-phase Sn abundances were found to decrease with increasing molecular fraction, a clear indication of the incorporation of Sn atoms into dust grains in higher density sight lines. In sight lines with low molecular fractions, Sofia et al.~(1999) found that Sn is entirely undepleted, and may even be enhanced in its abundance relative to the solar system. Much, if not most, of the synthesis of Cd and Sn occurs via the main $s$-process in low- and intermediate-mass AGB stars (Arlandini et al.~1999; Bisterzo et al.~2014), and the interstellar abundances of these elements do not appear to be deficient. The present-day interstellar abundance of Sn may even be supersolar (Sofia et al.~1999).

The only other $n$-capture elements that have been studied in a large number of diverse sight lines are gallium ($Z=31$) and germanium ($Z=32$). These elements are mainly produced by the weak $s$-process in massive stars (e.g., The et al.~2007; Pignatari et al.~2010), and both are depleted in the ISM relative to the solar system. However, both elements exhibit density-dependent depletion in a way consistent with the depletion behaviors of other more abundant elements (Cartledge et al.~2006; Ritchey et al.~2011, hereafter R11). Additional $n$-capture elements that have been detected in the ISM include arsenic ($Z=33$), which has been observed in three lines of sight (Cardelli et al.~1993; Federman et al.~2003), and lead ($Z=82$), which has been detected in just two individual sight lines (Cardelli 1994; Welty et al.~1995). With so few detections, the depletion behaviors of these elements are largely unknown, making it difficult to interpret their gas-phase abundances.

In this investigation, we seek to conduct a comprehensive examination of the gas-phase abundances and depletion behaviors of $n$-capture elements in the ISM. To this end, we carry out an extensive analysis of UV absorption lines arising from the dominant ions Ga~{\sc ii}, Ge~{\sc ii}, As~{\sc ii}, Kr~{\sc i}, Cd~{\sc ii}, Sn~{\sc ii}, and Pb~{\sc ii}, using archival data obtained with the Space Telescope Imaging Spectrograph (STIS) onboard \emph{HST}. The seven species that are the focus of our investigation represent the only dominant ions of $n$-capture elements with UV resonance lines accessible to STIS that have been detected in multiple lines of sight. The elements Ga, Ge, As, Kr, Cd, Sn, and Pb are among the only heavy elements with both relatively high cosmic abundances and relatively low condensation temperatures, which explains why their UV absorption lines are more readily detectable than those of other $n$-capture elements. While extensive studies using STIS data have already been published for Ga~{\sc ii}, Ge~{\sc ii}, and Kr~{\sc i} (Cartledge et al.~2006, 2008; R11), no such studies exist for the rarer species As~{\sc ii}, Cd~{\sc ii}, Sn~{\sc ii}, and Pb~{\sc ii}. (All of the currently published abundances for these latter species were derived from GHRS observations.) Moreover, it is desirable to examine the abundances of all seven $n$-capture elements in the same sample of interstellar sight lines (if possible), rather than compare studies that examined different sets of sight lines for different elements.

A thorough examination of the UV spectroscopic data available in the \emph{HST}/STIS archive allows us to discern a significant number of new detections of As~{\sc ii}, Cd~{\sc ii}, Sn~{\sc ii}, and Pb~{\sc ii}, and to expand on previously published results for Ga~{\sc ii}, Ge~{\sc ii}, and Kr~{\sc i}. This effort enables us to better constrain the depletion characteristics of these elements so that we can address claims of abundance deficiencies and enhancements in the ISM due to $s$- and $r$-process nucleosynthesis. The remainder of this paper is organized as follows. An overview of the archival survey is provided in Section 2, which also includes a description of the sample of sight lines and the steps involved in processing the data sets obtained from the STIS archive. Our column density determinations are presented in Section 3, where we also discuss recent updates to the oscillator strengths ($f$-values) relevant to our investigation. In Section 4, we derive the elemental abundances (Section 4.1), and determine depletion parameters for each element, adopting the methodology of J09 (Section 4.2). Since many sight lines yield only upper limits for some species, we also perform a survival analysis on the gas-phase abundance data (Section 4.3). The results of our analyses, and the implications for $s$- and $r$-process nucleosynthesis, are discussed in Section 5. We summarize our main conclusions in Section 6. Two appendices present a compilation of column density measurements from the literature (Appendix A) and an application of the J09 methodology to the rare light element boron (Appendix B).

\section{OBSERVATIONS AND DATA PROCESSING}
\subsection{Overview of the Archival Survey}
The primary aim of our extensive search of the \emph{HST}/STIS archive was the identification of sight lines showing significant absorption from As~{\sc ii}~$\lambda1263$, Cd~{\sc ii}~$\lambda2145$, Sn~{\sc ii}~$\lambda1400$, and Pb~{\sc ii}~$\lambda1433$. While these lines are the principal transitions used to study these species in diffuse interstellar clouds, they are only rarely seen in UV spectra (although Sn~{\sc ii}~$\lambda1400$ is somewhat more common than the others). To construct our sample, we started by examining the sight lines analyzed by R11 in their survey of B~{\sc ii} absorption in the diffuse ISM. The absorption features sought in the present survey are expected to have similar strengths compared to B~{\sc ii}~$\lambda1362$. Moreover, for each of the directions studied by R11, we have detailed knowledge of the line-of-sight component structure from moderately strong absorption lines of dominant ions such as O~{\sc i}~$\lambda1355$. This facilitates the search for weak features from other dominant ions that are expected to exhibit similar absorption profiles. We also expanded our search by examining the available STIS data for all other sight lines in the \emph{HST} archive with the necessary wavelength coverage, considering observations acquired using either the high-resolution gratings (E140H and E230H) or the medium-resolution gratings (E140M and E230M) of the STIS FUV and NUV Multi-Anode Microchannel Array (MAMA) detectors. We noted any sight lines that showed apparent absorption from one or more of the four principal lines of interest at a velocity similar to that of O~{\sc i}~$\lambda1355$. This process resulted in 14 sight lines being added to our initial list of 55 from the B~{\sc ii} survey, yielding our primary sample of 69 interstellar sight lines. (We included all 55 sight lines from R11 in our primary sample, even those that were not suspected of having absorption from As~{\sc ii}, Cd~{\sc ii}, Sn~{\sc ii}, or Pb~{\sc ii}, because the STIS data for these sight lines had already been processed, and thus it was straightforward to calculate upper limits for any non-detections.)

Each preliminary detection of As~{\sc ii}~$\lambda1263$, Cd~{\sc ii}~$\lambda2145$, Sn~{\sc ii}~$\lambda1400$, and Pb~{\sc ii}~$\lambda1433$ was scrutinized in detail, and, in some cases, we found that either the absorption was not significant enough (i.e., the equivalent width was smaller than twice the estimated uncertainty), or the velocity deviated too severely from the expected velocity (i.e., by more than 2$-$4 km s$^{-1}$ depending on the spectral resolution). In such cases, it may be that the apparent absorption feature is simply an artifact of noise or is of instrumental origin. In the final analysis (see Section 3.2), we found that 32 sight lines from our primary sample did not exhibit compelling evidence for absorption from at least one of the four main species of interest. We chose not to eliminate these sight lines from our sample, however, since upper limits on column densities can still prove useful, particularly for interpreting the abundances of elements with relatively few confirmed detections (as we demonstrate in Section 4.3). Moreover, while these sight lines do not yield detections for any of the rare $n$-capture species, they each provide information on at least one of the more common species (i.e., Ga~{\sc ii}, Ge~{\sc ii}, and/or Kr~{\sc i}), which constitute an important part of our overall survey.

Ultimately, through our extensive examination of STIS archival data, we identified 10 sight lines with secure detections of As~{\sc ii}~$\lambda1263$, 7 sight lines with detections of Cd~{\sc ii}~$\lambda2145$, 27 sight lines with detections of Sn~{\sc ii}~$\lambda1400$, and 8 sight lines with detections of Pb~{\sc ii}~$\lambda1433$. Heidarian et al.~(2015) recently reported the first detection in the ISM of the Pb~{\sc ii} transition at 1203.6~\AA{} from a composite spectrum obtained by co-adding high-resolution \emph{HST}/STIS spectra for over 100 sight lines. Their analysis indicated that the 1203.6~\AA{} line is about a factor of two stronger intrinsically than the line at 1433.9~\AA{}. However, in a typical interstellar sight line, where log~$N$(H~{\sc i})~$\sim$~21.2, the 1203.6~\AA{} line is positioned in a region of the spectrum where the flux is depressed by the damping wing of the nearby Ly$\alpha$ line, limiting the potential usefulness of this feature. Nevertheless, following the discovery of the Pb~{\sc ii}~$\lambda1203$ line in a composite spectrum by Heidarian et al.~(2015), we re-examined the archival STIS data to try to identify individual sight lines displaying this feature. Here, we report the detection of Pb~{\sc ii}~$\lambda1203$ in 4 individual sight lines, bringing our total number of Pb~{\sc ii} detections to 12. We also searched for the weaker line of the Cd~{\sc ii} doublet at 2265.7~\AA{}, and found this line in each of the 7 directions where Cd~{\sc ii}~$\lambda2145$ was detected.

In addition to searching for absorption from As~{\sc ii}, Cd~{\sc ii}, Sn~{\sc ii}, and Pb~{\sc ii}, we sought to incorporate available data on Ga~{\sc ii}, Ge~{\sc ii}, and Kr~{\sc i} into our analysis so that the abundances of all seven $n$-capture elements could be analyzed in a consistent manner. Column densities for Ga~{\sc ii} were previously reported by R11 for many of the sight lines in their B~{\sc ii} survey. Of the 14 additional sight lines added to our current sample, 5 have archival STIS data covering the Ga~{\sc ii} transition at 1414.4~\AA{}, and all 5 exhibit significant Ga~{\sc ii} absorption. Published column densities for Ge~{\sc ii} and Kr~{\sc i} are also available for many of the sight lines in our sample. Twenty-five of our sight lines have Ge~{\sc ii} column densities listed in Cartledge et al.~(2006), while a somewhat different subset of 25 sight lines has Kr~{\sc i} column densities given in Cartledge et al.~(2003, 2004, 2008). All of these Ge~{\sc ii} and Kr~{\sc i} measurements were deduced from STIS observations. One of the sight lines in our sample ($\zeta$~Per) has a published Kr~{\sc i} column density from GHRS observations (Cardelli \& Meyer 1997). To maintain consistency in how each sight line in our primary sample was analyzed across all of the elements of interest, we re-examined the STIS data on Ge~{\sc ii}~$\lambda1237$ and Kr~{\sc i}~$\lambda1235$ for each direction with a previously reported Ge~{\sc ii} or Kr~{\sc i} column density. We also sought to obtain Ge~{\sc ii} and Kr~{\sc i} column densities for any other sight lines in our sample without existing determinations in the literature.

For one of our primary targets (X~Per), the available STIS data do not cover the Ge~{\sc ii}~$\lambda1237$ or Kr~{\sc i}~$\lambda1235$ lines, but other archival data are available for these species. To derive the Ge~{\sc ii} column density toward X~Per, we examined GHRS data on Ge~{\sc ii}~$\lambda1237$, along with STIS data for the weaker Ge~{\sc ii} line at 1602.5~\AA{}. The Kr~{\sc i} column density that we derive for X~Per is based on an analysis of the Kr~{\sc i}~$\lambda1164$ line, which is available from observations made with the \emph{Far Ultraviolet Spectroscopic Explorer} (\emph{FUSE}) satellite. (While our intention in carrying out our archival survey was to focus exclusively on STIS data, the inclusion of the GHRS and \emph{FUSE} data sets toward X~Per meant that we could derive reliable abundances for all seven $n$-capture elements in this direction.) For another one of our sight lines (HD~99890), R11 reported column densities based on a single, relatively short STIS exposure obtained with the medium-resolution (E140M) grating. In the years since R11 published their study, new high-resolution STIS spectra of HD~99890 have been obtained under two separate observing programs, one of which (GO program 12191) was specifically designed to search for rare heavy elements in the ISM. Since the new high-resolution data are far superior to the spectrum analyzed by R11 (in terms of total exposure time and overall data quality), we redetermined column densities for this sight line using only the high-resolution data.

There are 30 additional sight lines not included in our primary sample --- because they showed no sign of absorption from As~{\sc ii}, Cd~{\sc ii}, Sn~{\sc ii}, or Pb~{\sc ii} and had not been analyzed previously by R11 --- that have published Ge~{\sc ii} and/or Kr~{\sc i} column densities from STIS observations (Cartledge et al.~2001, 2003, 2006, 2008; Welty 2007). There are another 19 sight lines with published column densities for Ga~{\sc ii}, Ge~{\sc ii}, As~{\sc ii}, Kr~{\sc i}, Cd~{\sc ii}, Sn~{\sc ii}, and/or Pb~{\sc ii} from GHRS observations (Savage et al.~1992; Cardelli et al.~1993; Hobbs et al.~1993; Cardelli 1994; Welty et al.~1995, 1999; Cardelli \& Meyer 1997; Sofia et al.~1999; Federman et al.~2003; Cartledge et al.~2008). All of these sight lines are included in the analysis described in Section 4, where we evaluate the depletion characteristics of the various elements, though we do not rederive any column densities in these directions. We do apply corrections to the column densities to account for any differences in the assumed $f$-values between the literature studies and our investigation (see Section 3.1 and Appendix A). When combined with our primary sample, this extended sample of sight lines with column density measurements from the literature brings the total number of sight lines considered in this investigation to 128. (This includes 10 additional sight lines with previously published column densities of O~{\sc i} only. These directions were added to our survey so that our sample for oxygen would be as complete as possible, just as for the other elements.)

\subsection{Characteristics of the Sight Lines}
Table 1 provides the relevant data for the 69 O and B-type stars that served as background targets for our primary sample of interstellar sight lines. Many of these stars were included in the investigation of element depletions by J09, who carefully compiled spectral types for the stars in his survey (and gives references to the original sources). From those spectral types, J09 determined distances and reddenings to his targets following a rigorous process based on spectroscopic parallax that was originally implemented by Bowen et al.~(2008, hereafter B08) in their survey of O~{\sc vi} absorption in the Galactic disk. In general, for any of the stars in our sample that were analyzed by J09 or B08, we adopt the same spectral types, distances, and reddenings given in those references. For stars not included in either investigation, we compute the distances and reddenings using the same set of procedures that those studies invoked, adopting intrinsic colors from Wegner (1994) and absolute visual magnitudes from B08 (see Appendix B of B08 for a detailed description of the method and its associated caveats). References to the sources of the chosen spectral types for these stars are given in Table 1. An exception to this general approach to obtaining distances to our targets is that if a star has an \emph{Hipparcos} parallax measurement that is significant at the 4$\sigma$ level or greater (based on the second reduction of \emph{Hipparcos} data; van Leeuwen 2007), we adopt the \emph{Hipparcos} distance to that star. (J09 had an acceptance threshold for \emph{Hipparcos} results of 10$\sigma$, while B08 had a threshold of 5$\sigma$.)

The Galactic coordinates and $V$ magnitudes listed in Table 1 were obtained from the SIMBAD database. We note, however, that these $V$ magnitudes are not necessarily the same as those used to determine the distances and reddenings for some stars since we follow the recommendation of B08 in using magnitudes from the \emph{Tycho} Starmapper catalog (ESA 1997)\footnote{Vizier Online Data Catalog I/239.} to obtain values for $B$ and $V$, after applying the appropriate transformation to the \emph{Tycho} magnitudes $B_T$ and $V_T$ (see Appendix B1 of B08). We did use the SIMBAD values of $B$ and $V$ for one of our stars (CPD$-$30 5410) as this star is not listed in the \emph{Tycho} catalog. Another one of our stars (HD~203338) has a composite spectrum, with both a hot (B1 V) component and a cool (M1 Ibep) component (Simonson 1968), making a determination of the distance and reddening for this sight line challenging. Fortunately, the \emph{Hipparcos} catalog lists values of $B_T$ and $V_T$ separately for the two components,\footnote{The individual magnitudes may be found in the \emph{Hipparcos Double and Multiples: Component Solutions} catalog (ESA 1997).} allowing us to derive an appropriate distance and reddening by focusing solely on the hot component. The results we obtain for this star, $E(\bv)=0.49$ and $d=1.4$~kpc, are similar to the results for other stars that (like HD~203338) are members of the Cep OB2 association, such as HD~207198, HD~207308, and HD~207538 (see Table 1). The distances and Galactic latitudes compiled for the stars in our sample were used to calculate the heights of the stars above or below the midplane of the Galaxy. These $z$ distances are given in the last column of Table 1.

Figure 1 presents a series of histograms that illustrates some of the properties of the 128 sight lines considered in this investigation (where we have included those sight lines with column density measurements from the literature). The three panels show distributions for the total hydrogen column densities, $N($H$_{\mathrm{tot}})=N($H~{\sc i}$)+2N($H$_2)$, the distances to the background stars, and the heights of the stars above or below the Galactic plane. (The sources of the adopted values of $N$(H~{\sc i}) and $N$(H$_2$) are discussed in Section 4.1.) The total hydrogen column densities range from log~$N$(H$_{\mathrm{tot}}$)~$\sim$~20.2 to 21.7, with a peak at approximately 21.3. The distances show a bimodal distribution reflecting the tendency for stars to be found either in the local arm of the Galaxy or in one of the two nearest spiral arms (the Sagittarius-Carina spiral arm or the Perseus arm; see also Figure 1 of R11). The $z$ distances (or rather their absolute magnitudes) have an asymmetric distribution, peaking at approximately log~$|z|$~$\sim$~1.8 and tapering off to about 3.5. These distributions are similar to the ones presented in J09 for the 243 sight lines included in that investigation (see Figures 2 and 3 of J09). While this similarity is not surprising, since there is a great deal of overlap between our sample and that of J09, it does suggest that our sample is large enough and diverse enough to be representative of the local Galactic ISM. As a result, the depletion parameters that we derive in Section 4 for the elements under consideration should be broadly applicable.

\begin{figure}
\centering
\includegraphics[width=1.0\textwidth]{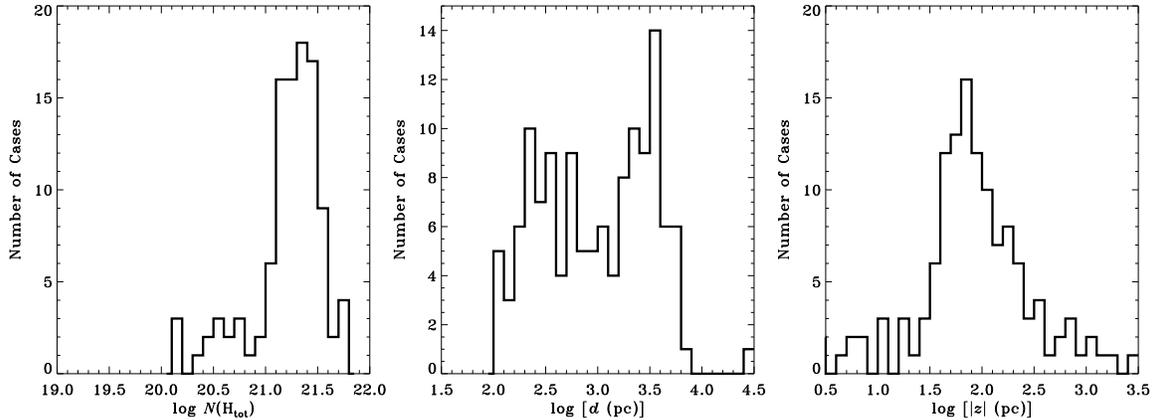}
\caption{Characteristics of the sample of sight lines included in this investigation. Left panel: Distribution of total hydrogen column densities along the lines of sight. Middle panel: Distribution of distances to the target stars. Right panel: Distribution of heights above or below the Galactic plane.}
\end{figure}

\subsection{Processing of the STIS Data}
The relevant STIS data for our primary sample of interstellar sight lines were processed in a manner analogous to that described in Section 2.2 of R11. All high-resolution (E140H and E230H) and medium-resolution (E140M and E230M) echelle observations of our targets covering lines of interest to our survey were downloaded from the Mikulski Archive for Space Telescopes (MAST). The STIS data sets used in this investigation are listed in Table 2, which also provides exposure times and other details concerning the observations. The high-resolution gratings of the STIS FUV and NUV MAMA detectors provide spectra at resolving powers that range from $R=82,000$ to 143,000 depending on the aperture. With the medium resolution gratings, resolving powers between 38,000 and 46,000 may be achieved (see, e.g., Table 2 of R11). Multiple exposures of a target acquired with the same echelle grating were co-added to improve the signal-to-noise (S/N) ratio in the final spectrum. When exposures that employed different apertures were co-added, the data were resampled at the dispersion of the lowest-resolution spectrum contributing to the sum. If a feature of interest appeared in adjacent echelle orders with sufficient continua on both sides of the line, then the overlapping portions of the two orders were averaged together (yielding an increase in the S/N ratio of approximately 40\% in most cases). Typical S/N ratios for this sample range from 20 to 100 (per pixel), although some sight lines have significantly higher values. (The highest S/N ratios are seen in the STIS spectra of X~Per, for which S/N~$\approx$~300 near 1400~\AA{} in the final combined spectrum.)

Essential information regarding the interstellar lines of interest to this investigation is given in Table 3. (We analyze the O~{\sc i}~$\lambda1355$ line in addition to those from $n$-capture elements because this feature is useful for defining the overall component structure along a particular line of sight. We further note that the Ge~{\sc ii}~$\lambda1602$ and Kr~{\sc i}~$\lambda1164$ lines were analyzed only for the line of sight to X~Per.) Small spectral segments centered on these lines were cut from the final co-added spectra. These segments were typically 2 \AA{} wide, although it was occasionally necessary to choose a smaller portion of the spectrum to analyze so as to avoid a strong stellar absorption line or an unrelated interstellar feature. (In cases where adjacent echelle orders were averaged together, only the overlapping portions of the spectra were retained.) Each of the individual segments was then normalized to the continuum by fitting a low-order polynomial to regions free of interstellar absorption. We normalized all of the absorption profiles for a given line of sight concurrently so that the profiles from stronger lines, such as O~{\sc i}~$\lambda1355$ and Ge~{\sc ii}~$\lambda1237$, could serve as guides for properly fitting the continua around the weaker absorption features.

In most instances, the process of normalizing the continuum proceded in a relatively straightforward manner. However, in the case of Sn~{\sc ii}~$\lambda1400$ toward HD~137595, there are upward deviations in the intensities on either side of an apparent absorption feature, which is at a velocity close to that expected for Sn~{\sc ii}. An unusually high order polynomial would be required to ``smooth out'' the continuum surrounding this absorption feature. Upon further examination of the overlapping echelle orders that cover the Sn~{\sc ii} transition toward this star, we find that one of these upward deviations is quite strong in one order and not seen in the other, indicating that the feature may be a cosmic ray hit or of instrumental origin. Since we cannot be certain that the absorption feature is unaffected by these upward deviations, we report only an upper limit on the column density of Sn~{\sc ii} toward HD~137595. In the case of As~{\sc ii}~$\lambda1263$ toward CPD$-$59~2603, a high negative velocity component of Si~{\sc ii}*~$\lambda1264$ overlaps with the expected position of the As~{\sc ii} line, preventing us from deriving even an upper limit on the column density of As~{\sc ii} in this direction. Finally, while there are medium-resolution (E230M) data that cover the Cd~{\sc ii}~$\lambda\lambda2145,2265$ lines toward 40~Per, strong and relatively narrow stellar features prevent us from obtaining a reliable Cd~{\sc ii} column density (or upper limit) for this sight line.

\section{RESULTS ON COLUMN DENSITIES}
\subsection{Updated Oscillator Strengths}
Oscillator strengths for the transitions relevant to our investigation were generally obtained from the compilations of Morton (2000, 2003), with a few notable exceptions (see Table 3). Heidarian et al.~(2015) recently reported the first experimentally determined oscillator strengths for the Pb~{\sc ii} transitions at 1203.6~\AA{} and 1433.9~\AA{}. Their results, from beam-foil spectroscopy, indicate $f$-values of $0.75\pm0.03$ and $0.321\pm0.034$ for Pb~{\sc ii}~$\lambda1203$ and $\lambda1433$, respectively. Previous studies of interstellar Pb~{\sc ii} abundances (e.g., Welty et al.~1995) relied on theoretical determinations of the oscillator strength of the Pb~{\sc ii}~$\lambda1433$ transition (Migdalek 1976; Cardelli et al.~1993), which Morton (2000) gives as 0.869. The new experimental $f$-value for Pb~{\sc ii}~$\lambda1433$ provided by Heidarian et al.~(2015), which we adopt here, is lower than the theoretical value by almost a factor of 3, yielding an increase in the Pb~{\sc ii} column density derived from the $\lambda1433$ line of 0.43 dex.

The oscillator strength for the Ge~{\sc ii} transition at 1237.1~\AA{} had likewise not been determined experimentally until recently. Heidarian et al.~(2017), again using beam-foil spectroscopy, report an $f$-value for Ge~{\sc ii}~$\lambda1237$ of $0.872\pm0.113$. Early studies of gas-phase Ge~{\sc ii} abundances in the ISM (e.g., Savage et al.~1992; Hobbs et al.~1993) relied on the oscillator strength given in Morton (1991) for the Ge~{\sc ii}~$\lambda1237$ transition.\footnote{A typographical error in Morton (1991) was subsequently corrected by Savage et al.~(1992). The corrected $f$-value from Morton (1991) for Ge~{\sc ii}~$\lambda1237$ is 0.8756.} Later studies of interstellar Ge~{\sc ii} (e.g., Welty et al.~1999; Cartledge et al.~2006) adopted the theoretical $f$-value for the $\lambda1237$ transition calculated by Bi{\'e}mont et al.~(1998) of 1.23. This is also the value listed by Morton (2000) for this transition. The new experimental $f$-value for Ge~{\sc ii}~$\lambda1237$ reported by Heidarian et al.~(2017), and adopted in this work, is in better agreement with the earlier theoretical value given in Morton (1991). (While the $f$-value for the Ge~{\sc ii} transition at 1602.5~\AA{} has not been determined experimentally, the column density we obtain from this line toward X~Per agrees with the value we derive from the $\lambda1237$ line at the 1$\sigma$ level.)

The only other $n$-capture element transition listed in Table 3 that does not have an experimentally determined $f$-value is As~{\sc ii}~$\lambda1263$. However, for this transition, most theoretical determinations agree with the value calculated by Bi{\'e}mont et al.~(1998), and adopted by Morton (2000), at the 10--20\% level (e.g., Bieron et al.~1991; Cardelli et al.~1993; Brage \& Leckrone 1995). Nevertheless, an experimental confirmation of the oscillator strength for As~{\sc ii}~$\lambda1263$ would be useful.

\begin{figure}
\centering
\includegraphics[width=1.0\textwidth]{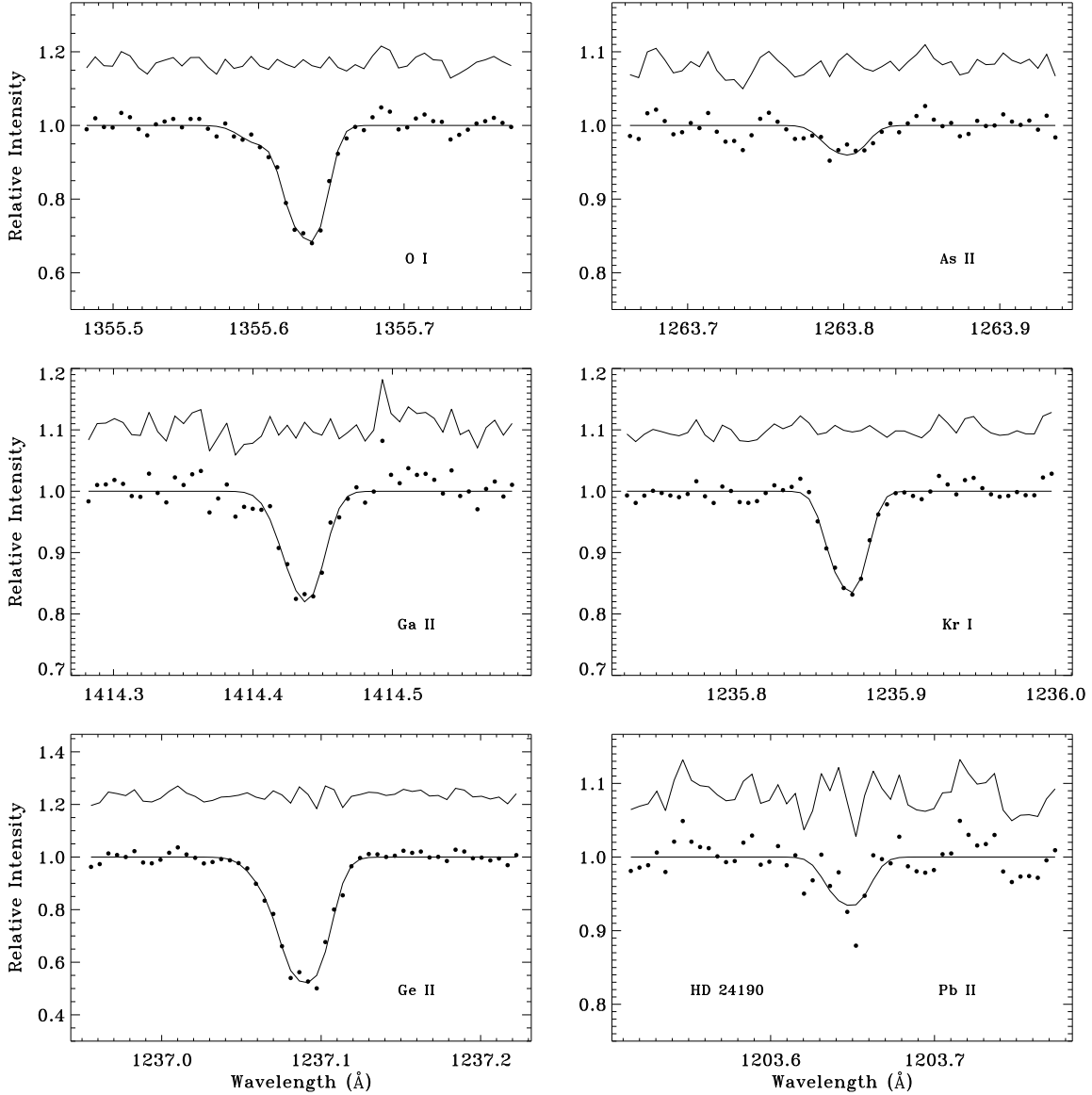}
\caption{Profile synthesis fits to the O~{\sc i}~$\lambda1355$, Ga~{\sc ii}~$\lambda1414$, Ge~{\sc ii}~$\lambda1237$, As~{\sc ii}~$\lambda1263$, Kr~{\sc i}~$\lambda1235$, and Pb~{\sc ii}~$\lambda1203$ lines toward HD~24190. Wavelengths are plotted in the local standard of rest (LSR) frame. The same range in velocity is displayed in each panel. The synthetic profiles are shown as solid lines passing through data points that represent the observed spectra. The fit residuals are plotted above each spectrum. A profile template based on O~{\sc i} was adopted in fitting the As~{\sc ii} and Pb~{\sc ii} lines.}
\end{figure}

\begin{figure}
\centering
\includegraphics[width=1.0\textwidth]{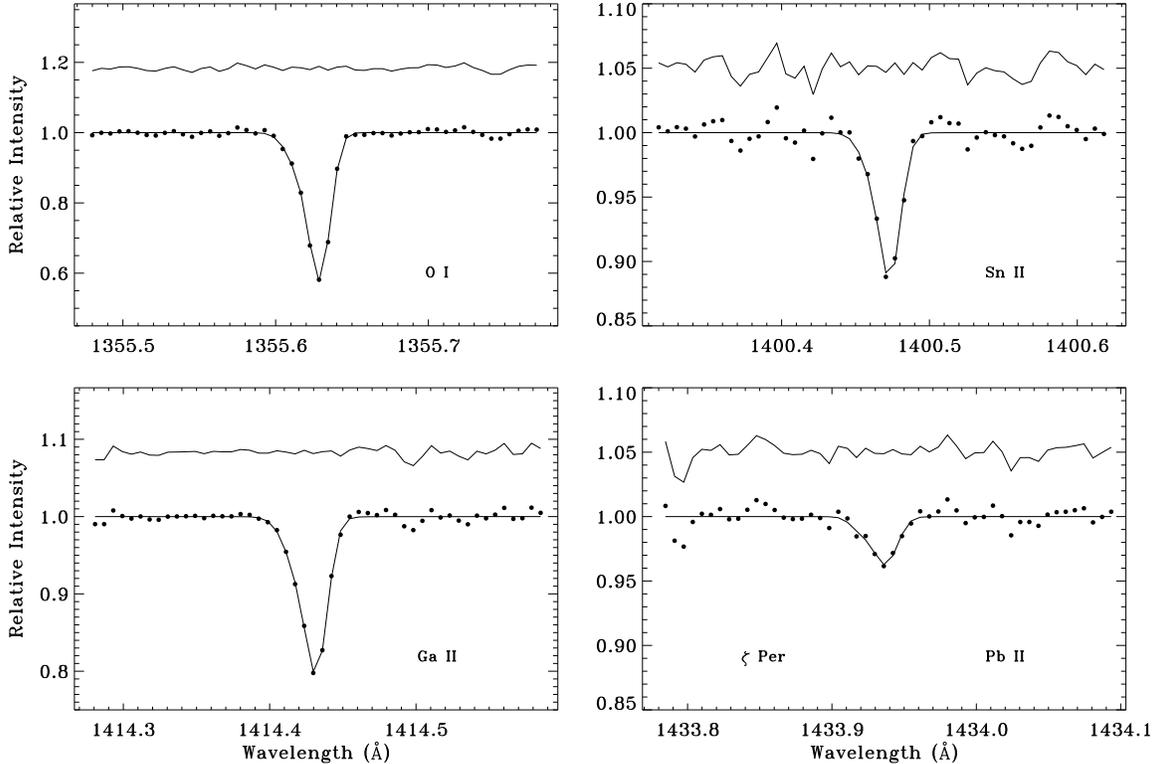}
\caption{Same as Figure 2 except for the O~{\sc i}~$\lambda1355$, Ga~{\sc ii}~$\lambda1414$, Sn~{\sc ii}~$\lambda1400$, and Pb~{\sc ii}~$\lambda1433$ lines toward $\zeta$~Per.}
\end{figure}

\begin{figure}
\centering
\includegraphics[width=1.0\textwidth]{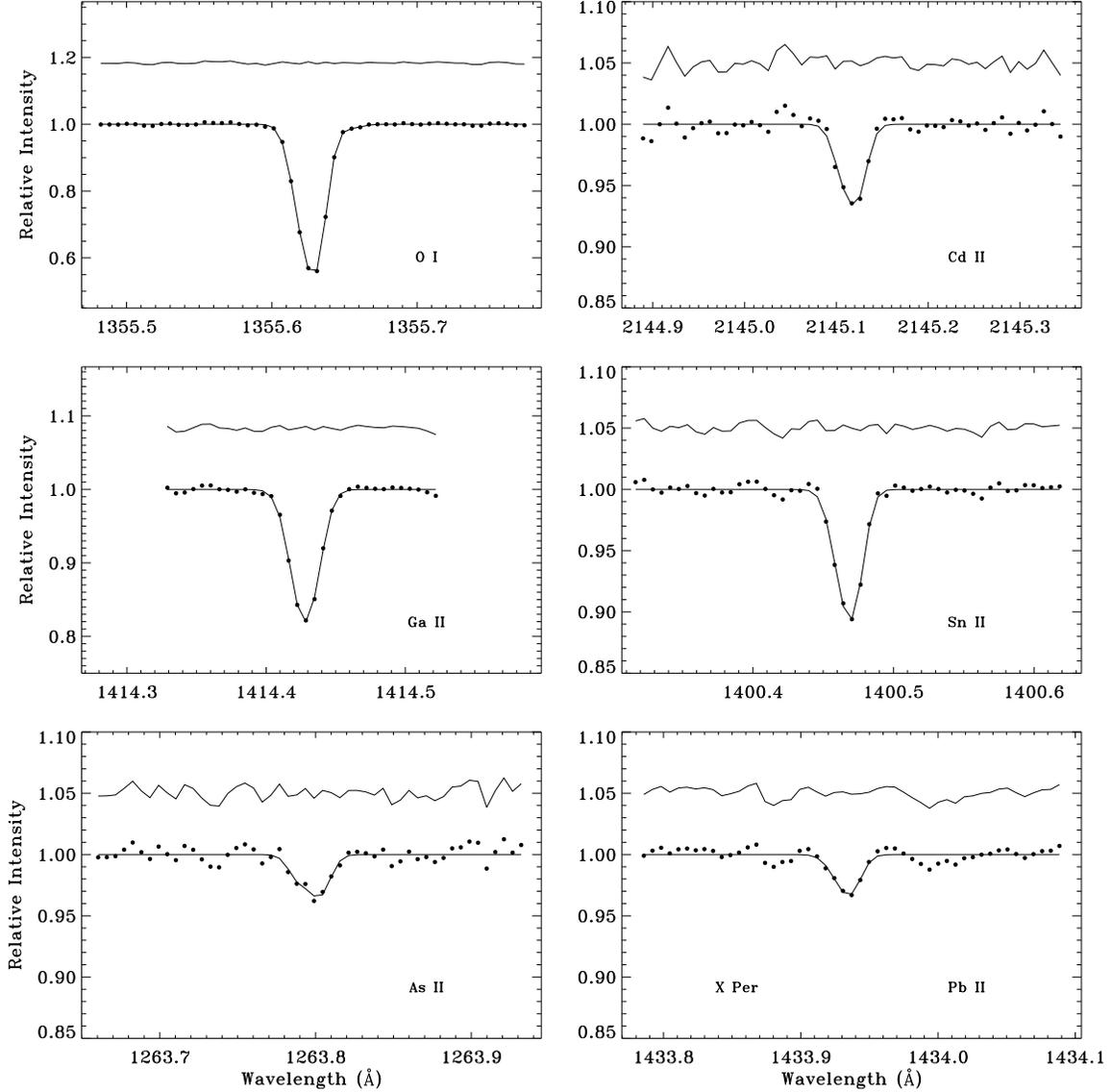}
\caption{Same as Figure 2 except for the O~{\sc i}~$\lambda1355$, Ga~{\sc ii}~$\lambda1414$, As~{\sc ii}~$\lambda1263$, Cd~{\sc ii}~$\lambda2145$, Sn~{\sc ii}~$\lambda1400$, and Pb~{\sc ii}~$\lambda1433$ lines toward X~Per.}
\end{figure}

\begin{figure}
\centering
\includegraphics[width=1.0\textwidth]{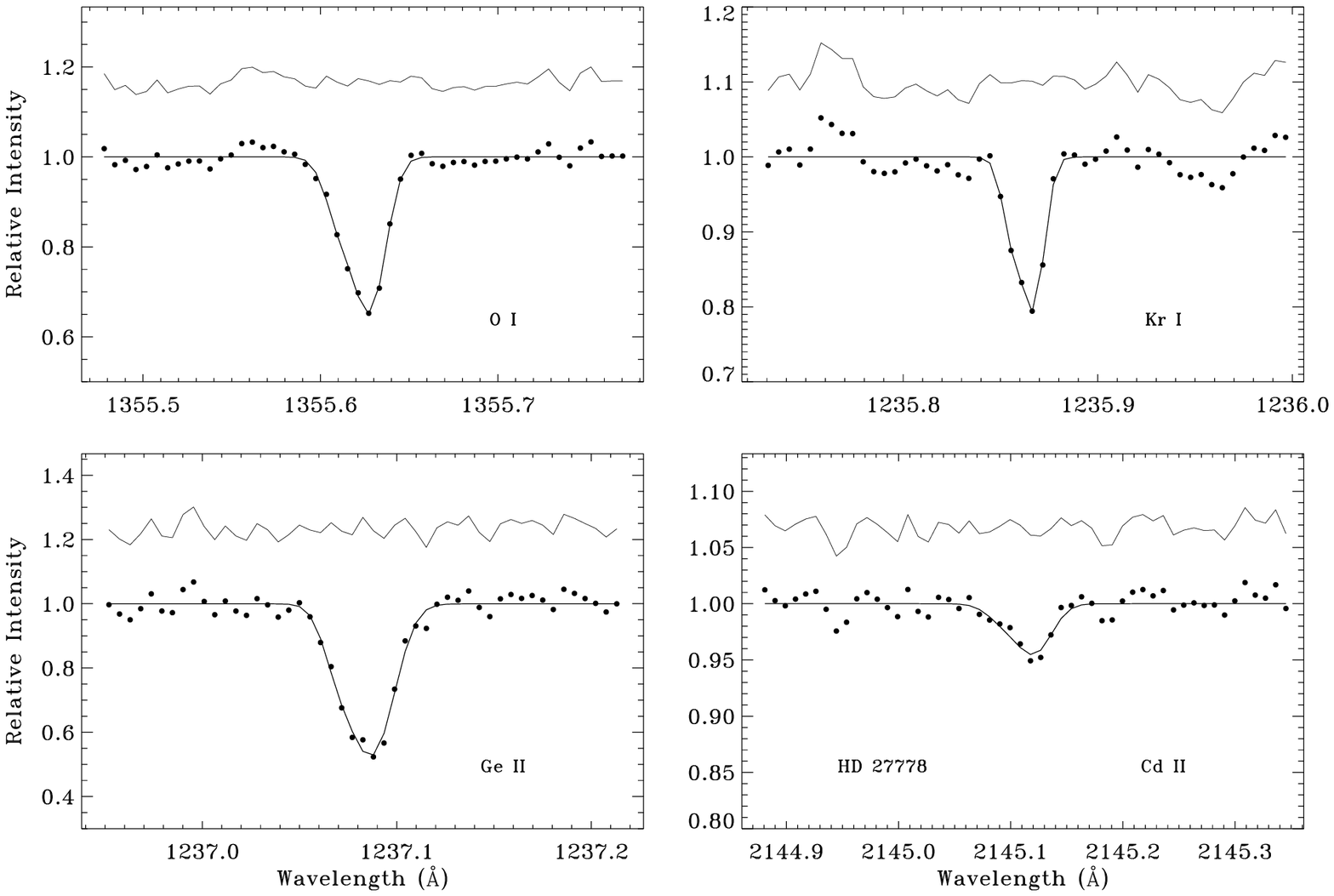}
\caption{Same as Figure 2 except for the O~{\sc i}~$\lambda1355$, Ge~{\sc ii}~$\lambda1237$, Kr~{\sc i}~$\lambda1235$, and Cd~{\sc ii}~$\lambda2145$ lines toward HD~27778. A template based on O~{\sc i} was adopted in fitting the Cd~{\sc ii} line.}
\end{figure}

\begin{figure}
\centering
\includegraphics[width=1.0\textwidth]{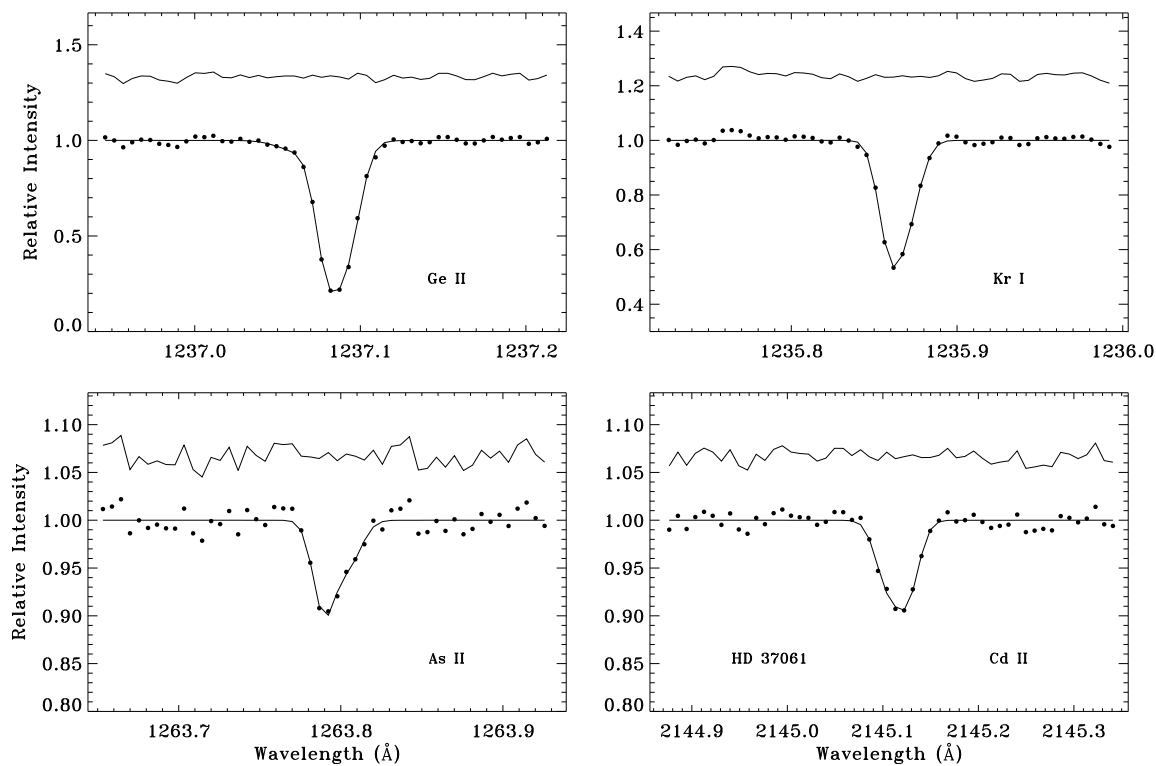}
\caption{Same as Figure 2 except for the Ge~{\sc ii}~$\lambda1237$, As~{\sc ii}~$\lambda1263$, Kr~{\sc i}~$\lambda1235$, and Cd~{\sc ii}~$\lambda2145$ lines toward HD~37061.}
\end{figure}

\begin{figure}
\centering
\includegraphics[width=1.0\textwidth]{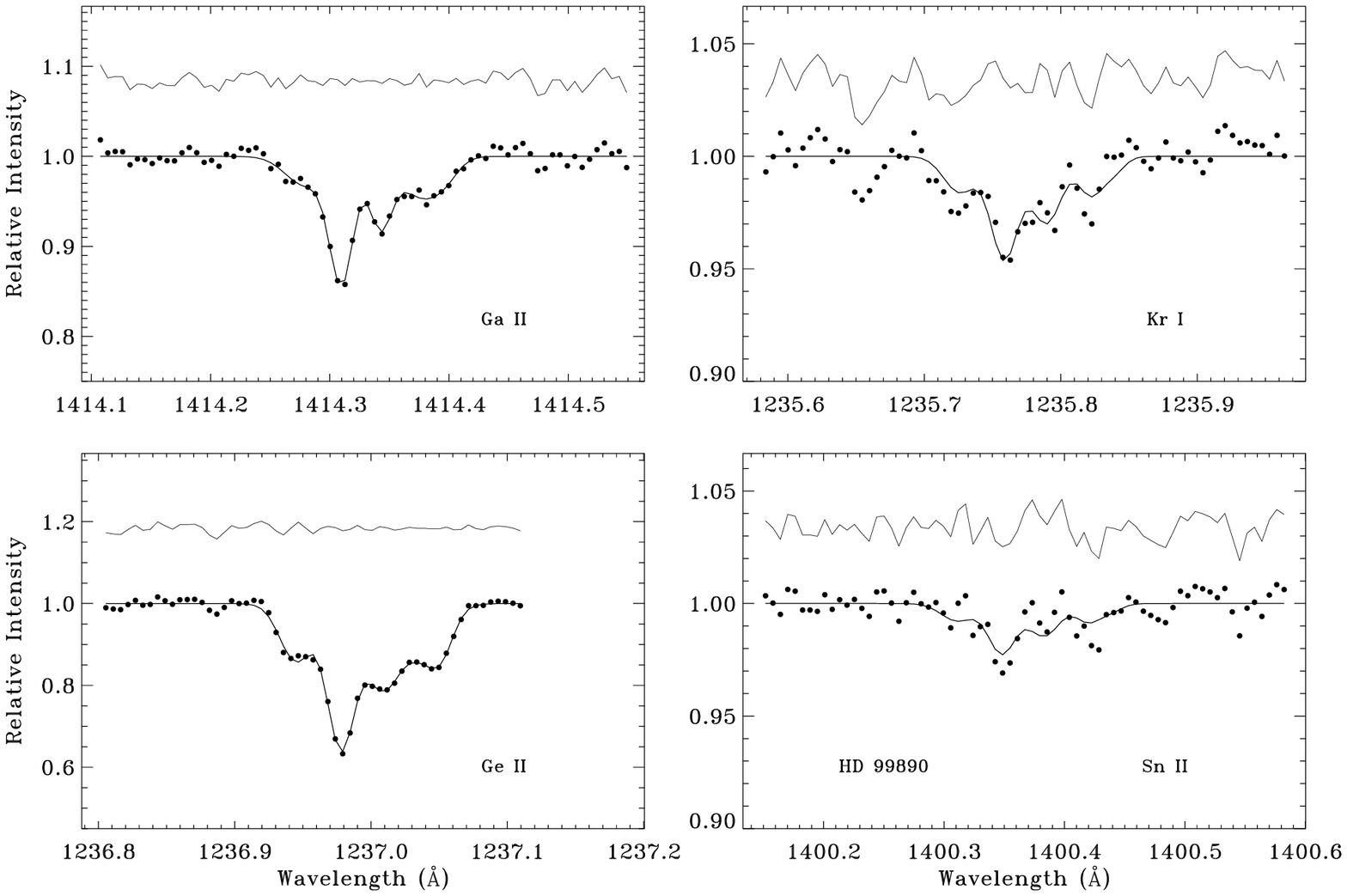}
\caption{Same as Figure 2 except for the Ga~{\sc ii}~$\lambda1414$, Ge~{\sc ii}~$\lambda1237$, Kr~{\sc i}~$\lambda1235$, and Sn~{\sc ii}~$\lambda1400$ lines toward HD~99890. A template based on O~{\sc i}~$\lambda1355$ (not shown) was adopted in fitting the Kr~{\sc i} and Sn~{\sc ii} lines.}
\end{figure}

\begin{figure}
\centering
\includegraphics[width=1.0\textwidth]{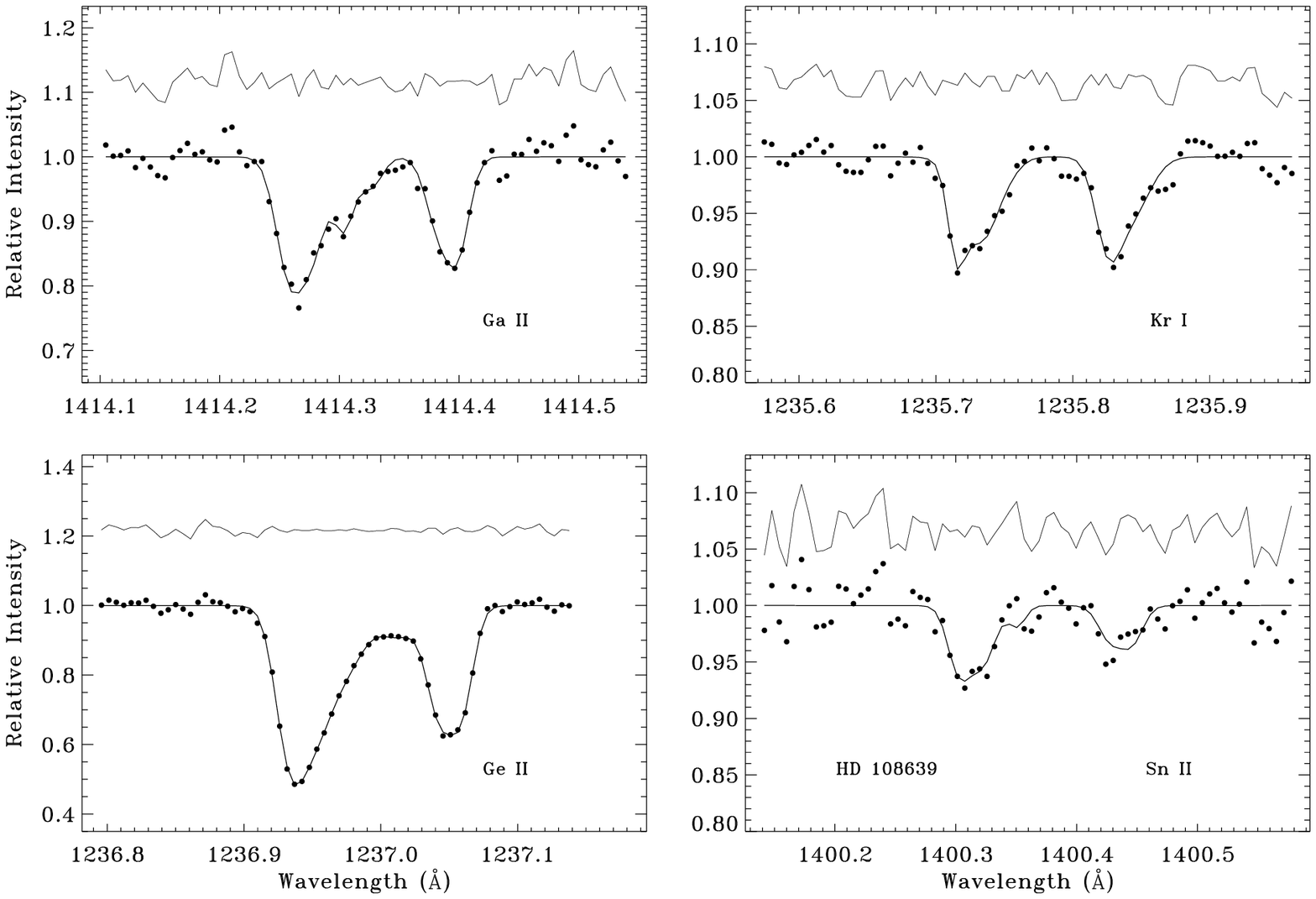}
\caption{Same as Figure 2 except for the Ga~{\sc ii}~$\lambda1414$, Ge~{\sc ii}~$\lambda1237$, Kr~{\sc i}~$\lambda1235$, and Sn~{\sc ii}~$\lambda1400$ lines toward HD~108639. The two absorption complexes apparent in the Sn~{\sc ii} profile were fit independently adopting templates based on  O~{\sc i}~$\lambda1355$ (not shown).}
\end{figure}

\begin{figure}
\centering
\includegraphics[width=1.0\textwidth]{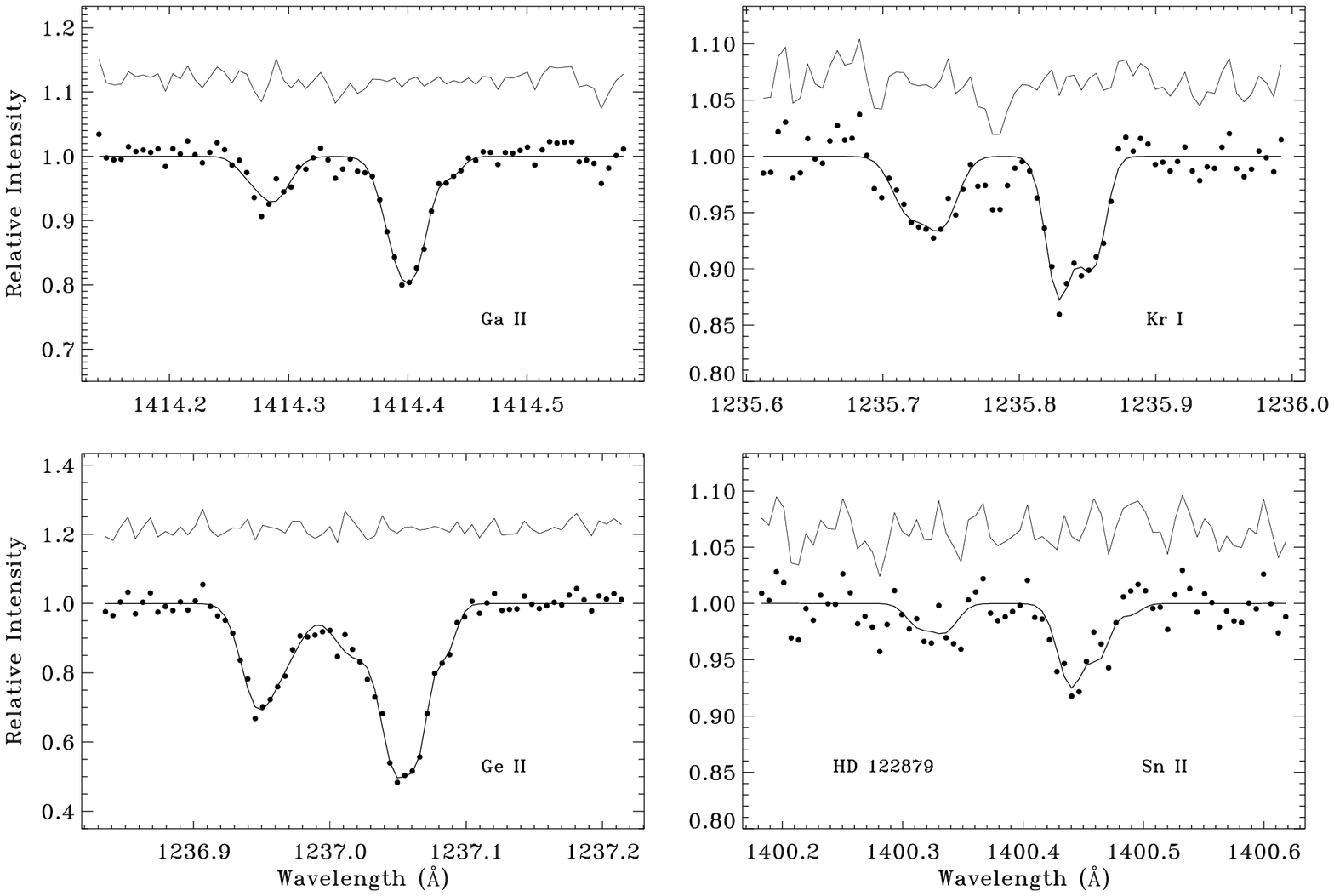}
\caption{Same as Figure 2 except for the Ga~{\sc ii}~$\lambda1414$, Ge~{\sc ii}~$\lambda1237$, Kr~{\sc i}~$\lambda1235$, and Sn~{\sc ii}~$\lambda1400$ lines toward HD~122879. The two absorption complexes apparent in the Sn~{\sc ii} profile were fit independently adopting templates based on  O~{\sc i}~$\lambda1355$ (not shown).}
\end{figure}

\begin{figure}
\centering
\includegraphics[width=1.0\textwidth]{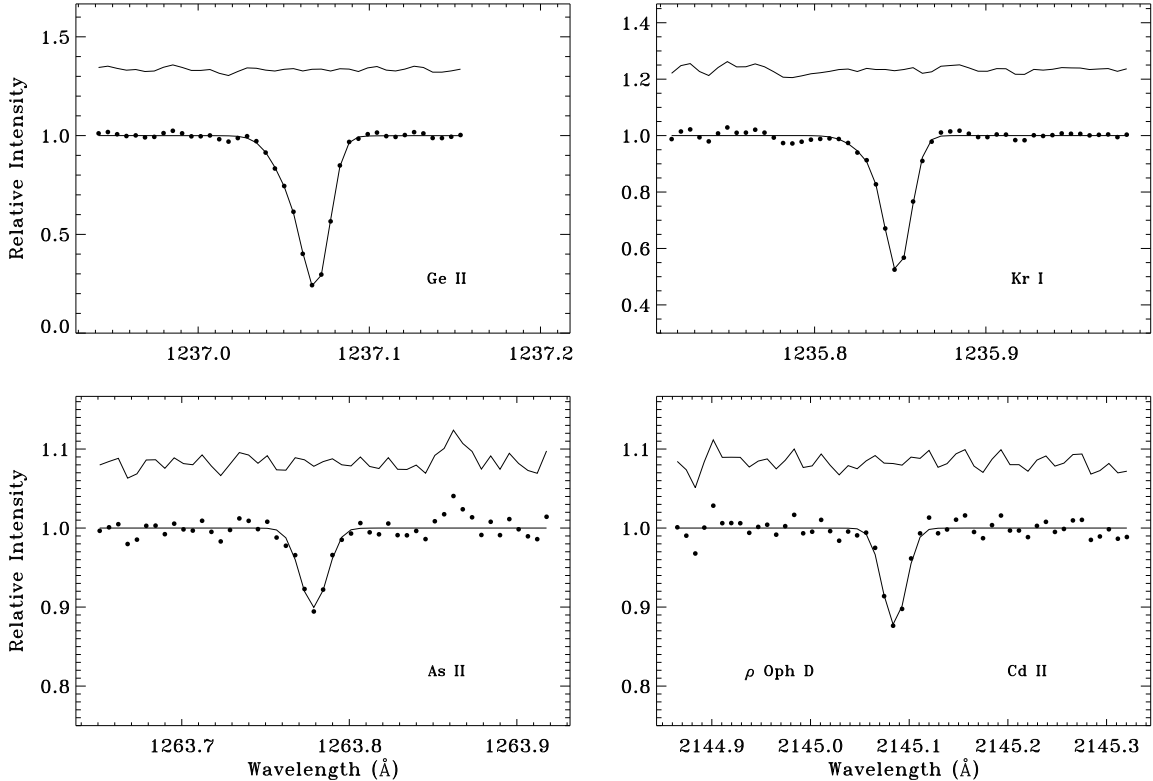}
\caption{Same as Figure 2 except for the Ge~{\sc ii}~$\lambda1237$, As~{\sc ii}~$\lambda1263$, Kr~{\sc i}~$\lambda1235$, and Cd~{\sc ii}~$\lambda2145$ lines toward $\rho$ Oph D.}
\end{figure}

\begin{figure}
\centering
\includegraphics[width=1.0\textwidth]{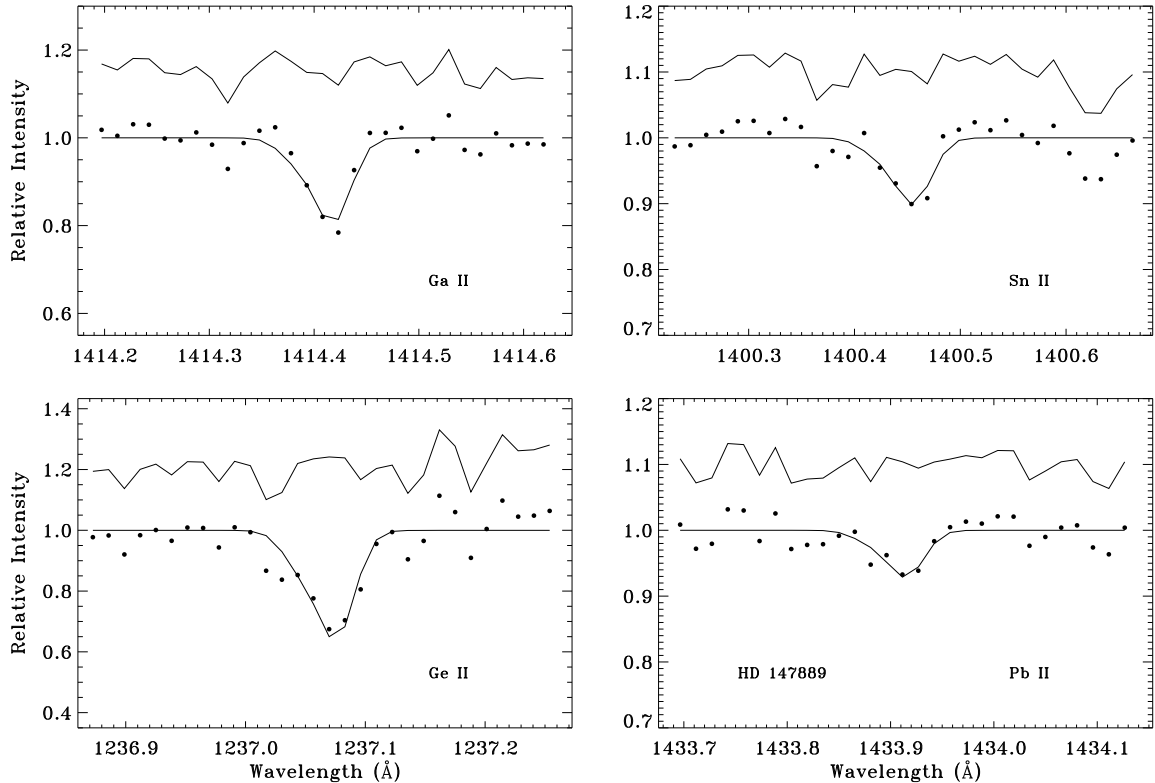}
\caption{Same as Figure 2 except for the Ga~{\sc ii}~$\lambda1414$, Ge~{\sc ii}~$\lambda1237$, Sn~{\sc ii}~$\lambda1400$, and Pb~{\sc ii}~$\lambda1433$ lines toward HD~147889. These data were acquired at medium resolution. A profile template based on O~{\sc i}~$\lambda1355$ (not shown) was adopted in each case.}
\end{figure}

\begin{figure}
\centering
\includegraphics[width=1.0\textwidth]{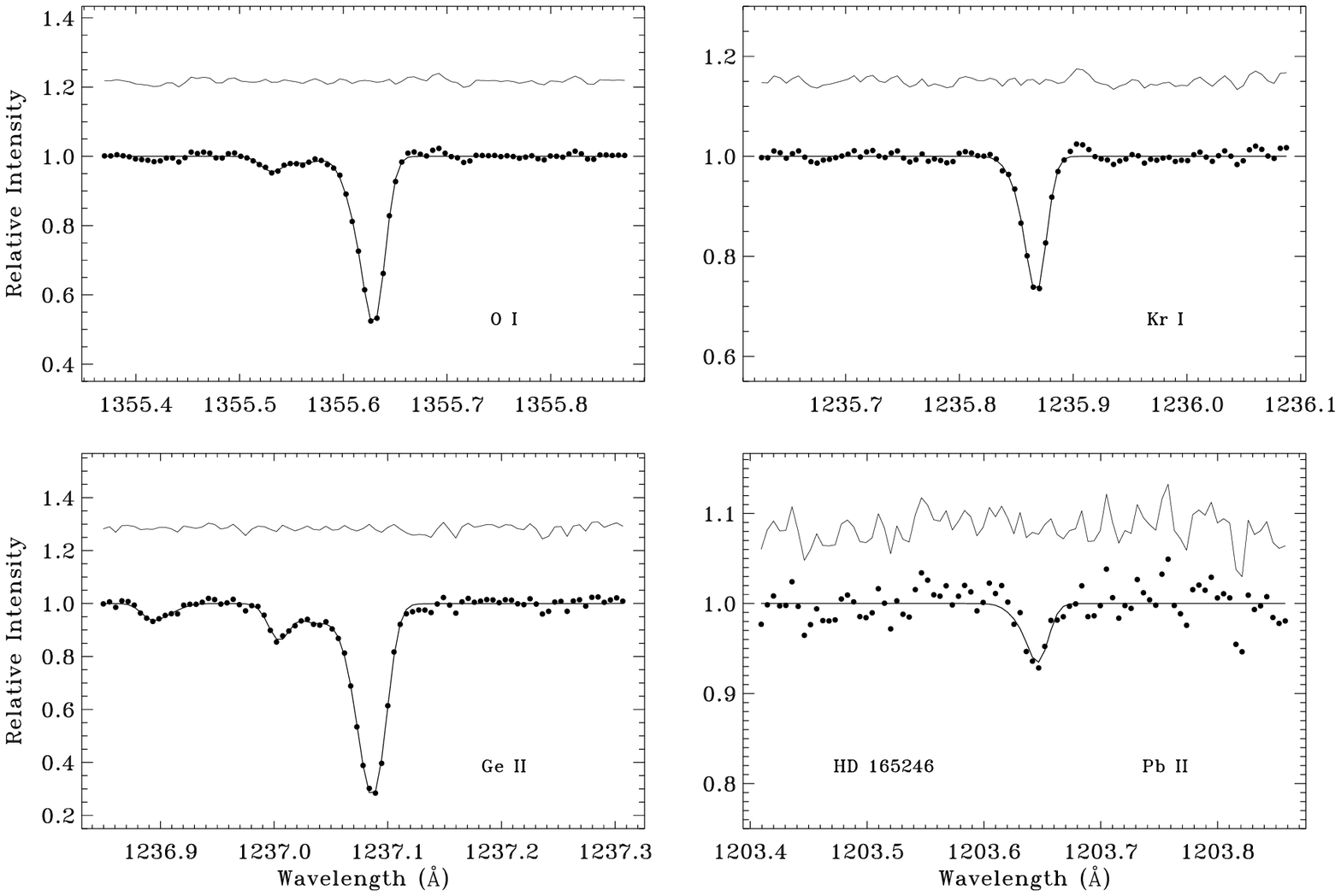}
\caption{Same as Figure 2 except for the O~{\sc i}~$\lambda1355$, Ge~{\sc ii}~$\lambda1237$, Kr~{\sc i}~$\lambda1235$, and Pb~{\sc ii}~$\lambda1203$ lines toward HD~165246. A template based on O~{\sc i} was adopted in fitting the Pb~{\sc ii} line.}
\end{figure}

\begin{figure}
\centering
\includegraphics[width=1.0\textwidth]{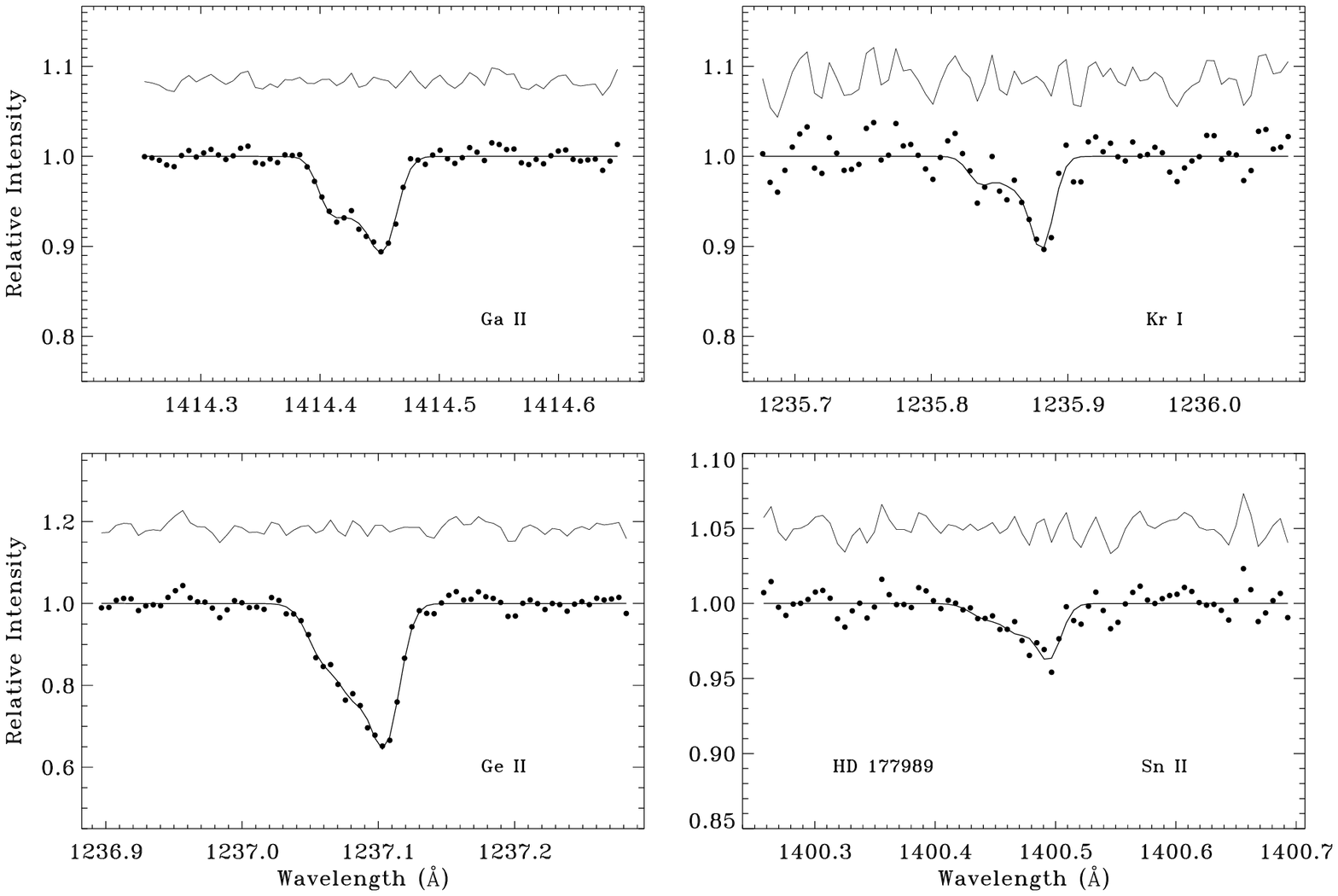}
\caption{Same as Figure 2 except for the Ga~{\sc ii}~$\lambda1414$, Ge~{\sc ii}~$\lambda1237$, Kr~{\sc i}~$\lambda1235$, and Sn~{\sc ii}~$\lambda1400$ lines toward HD~177989. A template based on O~{\sc i}~$\lambda1355$ (not shown) was adopted in fitting the Sn~{\sc ii} line.}
\end{figure}

\begin{figure}
\centering
\includegraphics[width=1.0\textwidth]{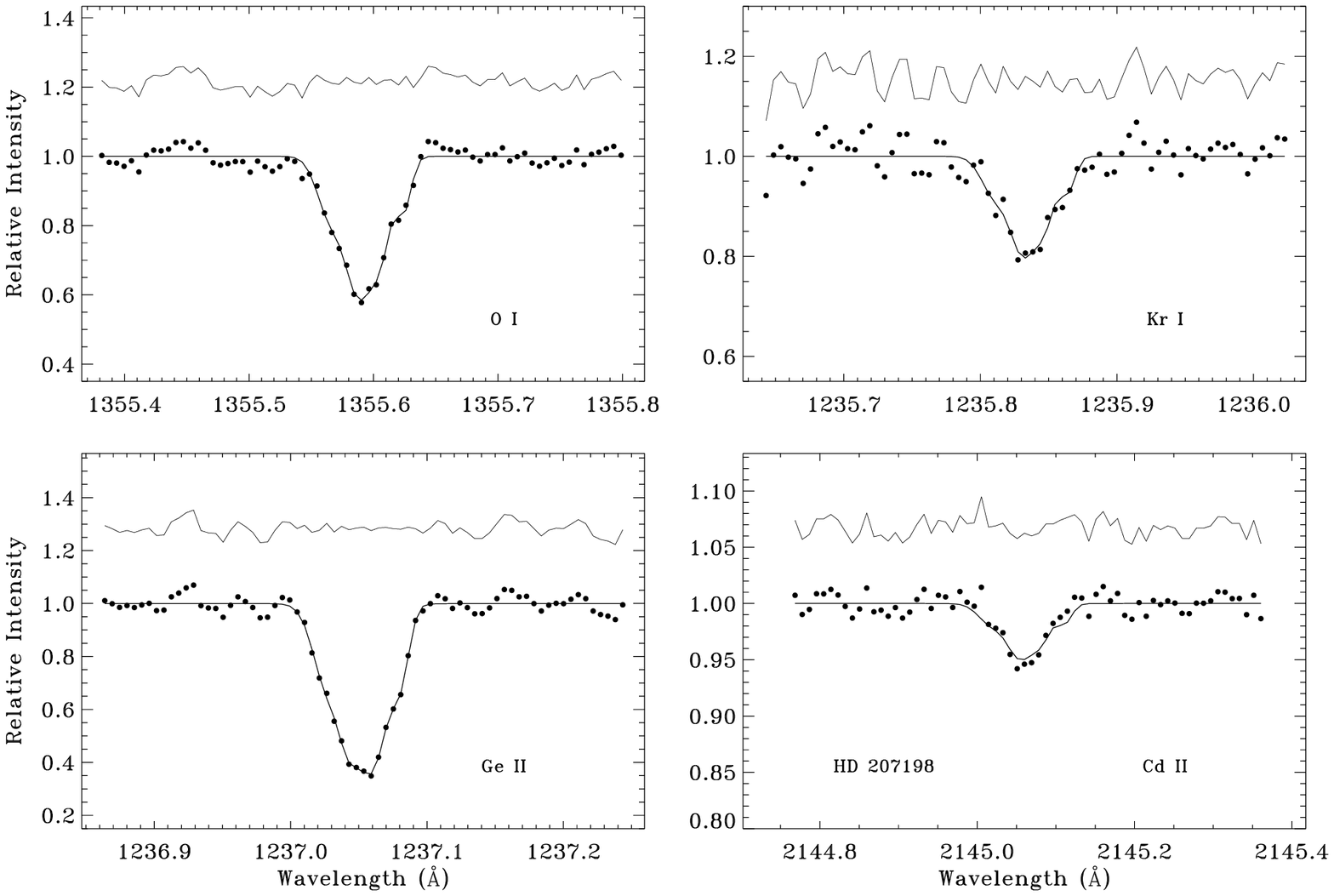}
\caption{Same as Figure 2 except for the O~{\sc i}~$\lambda1355$, Ge~{\sc ii}~$\lambda1237$, Kr~{\sc i}~$\lambda1235$, and Cd~{\sc ii}~$\lambda2145$ lines toward HD~207198. A template based on O~{\sc i} was adopted in fitting the Kr~{\sc i} and Cd~{\sc ii} lines.}
\end{figure}

\begin{figure}
\centering
\includegraphics[width=1.0\textwidth]{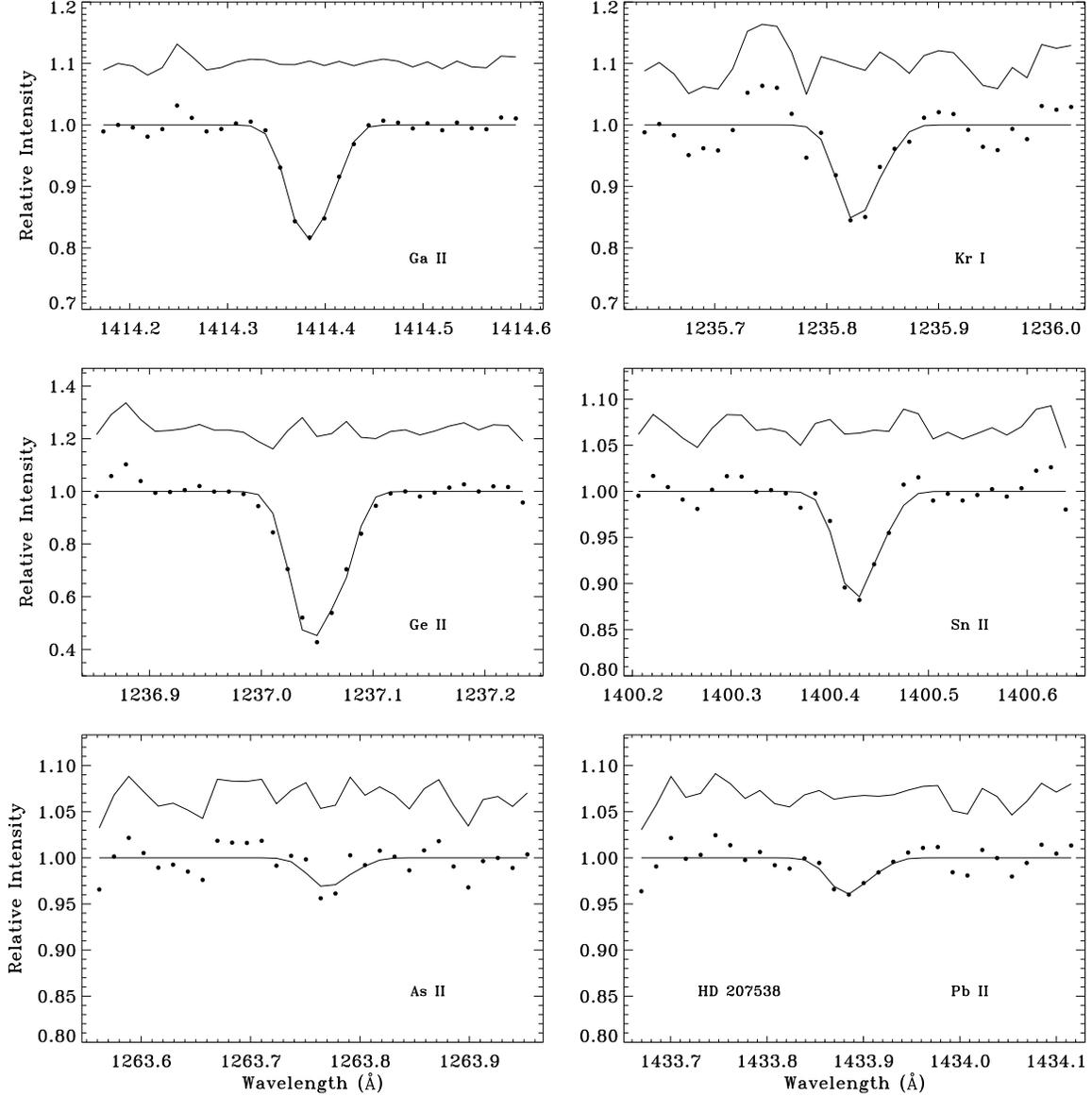}
\caption{Same as Figure 2 except for the Ga~{\sc ii}~$\lambda1414$, Ge~{\sc ii}~$\lambda1237$, As~{\sc ii}~$\lambda1263$, Kr~{\sc i}~$\lambda1235$, Sn~{\sc ii}~$\lambda1400$, and Pb~{\sc ii}~$\lambda1433$ lines toward HD~207538. These data were acquired at medium resolution.  A template based on O~{\sc i}~$\lambda1355$ (not shown) was adopted in fitting the As~{\sc ii}, Kr~{\sc i}, Sn~{\sc ii}, and Pb~{\sc ii} lines.}
\end{figure}

\begin{figure}
\centering
\includegraphics[width=1.0\textwidth]{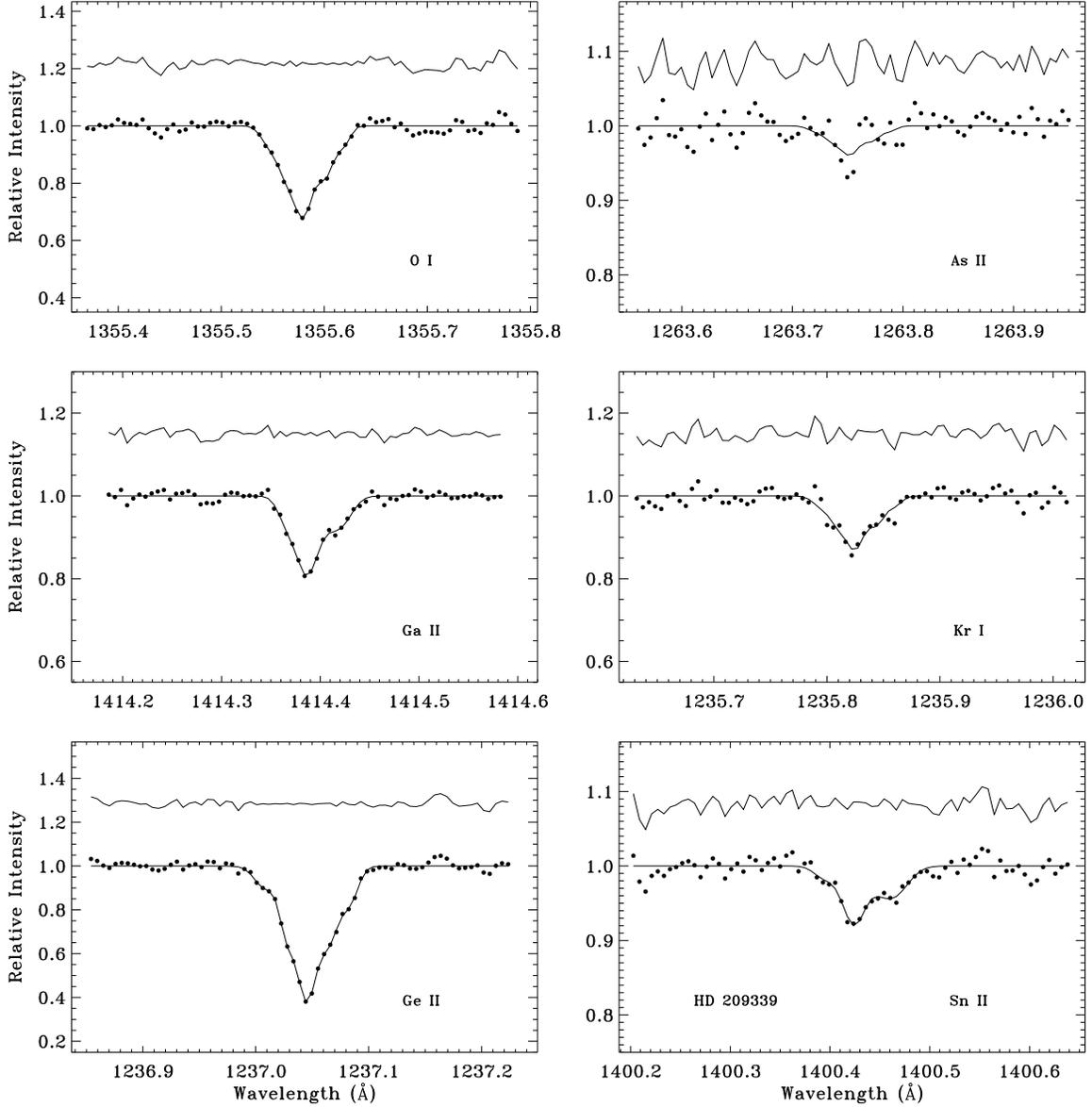}
\caption{Same as Figure 2 except for the O~{\sc i}~$\lambda1355$, Ga~{\sc ii}~$\lambda1414$, Ge~{\sc ii}~$\lambda1237$, As~{\sc ii}~$\lambda1263$, Kr~{\sc i}~$\lambda1235$, and Sn~{\sc ii}~$\lambda1400$ lines toward HD~209339. A template based on O~{\sc i} was adopted in fitting the As~{\sc ii} and Kr~{\sc i} lines.}
\end{figure}

\begin{figure}
\centering
\includegraphics[width=1.0\textwidth]{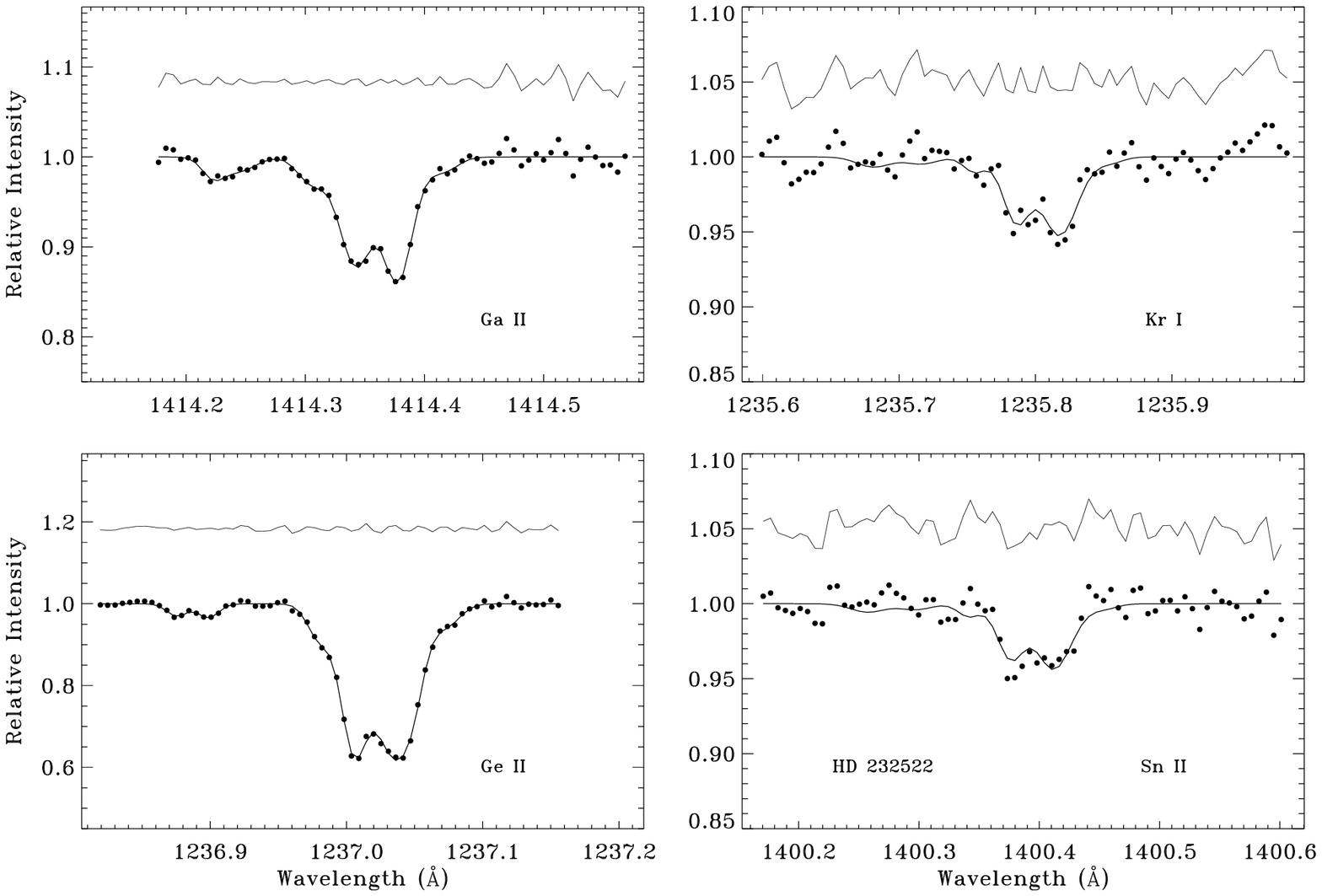}
\caption{Same as Figure 2 except for the Ga~{\sc ii}~$\lambda1414$, Ge~{\sc ii}~$\lambda1237$, Kr~{\sc i}~$\lambda1235$, and Sn~{\sc ii}~$\lambda1400$ lines toward HD~232522. A template based on O~{\sc i}~$\lambda1355$ (not shown) was adopted in fitting the Kr~{\sc i} and Sn~{\sc ii} lines.}
\end{figure}

\subsection{Profile Synthesis Fits}
Column densities were derived through multi-component profile synthesis fits to the normalized spectra, using the code ISMOD (see Sheffer et al.~2008). The profile fitting routine treats the column densities, $b$-values, and velocities of the individual components as free parameters, while minimizing the rms of the fit residuals. For a given sight line, we started by analyzing the O~{\sc i}~$\lambda1355$ profile so that we could develop a basic understanding of the component structure in that direction. Many of the sight lines in our primary sample already had detailed O~{\sc i} (and Ga~{\sc ii}) component structures from R11. For the others, we sought to produce similarly high quality component decompositions following the same basic procedures as in R11. In most cases, direct fits to the O~{\sc i}~$\lambda1355$ profiles yielded good constraints on the velocities and $b$-values of the individual components, particularly when high-resolution (E140H) data were being analyzed. When analyzing medium-resolution (E140M) spectra, it was occasionally necessary to impose certain restrictions on the fits, such as requiring the $b$-values to fall within a particular range as determined either directly from the data or from other species observed at higher resolution.\footnote{A significant fraction of the sight lines in R11 have auxiliary ground-based data on Ca~{\sc ii} and K~{\sc i} obtained at high resolution ($R>150,000$). When available, those species offered an additional means of constraining the line-of-sight component structure.}

Once a suitable component decomposition of the O~{\sc i}~$\lambda1355$ profile was in hand, we proceeded to fit the profiles of the other species of interest that were available for that particular sight line. We began by adopting the O~{\sc i} component structure (i.e., the fractional column densities, relative velocities, and $b$-values of the O~{\sc i} components) as input parameters to the profile fitting routine, allowing all of the parameters to vary freely. However, due to the weakness of many of the lines of interest (particularly those of As~{\sc ii}, Sn~{\sc ii}, and Pb~{\sc ii}), it was often necessary to constrain the fits in some way so as to achieve realistic results while mitigating the effects of noise in the profiles. In these cases, depending on the quality of the data, we would either fix the relative velocities of the components, allowing the individual column densities and $b$-values to vary, or we would fix the fractional column densities, relative velocities, and $b$-values, allowing only the total column density and the absolute velocity of the profile template to vary freely. When multiple absorption complexes were evident in the profile, a template was created for each complex and fitted to that portion of the profile independently. A similar approach was adopted by R11, who demonstrated the reliability and effectiveness of using profile templates to fit weak interstellar features. For the present results, one can judge the degree to which the templates match the observed spectra by examining the examples presented in Figures~2 through 17. (In Table 4, we present the final component structures obtained for O~{\sc i}, Ga~{\sc ii}, and Ge~{\sc ii} for sight lines not analyzed by R11. The table also includes component results for HD~99890 derived from high-resolution spectra newly analyzed in this work.)

Table 5 presents the total (line-of-sight) column densities of O~{\sc i}, Ga~{\sc ii}, Ge~{\sc ii}, As~{\sc ii}, Kr~{\sc i}, Cd~{\sc ii}, Sn~{\sc ii}, and Pb~{\sc ii} for the 69 sight lines in our primary sample. (A compilation of column density measurements from the literature for an additional 59 sight lines with STIS or GHRS data is presented in Appendix A.) Uncertainties in total column density for the sight lines in our primary sample were calculated as the quadrature sums of the uncertainties in the column densities of the individual components contributing to the absorption profiles. For the individual components, the column density uncertainties are proportional to the corresponding equivalent width uncertainties, which are based on the widths of the components (i.e., the full widths at half maximum) and the rms variations in the continua surrounding the profiles. This approach of basing the column density uncertainties on equivalent width uncertainties is justified since the majority of absorption lines in our survey fall on the linear portion of the curve of growth. The Ge~{\sc ii}~$\lambda1237$ lines, however, are stronger than any of the other lines we examine, and may be somewhat saturated in many cases. We therefore added (in quadrature) an extra amount of uncertainty to the error for Ge~{\sc ii} based on an estimate of the degree of saturation in the line profile. To estimate the degree of saturation, we considered the difference between the fitted column density and the column density one would obtain under the assumption that the line is optically thin. Our method is equivalent to varying the effective $b$-value by about 10\% using a traditional curve-of-growth analysis. (This process increased the uncertainty in the Ge~{\sc ii} column density by only 0.01 dex, on average. However, for the more heavily saturated sight lines, the increase in uncertainty was as high as 0.04 dex.) For all non-detections, and cases where the derived column density was less than twice the calculated uncertainty, we determined 3$\sigma$ upper limits by calculating what the error in column density would be if the undetected line had a component structure identical to that of O~{\sc i}~$\lambda1355$.

In situations where we had both high-resolution and medium-resolution spectra covering a particular absorption feature, we analyzed the two absorption profiles separately as a check on consistency. In most of these cases, the column densities derived from the high and medium-resolution profiles agree with each other at the 1$\sigma$ level or better. For HD~1383 and HD~210809, the Ge~{\sc ii} column densities from high and medium-resolution data covering the $\lambda1237$ line agree at about the 2$\sigma$ level. These small discrepancies are likely the result of differences in continuum placement since, in both cases, the absorption profiles have multiple complexes that span a large range in velocity (55 to 75 km~s$^{-1}$), and the medium-resolution profiles have rather low S/N values (S/N~$\sim$~25), making it difficult to judge the proper location of the continuum. That, even in these more difficult cases, the results agree at the 2$\sigma$ level suggests that our error estimates adequately account for uncertainties due to both noise and continuum placement. We also analyzed the Cd~{\sc ii}~$\lambda2145$ and $\lambda2265$ profiles separately (adopting the same component structure for both lines, determined either from the stronger Cd~{\sc ii} line or from O~{\sc i}~$\lambda1355$). For all seven sight lines with Cd~{\sc ii} detections, the column densities derived from the $\lambda2145$ and $\lambda2265$ lines agree at the 1$\sigma$ level or better. Final column densities in cases where both high and medium-resolution spectra yielded reliable determinations were obtained by taking the weighted mean of the two results. The same approach was used to derive final Cd~{\sc ii} column densities from our independent measurements of the $\lambda2145$ and $\lambda2265$ lines.

\subsection{Comparison with Previous Studies}
As noted earlier, many of the sight lines in our primary sample have been analyzed previously for the purposes of deriving column densities of Ge~{\sc ii}, Kr~{\sc i}, or both (e.g., Cartledge et al.~2003, 2006, 2008). A comparison between our determinations for Ge~{\sc ii} and Kr~{\sc i} for these sight lines and those from the literature is presented in Table 6. Overall, we find good agreement between the column densities we derive and those found in previous investigations, most of which relied on the same STIS data that we examine here. If we consider all of the Ge~{\sc ii} and Kr~{\sc i} comparisons together, we find that in 62\% of the cases the column densities agree at the 1$\sigma$ level or better, while in 96\% of the cases the determinations agree within 2$\sigma$. In only two cases do the column densities disagree by more than 2$\sigma$. One of these is the Ge~{\sc ii} column density toward HD~210809, which we find to be 0.24 dex lower than that given in Cartledge et al.~(2006).\footnote{The Ge~{\sc ii} column densities listed in Cartledge et al.~(2006) have been revised in Table 6 to reflect the new experimental $f$-value we adopt for Ge~{\sc ii}~$\lambda1237$ from Heidarian et al.~(2017).} The difference in this case does not appear to be related to any optical depth effects since the equivalent widths disagree by approximately the same amount as the column densities. Rather, the discrepancy seems to be related to data quality issues. In the time since Cartledge et al.~(2006) published their value, new high-resolution STIS spectra have been obtained for HD~210809 on two separate occasions (under GO programs 11737 and 12192). The new data, when co-added with the original spectrum, yield a factor of nine increase in exposure time for the region near Ge~{\sc ii}~$\lambda1237$, corresponding to a factor of three decrease in the uncertainty of the Ge~{\sc ii} equivalent width.

The other case in which our column density determination disagrees with a previously published value by more than 2$\sigma$ is the Kr~{\sc i} column density toward HD~152590. Our result is 0.15 dex lower than the value listed in Cartledge et al.~(2008). Here, again, the equivalent widths disagree by approximately the same amount as the column densities, indicating that the issue lies in the placement of the continuum. Indeed, as Cartledge et al.~(2008) also point out, the underlying stellar spectrum of HD~152590 exhibits an unusually steep curvature in the immediate vicinity of the interstellar Kr~{\sc i}~$\lambda1235$ feature. HD~152590 is one of two sight lines (along with HD~116852) that were previously identified by Cartledge et al.~(2003) as potentially having elevated Kr abundances. That conclusion for HD~152590 was based on the Kr~{\sc i} column density originally published by Cartledge et al.~(2001). In their subsequent work, Cartledge et al.~(2008) revised their column density determination for this sight line downward by 0.17 dex after acquiring additional STIS observations of HD~152590 that substantially increased the exposure time near Kr~{\sc i}~$\lambda1235$. With this revision, HD~152590 no longer appeared to exhibit a Kr abundance that deviated significantly from the mean interstellar value of log~(Kr/H)~$\approx$~$-$9.0 (Cartledge et al.~2008).\footnote{Cartledge et al.~(2008) acquired additional STIS spectra of HD~116852 as well, and used those observations to derive a Kr~{\sc i} column density that was 0.23 dex lower than the value given in Cartledge et al.~(2003). As with HD~152590, this revision eliminated the appearance of a Kr enhancement toward HD~116852.} Our determination for Kr~{\sc i} toward HD~152590 further indicates that the Kr abundance is likely not enhanced in this direction.

Two other comparisons between our results and those of Cartledge et al.~(2008) bear mentioning. The Kr~{\sc i} column densities we find toward HD~104705 and HDE~303308 are larger than the Cartledge et al.~(2008) values by 0.18 dex and 0.12 dex, respectively. While these are only $\sim$1$\sigma$ discrepancies (and thus not especially concerning), the reason for the discrepancies illustrates an important aspect of our approach to profile fitting. Both HD~104705 and HDE~303308 are extended sight lines (with path lengths of 5.0 and 3.8 kpc, respectively) that cross the Sagittarius-Carina spiral arm. The interstellar absorption profiles of moderately strong lines like O~{\sc i}~$\lambda1355$ and Ge~{\sc ii}~$\lambda1237$ in these directions exhibit a strong complex of absorption components near $v_{\mathrm{LSR}}=0$~km~s$^{-1}$ (presumably tracing gas in the local ISM) along with weaker components at more negative velocities (which likely probe gas in the Sagittarius-Carina spiral arm). In Kr~{\sc i}~$\lambda1235$, however, only the local ISM components are readily detectable. It appears that Cartledge et al.~(2008), in their analysis of Kr~{\sc i} toward HD~104705 and HDE~303308, included only these readily detectable components in their profile fits. (We find essentially the same column densities if we include only these more obvious features in our fits.) However, such an approach would tend to underestimate the Kr abundance since the total hydrogen column density applies to the entire line of sight. With our approach of fitting weak interstellar features with profile templates (constructed from the components seen in O~{\sc i}~$\lambda1355$ in these cases), we are able to include all of the relevant line-of-sight components in our fits even when some of those components would not be detected significantly on their own.

\section{DETERMINATION OF ELEMENT DEPLETION PARAMETERS}
\subsection{Elemental Abundances}
Since the ionization potentials for all of the neutral and singly-ionized species we examine are greater than (or approximately equal to) that of neutral hydrogen, we can be reasonably well assured that the column densities we obtain for these species represent the total column densities of the elements for the lines of sight included in our survey. As demonstrated in Figure 1, none of the sight lines we examine have total hydrogen column densities below log~$N$(H$_{\mathrm{tot}}$)~$\approx$~20.0, where ionization effects begin to become important. We can thus derive the elemental abundances for these sight lines in the usual way. That is, we define the (logarithmic) abundance of element $X$ as log~($X$/H)~$\equiv$~log~$N$($X$)~$-$~log~$N$(H$_{\mathrm{tot}}$), where $N$($X$) refers to the column density of the element in its preferred stage of ionization for neutral diffuse gas. For most of the sight lines, we were able to calculate total hydrogen column densities from the values of $N$(H~{\sc i}) and $N$(H$_2$) listed in Table 2 of J09 (where references to the original sources of those values may also be found). For HD~148937, which was not included in J09, we obtained $N$(H~{\sc i}) from Diplas \& Savage (1994) and $N$(H$_2$) from Sheffer et al.~(2007). We were unable to determine values of $N$(H$_{\mathrm{tot}}$) for 17 of the 69 sight lines in our primary sample, however, due to a lack of available information on H~{\sc i} and/or H$_2$. For the other 52 sight lines, we list in Table 7 the total hydrogen column densities along with the corresponding elemental abundances based on the column density measurements presented in Table 5. Uncertainties in these quantities were determined through standard error propogation techniques. (Total hydrogen column densities for sight lines with column density measurements from the literature are provided in Appendix A.)

In compiling values of $N$(H~{\sc i}) from the literature for the sight lines in his survey, J09 was careful to make corrections for stellar Ly$\alpha$ absorption in cases where the background stars had spectral types B1 or cooler. In two cases relevant to our investigation, however, those corrections carried large uncertainties, and the nominal corrections appear to have been too large. For HD~27778 and HD~203532, J09 obtained interstellar H~{\sc i} column densities of log~$N$(H~{\sc i})$_{\mathrm{ISM}}$~=~20.35 and 20.22, respectively. In both cases, however, the error associated with the determination of $N$(H~{\sc i})$_{\mathrm{ISM}}$ was larger than the value itself. The observed H~{\sc i} column densities toward HD~27778 and HD~203532 are log~$N$(H~{\sc i})$_{\mathrm{obs.}}$~=~$21.10\pm0.12$ and $21.27\pm0.09$ (Cartledge et al.~2004). These uncorrected H~{\sc i} column densities, when combined with the corresponding H$_2$ column densities from Cartledge et al.~(2004), yield total hydrogen column densities of log~$N$(H$_{\mathrm{tot}}$)~=~21.36$^{+0.08}_{-0.09}$ for HD~27778 and 21.44$^{+0.07}_{-0.08}$ for HD~203532. If the J09 corrections were adopted instead, the total hydrogen column densities would be log~$N$(H$_{\mathrm{tot}}$)~=~21.10 and 21.02. Using the results of his depletion analysis, J09 also derived ``synthetic'' values of $N$(H$_{\mathrm{tot}}$) for his sight lines based on the relative gas-phase abundances of the elements observed (and the pattern of depletions expected for those elements). For HD~27778 and HD~203532, the synthetic values of $N$(H$_{\mathrm{tot}}$) were found to be log~$N$(H$_{\mathrm{tot}}$)$_{\mathrm{syn.}}$~=~$21.38\pm0.07$ and $21.37\pm0.08$. The good agreement between these synthetic values and the values obtained when the uncorrected H~{\sc i} column densities are adopted suggests that the corrections for stellar Ly$\alpha$ absorption derived by J09 may be too large for these particular stars. We therefore use the uncorrected values when calculating total hydrogen column densities and abundances for these sight lines.

In two other instances, we were able to obtain values for $N$(H$_{\mathrm{tot}}$) despite having incomplete information. No molecular hydrogen column densities are available for HD~37021 or HD~37061 (primarily because these stars were never observed by the \emph{FUSE} satellite). However, as first pointed out by Cartledge et al.~(2001), these sight lines show very little, if any, Cl~{\sc i} absorption in the relatively strong line at 1347.2~\AA, suggesting that little H$_2$ is present, given the close correspondence between neutral chlorine and molecular hydrogen (Jura \& York 1978; Moomey et al.~2012). To derive more quantitative estimates for $N$(H$_2$) in these directions, we examined the available STIS spectra of the two stars in the vicinity of Cl~{\sc i}~$\lambda1347$. Toward HD~37021, there is no discernible absorption feature at the expected position of the Cl~{\sc i} line, and we derive a 3$\sigma$ upper to the Cl~{\sc i} column density of log~$N$(Cl~{\sc i})~$<$~11.6. Weak absorption from Cl~{\sc i}~$\lambda1347$ is detected toward HD~37061 with an equivalent width of $9.1\pm0.8$~m\AA{} and a column density of log~$N$(Cl~{\sc i})~=~$12.60\pm0.04$. These Cl~{\sc i} column densities indicate that the H$_2$ column densities are not likely to be greater than log~$N$(H$_2$)~$\approx$~19.0 (e.g., Moomey et al.~2012). Since the H~{\sc i} column densities toward HD~37021 and HD~37061 are log~$N$(H~{\sc i})~=~$21.65\pm0.13$ and $21.73\pm0.09$ (J09), such small contributions from H$_2$ would not alter the total hydrogen column densities in any significant way. We therefore accept the H~{\sc i} column densities as being representative of the total amounts of hydrogen along the lines of sight.

\subsection{Depletion Parameters}
One of our primary objectives in investigating the abundances of $n$-capture elements in the ISM is to uncover the extent to which the elements are depleted onto interstellar dust grains and also to examine how the depletions change with changing environmental conditions from one sight line to another. Without this knowledge, it would be difficult to draw any definitive conclusions regarding the ways in which $s$- and $r$-process nucleosynthesis might be affecting interstellar abundances in the current epoch. In a landmark study of interstellar depletions, J09 examined the depletion characteristics of 17 different elements (including Ge and Kr) in a sample of 243 sight lines probing the local Galactic ISM. We seek to perform a similar analysis for the elements studied in this investigation so that their depletion properties may be directly compared with those of more abundant elements. Presumably, this will allow us to ascertain whether the $n$-capture elements follow ``normal'' depletion patterns, or whether their abundances have been affected in some way by nucleosynthetic processes. The unified framework that J09 developed was predicated on the empirical observation that, while different elements exhibit different degrees of depletion, the depletions of most elements tend to increase in a systematic way as the overall strength of depletions increases from one sight line to the next. Sight lines with stronger depletions are thought to contain higher proportions of denser and/or colder gas. This phenomenon of ``density-dependent'' depletions has been studied by others (e.g., Cartledge et al.~2006; R11), and is generally taken as evidence of grain growth in the diffuse ISM.

\subsubsection{Definitions and Methodology}
As a first step toward developing a unified framework within which to characterize the depletions of different elements along different lines of sight, J09 defined a line-of-sight depletion strength factor, denoted $F_*$, which indicates for a given direction the extent to which depletion processes have succeeded in removing atoms from the gas phase. The value of $F_*$ for a specific sight line is based on a weighted average of the available observed depletions for that direction (see Equation (4) in J09). Sight lines showing strong depletions, such as those seen in the low velocity ($v_{\mathrm{LSR}}=-1$~km~s$^{-1}$) component toward $\zeta$~Oph, have depletion factors near $F_*=1$, while sight lines showing only very modest depletions have values of $F_*$ closer to zero. In an idealized situation, the depletion of element $X$, defined in logarithmic terms as [$X$/H]~=~log~($X$/H)~$-$~log~($X$/H)$_{\sun}$, is related to the sight-line depletion factor $F_*$ according to:

\begin{equation}
[X/\mathrm{H}]=B_X+A_X(F_*-z_X),
\end{equation}

\noindent
where the depletion parameters, $A_X$, $B_X$, and $z_X$, are unique to element $X$. Among these element-specific parameters, the depletion slope $A_X$ is the most fundamental. It indicates how much the depletion of a particular element changes as the growth of dust grains progresses within interstellar clouds. (Note that the value of the $A_X$ parameter does not depend on the choice of the solar reference abundance to which the measured gas-phase abundances are compared, and that it is largely insensitive to the adopted $f$-values.) The intercept parameter $B_X$ indicates the expected depletion of element $X$ at $F_*=z_X$, where $z_X$ represents a weighted mean value of $F_*$ for the particular set of depletion measurements under consideration. Values of the coefficients $A_X$ and $B_X$ may be obtained for each element through the evaluation of a least-squares fit, with [$X$/H] as the dependent variable and $F_*$ the independent variable. (The reason for the additional term involving $z_X$ in Equation (1) is that for a particular choice of $z_X$ there is a near zero covariance between the formal fitting errors for the solutions of $A_X$ and $B_X$; see J09.)

Since neither the element-specific depletion parameters nor the sight-line depletion factors are known \emph{a priori}, J09 used an iterative procedure to evaluate these parameters and determine fiducial values of $F_*$ for the sight lines in his survey. We do not re-evaluate the depletion factors for the sight lines in common with our survey (even though new depletion measurements are now available for those sight lines for elements not considered by J09) since any changes to those $F_*$ values would likely be small. Rather, we simply adopt the values (and their associated uncertainties) already calculated by J09 (accepting only those values determined directly from the observed depletions, and not the so-called ``synthetic'' values derived from least-squares fits to the relative gas-phase abundances, even in cases where an observed value is not available). This choice allows us to more easily compare the depletion trends exhibited by the $n$-capture elements examined here with those found for the more abundant elements studied by J09. The adopted values of $F_*$ for the sight lines in our primary sample are listed along with the total hydrogen column densities in Table 7. ($F_*$ values for sight lines with column density measurements from the literature are given in Appendix A.)

\subsubsection{Solar Reference Abundances}
Having established suitable values for the total hydrogen column densities and sight-line depletion factors for as many lines of sight as possible from our survey, the only remaining component required for our analysis is the set of solar reference abundances against which to measure the interstellar depletions.\footnote{While the abundances in young F and G-type stars of the solar neighborhood might serve as a better set of reference abundances for the present-day ISM, these abundances are not very much different from those of the Sun (e.g., Lodders et al.~2009). Moreover, the $n$-capture elements that are the focus of our investigation have generally not been observed in such stars.} Following J09, we adopt the recommended solar system abundances from Lodders (2003), which we list in Table 8 for the elements of interest to our survey. While there are more modern references for the chemical composition of the Sun and solar system (e.g., Asplund et al.~2009; Lodders et al.~2009; Grevesse et al.~2015), use of the Lodders (2003) abundances allows us to place the element depletion parameters that we derive on the same scale as those in J09. The differences between the abundances in Lodders (2003) and those in more recent compilations for the $n$-capture elements in our investigation are minimal regardless. Solar abundances for the elements Ga, Ge, As, Cd, Sn, and Pb tend to be based on meteoritic abundances since these are generally more accurate than the corresponding photospheric abundances. Lodders et al.~(2009) compile a set of updated meteoritic abundances, yet these differ from the ones in Lodders (2003) by at most 0.02 dex for the elements of interest. (The solar Kr abundance from Lodders (2003) is based on theoretical $s$-process production rates; the value has not changed significantly in more recent compilations.)

A potentially more significant systematic effect involves the correction applied to the solar photospheric (and meteoritic) abundances to account for the gravitational settling that is thought to have occurred between the formation of the protosolar nebula and the present-day Sun. In deriving her recommended solar system (i.e., protosolar) abundances, Lodders (2003) applied a correction of +0.074~dex to the photospheric abundances of all elements heavier than He. In Lodders et al.~(2009), that correction was reduced to +0.053~dex. Asplund et al.~(2009) recommend a correction of +0.04~dex. While we adopt the protosolar abundances from Lodders (2003), for consistency with J09, we acknowledge that the upward correction of 0.074 dex applied to the present-day photospheric abundances may be too large. Of course, this is a systematic correction that affects all depletion measurements in the same way, and thus can be easily adjusted if a different overall correction proves to be more appropriate.

\subsubsection{Least-Squares Fits}
Depletion parameters for the elements O, Ga, Ge, As, Kr, Cd, Sn, and Pb were determined through linear least-squares fits to trends of [$X$/H] versus $F_*$, adopting the functional form expressed in Equation (1). The least-squares fits were performed using the Interactive Data Language (IDL) procedure FITEXY, which properly accounts for errors in both the $x$ and $y$ coordinates (see Press et al. 2007). The resulting values of $A_X$, $B_X$, and $z_X$ for each element are presented in Table 8, while the fits themselves are shown in Figures 18 through 20.\footnote{Note that the ordinate of the plots in Figures 18$-$20 is log~($X$/H) rather than [$X$/H]. This was done so that the raw abundance measurements could be displayed, making it easy to see how the depletions would change if, for example, a different solar reference abundance were adopted.} (The least-squares fits described in this section are depicted by the solid orange lines in the plots for the different elements.) Since the values we obtain for $B_X$ depend on our choice of reference abundances, the errors associated with the adopted solar system abundances were added in quadrature to the formal fitting errors for $B_X$ to derive the $B_X$ uncertainties given in Table 8. (The same is not true of the $A_X$ uncertainties since the derived values of $A_X$ are entirely independent of the adopted reference abundances.)

\begin{figure}
\centering
\includegraphics[width=0.99\textwidth]{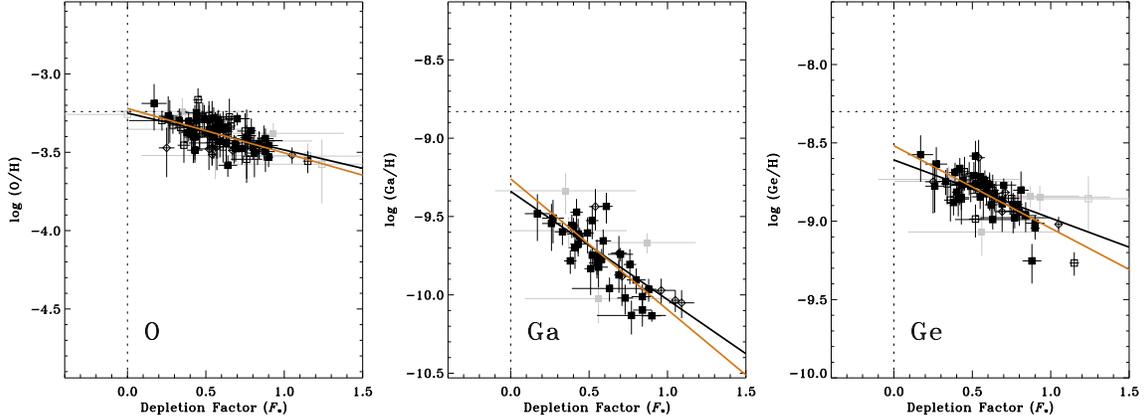}
\caption{Gas-phase abundances as a function of the sight line depletion factor ($F_*$) from J09 for the elements O, Ga, and Ge. Solid symbols represent abundances derived in this work; open symbols are used for results obtained from the literature (squares: STIS; diamonds: GHRS). Grey symbols denote sight lines where $\sigma(F_*)\ge0.30$. The solid orange line shows the linear fit based on the methodology of J09, with parameters given in Table 8. The solid black line shows the fit based on a survival analysis, with parameters given in Table 9. The horizontal dotted line in each panel gives the adopted solar system abundance from Lodders (2003).}
\end{figure}

\begin{figure}
\centering
\includegraphics[width=0.99\textwidth]{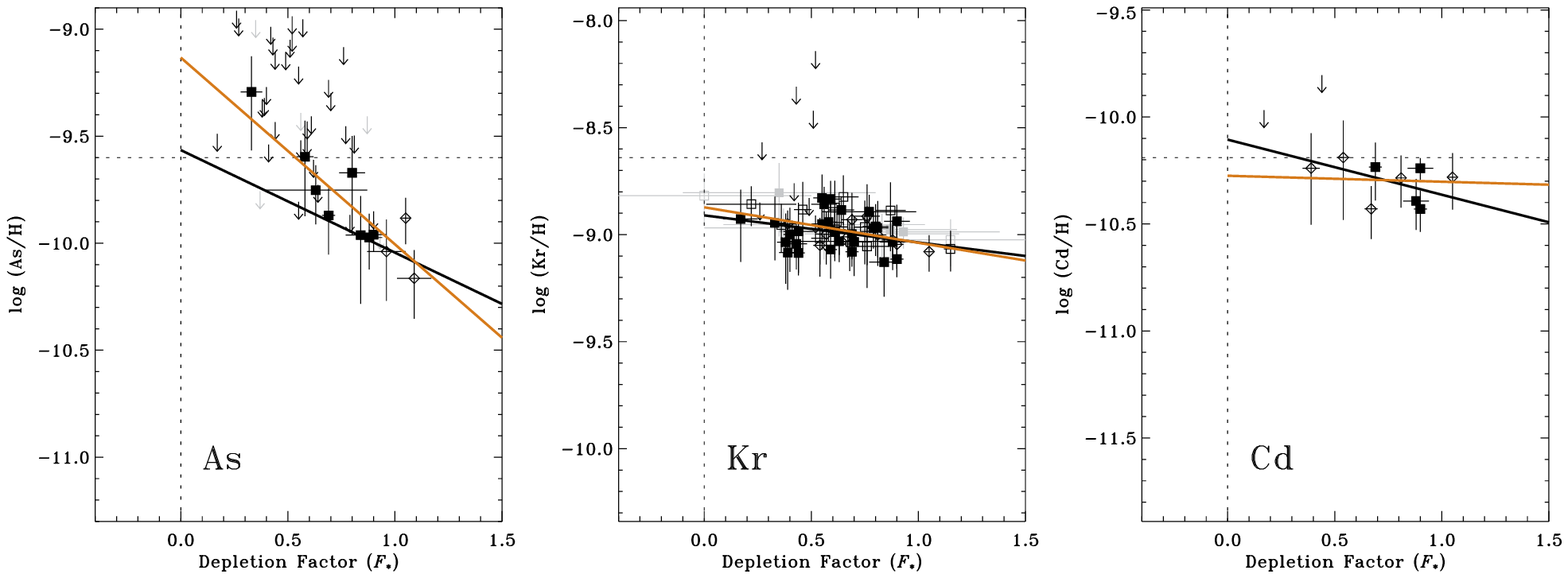}
\caption{Same as Figure~1 except for the elements As, Kr, and Cd. All upper limits are our determinations.}
\end{figure}

\begin{figure}
\centering
\includegraphics[width=0.66\textwidth]{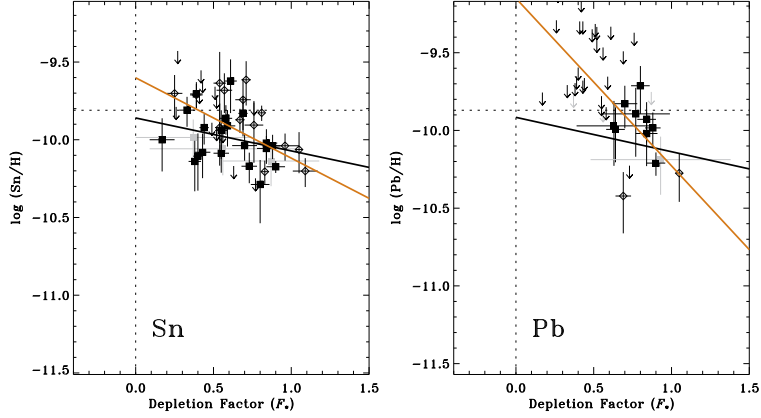}
\caption{Same as Figure~1 except for the elements Sn and Pb. All upper limits are our determinations.}
\end{figure}

The $B_X$ parameters are not fundamental quantities intrinsic to the elements since they are sensitive to the specific set of depletion measurements available in each case. If different sets of observations were considered, particularly if the sight lines had much different distributions of $F_*$ values, then the derived values of $B_X$ (and $z_X$) could be much different from the values listed in Table 8 (even if the underlying depletion trends were the same). It is therefore important to evaluate two additional depletion parameters, which are insensitive to the particular sample of observations under consideration, so that the depletions of many different elements may be compared in a consistent manner. Following J09, we evaluate for each element the parameters:

\begin{equation}
[X/\mathrm{H}]_0=B_X-A_Xz_X
\end{equation}

\noindent
and

\begin{equation}
[X/\mathrm{H}]_1=B_X+A_X(1-z_X),
\end{equation}

\noindent
which yield the expected depletions at $F_*=0$ and $F_*=1$, respectively. The former indicates the initial amount of depletion present in the diffuse ISM before significant grain growth has occurred (or after the outer portions of the grains have been destroyed by the passage of an interstellar shock, for example), while the latter represents the depletion seen in a relatively dense and/or cold interstellar cloud, such as the $v_{\mathrm{LSR}}=-1$~km~s$^{-1}$ cloud toward $\zeta$ Oph. The values we obtain for [$X$/H]$_0$ and [$X$/H]$_1$ for the elements in our survey, calculated from the corresponding values of $A_X$, $B_X$, and $z_X$, are given in Table 8. (The uncertainties in these quantities were determined according to the relations given in J09.)

The last three columns of Table 8 list for each element the $\chi^2$ value associated with the least-squares fit, the degrees of freedom in the fit (i.e., the number of observations minus two), and the probability of obtaining a worse fit than that which we obtain. Exceptionally low probabilities of obtaining a worse fit are found for the elements Ga, Sn, and Pb, which could indicate that the model developed by J09 does not provide a complete description of the variations seen in the gas-phase abundances of these elements. (This could be the case if, for example, there are true abundance variations that are superimposed onto the general trend due to depletion.) Alternatively, the low probabilities could be an indication that the uncertainties in the measured column densities have been underestimated. The opposite situation is seen in our results for the elements O and Kr, where the relatively low $\chi^2$ values yield high probabilities of a worse fit. That we obtain such high values for these probabilities may indicate that the column density uncertainties have been overestimated in these cases.

\subsection{Survival Analysis}
In deriving column densities for the sight lines in our primary sample, considerable effort was made to determine upper limits for any species covered by the observations that were not detected at the 2$\sigma$ level or greater. Moreover, a number of sight lines were included in our sample even though they did not show any evidence of absorption from As~{\sc ii}, Cd~{\sc ii}, Sn~{\sc ii}, or Pb~{\sc ii}, which were the primary objectives of our archival search. This was done in recognition of the fact that upper limits on column densities would still be useful, particularly in cases where few confirmed detections are available. An illustration of this fact is provided by the plot showing the depletion trend for Pb in Figure 20. The least-squares fit for Pb (represented by the solid orange line in the plot) indicates that the depletion slope $A_{\mathrm{Pb}}$ is quite steep, and that the initial value of the depletion [Pb/H]$_0$ is \emph{positive}, suggesting a gas-phase abundance considerably higher than the adopted solar abundance. However, there is reason to suspect that this fit does not truly represent the depletion properties of Pb in the ISM. The sight lines yielding Pb~{\sc ii} detections all fall within a very narrow range in $F_*$ (between about 0.6 and 1.0), meaning that there is very little leverage for determining the slope of the depletion trend. Indeed, the $A_{\mathrm{Pb}}$ parameter has the highest uncertainty among all of the $A_X$ values in Table 8. Furthermore, there are a number of upper limits at small values of $F_*$ that lie below the best-fit line representing the least-squares fit. This is a clear indication that the depletion slope for Pb is most likely not as steep as the least-squares solution would indicate. A similar situation may be seen in the case of As (Figure 19), where there are a handful of upper limits at low $F_*$ that appear to be inconsistent with the least-squares fit.

The methods typically employed for analyzing observational data containing nondetections (or ``censored'' data points) are collectively known as survival analysis (e.g., Feigelson \& Nelson 1985; Isobe et al.~1986). For our purposes, we wish to perform a linear regression between the values of [$X$/H] and $F_*$ that does not completely ignore the presence of the upper limits that were derived in cases where the depletions could not be measured. To accomplish this, we employ Schmitt's binned method (Schmitt 1985) to derive the regression coefficients (i.e, the slope and intercept of the regression line), using the task SCHMITTBIN, which is part of the STATISTICS package within the Space Telescope Science Data Analysis System (STSDAS). Along with estimates for the slope and intercept, the SCHMITTBIN task outputs the standard deviations of these estimates using Schmitt's bootstrap error analysis (Schmitt 1985). The linear regression coefficients resulting from our application of this method are presented in Table 9, which also lists for each element the number of detections and nondetections considered in the analysis. (The corresponding regression lines are shown on the depletion plots presented in Figures 18$-$20 as solid black lines.)

While our motivation in performing this analysis was to better understand the depletion trends for elements like As and Pb, where the observational data are heavily censored, we applied this method to all elements so that the results for elements like O and Ge, which have zero nondetections, could be compared with the results of the least-squares fits described above to check for consistency. In making these comparisons, we note that the survival analysis does not account for errors in either the independent or dependent variable. Thus, differences in the regression coefficients derived using the two methods might be expected, even in cases where all data points represent detections. Still, by comparing the $A_X$ values from Table 8 with the corresponding slope parameters in Table 9, we find that, in most cases, the results agree within their mutual uncertainties (at about the 2$\sigma$ level or better). The largest differences are seen for the elements As, Sn, and Pb. In all three cases, the survival analysis regression suggests that the depletion slope may not be as steep as indicated by the least-squares fit. Furthermore, the intercept parameters derived through survival analysis for As, Sn, and Pb (Table 9) suggest that the initial depletions of these elements are all approximately zero. These findings contrast with the high \emph{positive} values of [As/H]$_0$, [Sn/H]$_0$, and [Pb/H]$_0$ derived from the least-squares fits (Table 8), which seem to indicate that the gas-phase abundances of As, Sn, and Pb are supersolar in low-depletion sight lines.

To further investigate the differences in slope between the least-squares linear fits and the survival analysis regressions for As, Sn, and Pb, we reevaluated the survival analysis fits after removing the upper limits from consideration. We did this to test whether the differences in slope are driven by the inclusion of the upper limits or are simply a result of the fact that the survival analysis regressions do not account for errors in the measured quantities. For As, it seems that the difference in slope between the orange and black lines in Figure 19 is entirely driven by the inclusion of the upper limits since the survival analysis regression is nearly identical to the least-squares linear fit when the upper limits are not considered. For Sn and Pb, however, much of the difference in the slopes of the lines presented in Figure 20 is due to the fact that the survival analysis regressions do not account for errors in the abundances or depletion factors. Ultimately, more precise abundance measurements are needed, particularly at small values of $F_*$, to better constrain the slopes of the depletion trends for these elements.

\section{DISCUSSION}
\subsection{Extending the Analysis of Interstellar Depletions to Rare $n$-Capture Elements}
In the preceding sections, we have described our efforts to examine the gas-phase abundances and depletion behaviors of $n$-capture elements in a comprehensive and consistent manner. Such efforts are necessary for developing a more complete picture of element depletions in the local Galactic ISM. In his analysis of the depletion patterns of 17 different elements, J09 demonstrated that the depletions of most elements are highly correlated, meaning that the depletion of any individual element tends to increase as the overall strength of depletions increases from one sight line to another. (Only for the element N did J09 find that the observed depletions remain constant even as the value of $F_*$ increases.) Our investigation, and in particular our derivation of the element depletion parameters in Section 4.2, allows us to extend the analysis of J09 to include the elements Ga, As, Cd, Sn, and Pb. (In Appendix B, we present depletion parameters derived using the same methods for the element B based on the observational data reported by R11.) These results allow us to better understand the depletion behaviors of elements with low-to-moderate condensation temperatures, specifically those with $T_C$ in the range 600 to 1100~K, where the transition from relatively mild to relatively strong depletions begins to occur (Savage \& Sembach 1996; J09). A central question to ask is whether the elements studied in this investigation participate in the same collective depletion behavior exhibited by the other more abundant elements studied by J09.

Before addressing that question specifically, we first compare our results to those reported in J09 for the elements O, Ge, and Kr since these elements were considered in both investigations. We focus mainly on the derived values of the parameters $A_X$ and [$X$/H]$_0$ since together these two parameters fully characterize the depletion trend for a given element. By comparing the $A_X$ and [$X$/H]$_0$ values from Table 8 for O, Ge, and Kr with the corresponding values from Table 4 of J09, we find that in each case the results agree at the 1$\sigma$ level or better,\footnote{There is a systematic offset between our values for $B_{\mathrm{Ge}}$, [Ge/H]$_0$, and [Ge/H]$_1$ and those of J09 due to our use of a new experimental $f$-value for the Ge~{\sc ii}~$\lambda1237$ transition (Heidarian et al.~2017). The J09 values must be increased by +0.149 for comparison with our values.} although our analysis has led to a reduction in the uncertainties associated with each of these parameters. The smaller uncertainties reflect the fact that our samples are larger\footnote{Our O, Ge, and Kr samples are larger than those in J09 both because we have measurements for more sight lines and because we do not place any restrictions on the samples in terms of which sight lines we include in the analysis. Among the restrictions imposed in his analysis, J09 considered only those stars with Galactocentric distances in the range $7<R_{\mathrm{GC}}<10$~kpc out of concern that an abundance gradient could distort the results for stars outside the solar circle. Later, he found that no abundance gradient was evident even when all sight lines were considered.} and may also be due to our having redetermined the O~{\sc i}, Ge~{\sc ii}, and Kr~{\sc i} column densities in a consistent way for all of the sight lines in our primary sample. (Recall the discussion in Section 3.3 concerning Kr~{\sc i} toward HD~104705 and HDE~303308. If we had simply adopted the Kr~{\sc i} column densities from Cartledge et al.~(2008) for these sight lines, the Kr depletions would have seemed unusually large compared to their respective values of $F_*$.) The largest reductions in uncertainty are seen for the depletion slope parameters $A_{\mathrm{Ge}}$ and $A_{\mathrm{Kr}}$. The errors we find for these quantities are lower than their counterparts in J09 by $\sim$40\%. This is particularly significant in the case of Kr, which as a noble gas would generally not be expected to participate in the collective depletion behavior exhibited by many other elements. While J09 found that Kr does seem to show progressively stronger depletions as $F_*$ increases (i.e., $A_{\mathrm{Kr}}=-0.166\pm0.103$; J09), the significance of that result was only 1.6$\sigma$. Our analysis yields the same negative slope for Kr ($A_{\mathrm{Kr}}=-0.166\pm0.059$) but at a significance level of 2.8$\sigma$.

We now turn to the issue of how the depletion properties of the elements examined in this investigation (particularly those not considered by J09) compare to the properties of the other more abundant elements that have been studied previously. In Figure 21, we plot the [$X$/H]$_0$ and [$X$/H]$_1$ values we derive (Table 8), along with the values derived by J09 for elements not considered here, against the condensation temperatures of the elements. The adopted values of $T_C$ correspond to the 50\% condensation temperatures computed by Lodders (2003) for a solar system (i.e., protosolar) composition at 10$^{-4}$ bar total pressure. While calculations like these strictly apply only in situations involving chemical equilibrium, the resulting condensation temperatures may still be useful indicators of the tendencies of different elements to form solid compounds within interstellar dust grains. In Figure 22, we present a similar plot of the depletion slopes $A_X$, derived here (Table 8) and in J09, as a function of $T_C$. (Note that our results for B, described in Appendix B, are included among the results shown in Figures 21 and 22.)

\begin{figure}
\centering
\includegraphics[width=0.8\textwidth]{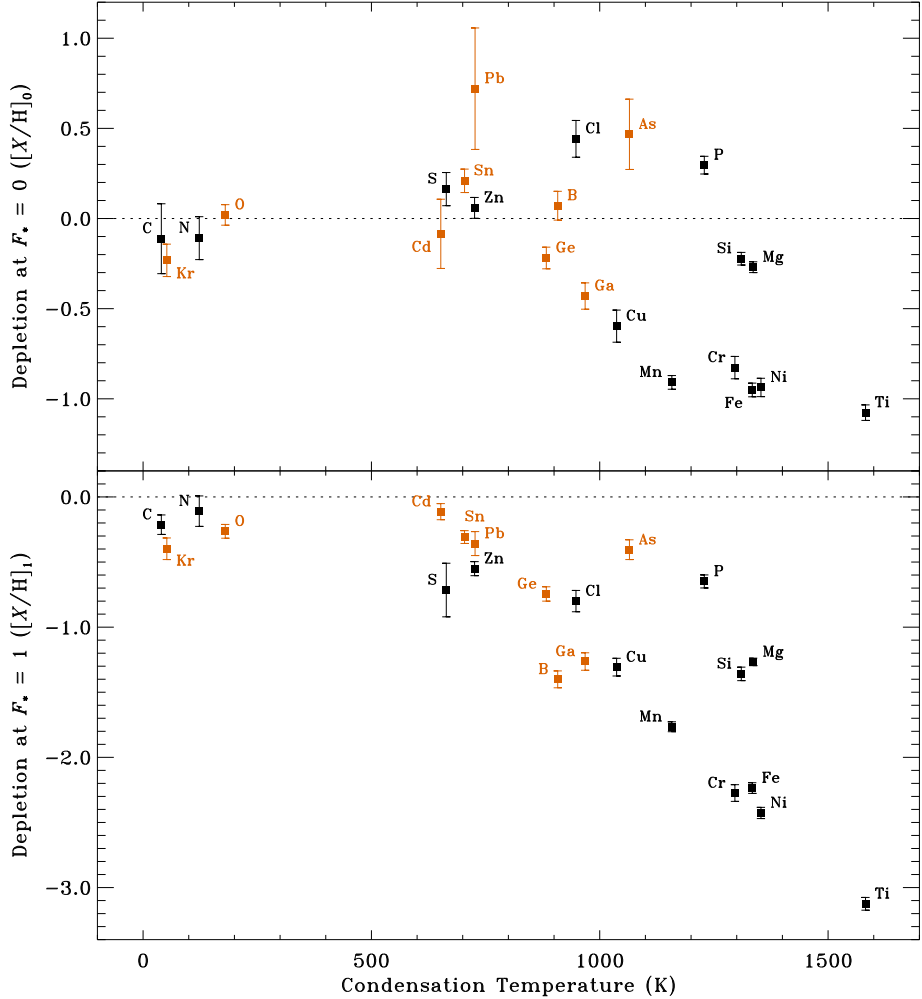}
\caption{Element depletions as a function of the condensation temperature ($T_C$) from Lodders (2003). Orange symbols are used for the results obtained in this work; black symbols show results for additional elements examined by J09. Upper panel: Depletion at $F_*=0$ (denoted [$X$/H]$_0$ in Table 8) versus $T_C$. Lower panel: Depletion at $F_*=1$ (denoted [$X$/H]$_1$ in Table 8) versus $T_C$.}
\end{figure}

\begin{figure}
\centering
\includegraphics[width=0.9\textwidth]{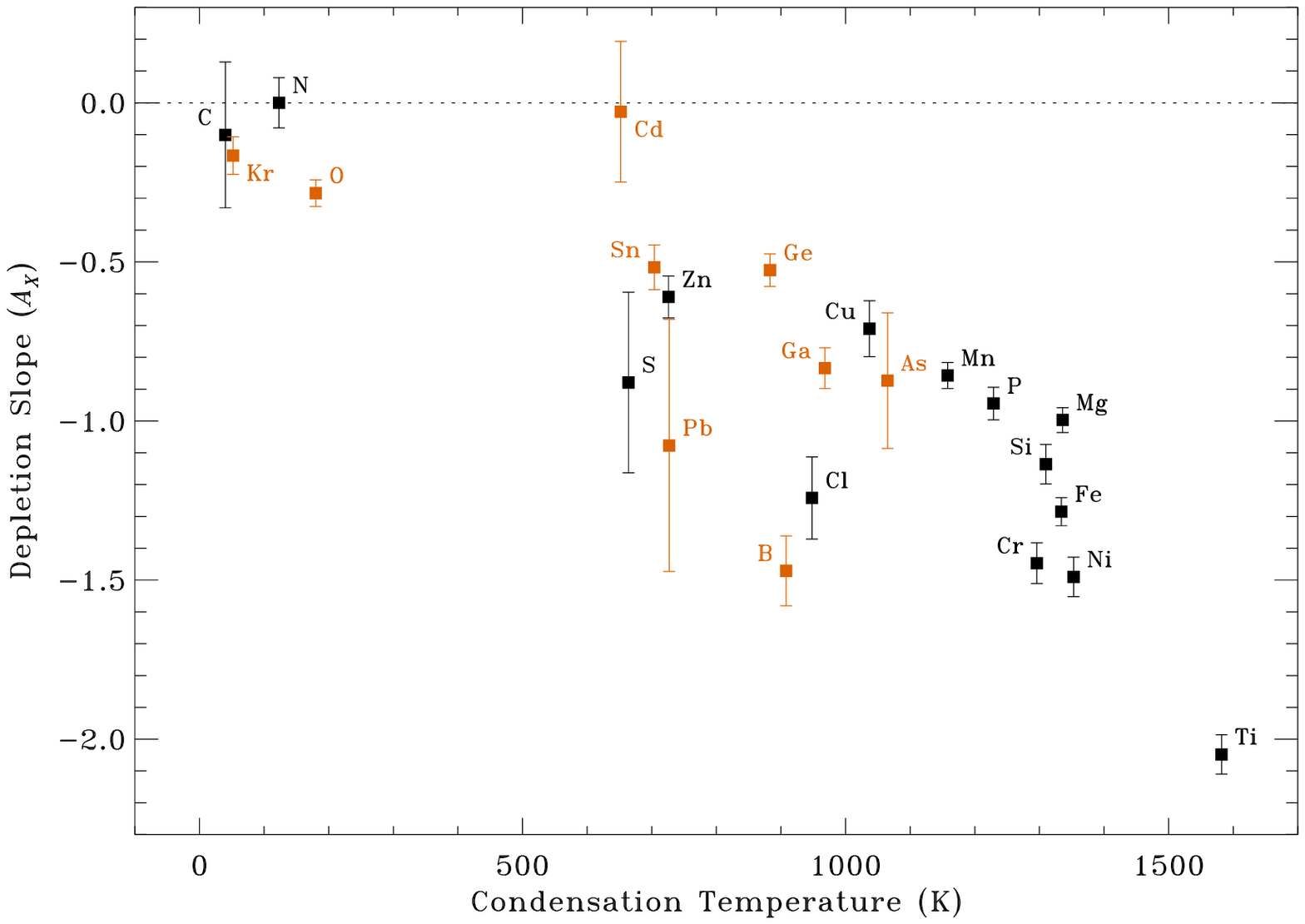}
\caption{Depletion slope ($A_X$) as a function of the condensation temperature of the element. Orange symbols are used for the results obtained in this work; black symbols show results for additional elements examined by J09.}
\end{figure}

The [$X$/H]$_0$ values (plotted in the upper panel of Figure 21) are indicators of the depletions seen along sight lines showing the lowest levels of depletion overall (among the 243 sight lines analyzed by J09). Presumably, these sight lines contain gas at low enough densities and/or at high enough temperatures that significant grain growth (or mantling) has not yet commenced. Or, perhaps these sight lines probe regions where the grains have been disrupted by shocks, which strip the grains of any mantles they may initially have had. In either case, these initial depletions should provide a fair representation of the composition of dust grains (or, more specifically, the resilient cores of dust grains) that emerge from various stellar sources (e.g., evolved low- and high-mass stars and Type II supernovae). It can be seen from the figure that most elements with $T_C<800$~K remain undepleted at this stage (i.e., that their [$X$/H]$_0$ values are consistent with zero). However, Kr appears to be depleted below the solar system value (at the 2.6$\sigma$ level), while Sn seems overabundant (at the 3.2$\sigma$ level). (The initial depletion level for Pb should be considered highly uncertain since the slope of the depletion trend for this element is not very well constrained.) As for elements with $T_C>800$~K, more than half show a fairly regular trend of increasing initial depletion with increasing $T_C$. The [$X$/H]$_0$ values that we derive for Ga and Ge, for example, seem to be natural extensions of the trend seen in the initial depletions for elements from Cu to Ti. If this trend represents a dust condensation sequence, then the temperature at which the sequence seems to terminate (between 800 and 900 K) could be related to the dust formation temperature. Indeed, dust shells around AGB stars are observed to have temperatures in the range 800 to 1100~K (e.g., Gail et al.~2013).

A significant number of elements with fairly high condensation temperatures show much weaker depletions at $F_*=0$ than would be expected from the above trend, however. Indeed, the initial gas-phase abundances of Cl, As, and P appear to be supersolar (at significance levels of 4.3$\sigma$, 2.4$\sigma$, and 6.0$\sigma$, respectively). In the case of Cl, at least, the high \emph{positive} value for the initial depletion most likely reflects the fact that J09 considered only Cl~{\sc ii} measurements in deriving gas-phase Cl abundances, neglecting any contributions from Cl~{\sc i}. In fact, there are a number of sight lines where $N$(Cl~{\sc i})~$>$~$N$(Cl~{\sc ii}), and these sight lines tend to have high molecular fractions due to the close correspondence between Cl~{\sc i} and H$_2$ (A.~M.~Ritchey et al., in preparation; see also Moomey et al.~2012). Since these sight lines preferentially have high values of $F_*$ as well, the depletion slope for Cl derived by J09 is most likely too steep (and the initial depletion value too high). We mentioned already that the depletion slope for As may be too steep since the survival analysis suggests a somewhat shallower trend. Most of the As depletion measurements were made along sight lines with relatively high values of $F_*$. Only one sight line with an As~{\sc ii} detection has $F_*<0.5$ (HD~104705), and that one measurement exerts considerable influence over the slope of the depletion trend. Still, the depletion slope for As does not seem to be unusual compared to those of elements with similar condensation temperatures (Figure 22). Likewise, the depletion slope for P does not seem to be unusual, falling intermediate between the slopes for Mn and Si. Moreover, the P depletion trend appears to be quite well defined, with little ambiguity in the derived parameters (see Figure 6 in J09).

If the depletion slopes for As and P are not in error, then the fact that the initial abundances of these elements appear to be higher than the adopted solar reference abundances could indicate either that the solar abundances are incorrect (or are not appropriate as reference abundances), or that the oscillator strengths of the As~{\sc ii} and P~{\sc ii} transitions used to derive the interstellar abundances are not accurate. (Recall that Lodders (2003) applied a correction of +0.074 dex to the photospheric and meteoritic abundances of all elements heavier than He to arrive at her recommended solar system abundances. If such a correction had not been applied, the discrepancies with the initial interstellar abundances for As and P would be even greater.) Lodders (2003) adopts the average of the photospheric P abundance\footnote{The abundances discussed in this paragraph are given on the standard logarithmic scale used in stellar astronomy where the abundance is equal to log~$\epsilon$($X$)~=~log~($X$/H)~+~12.} ($5.49\pm0.04$; Berzinsh et al.~1997) and the meteoritic P abundance ($5.43\pm0.05$; Wolf \& Palme 2001) to arrive at her value of log~$\epsilon$(P)~=~$5.46\pm0.04$ for the solar photosphere. A more modern reference for the solar photospheric abundance of P gives $5.41\pm0.03$ (Scott et al.~2015), which would only exacerbate the disagreement with the initial interstellar abundance implied by the depletion trend for P. (The interstellar P abundance at $F_*=0$ from the fit presented in J09 is equivalent to log~$\epsilon$(P)~=~$5.84\pm0.05$.) The solar abundance of As, which is based solely on the meteoritic abundance since there are no usable As lines in the solar spectrum, has not changed between Lodders (2003) and Lodders et al.~(2009). The latter report a value of log~$\epsilon$(As)~=~$2.32\pm0.04$ from an average of 20 analyses. (For comparison, the interstellar abundance of As at $F_*=0$ from our least-squares fit is equivalent to log~$\epsilon$(As)~=~$2.87\pm0.20$.)

If we assume that the solar As and P abundances are accurate, then the next question to ask is whether they are appropriate as reference abundances for the present-day ISM. In the case of P, at least, the answer seems to be that the solar abundance is an approriate benchmark for local gas. Recent abundance determinations for local stars using near-infrared lines (Caffau et al.~2011) and ultraviolet lines (Roederer et al.~2014a) give [P/Fe] ~$\approx$~0 and [P/S]~$\approx$~0 for stars with metallicities close to that of the Sun. The situation for As is less clear as this element has been detected in only seven metal-poor stars (Roederer 2012; Roederer et al.~2014b). The [As/Fe] ratios in these stars are generally supersolar (with a mean value of +0.28 dex; Roederer et al.~2014b) and show little evolution across the entire metallicity range probed. Still, the implications of these results for the As abundance in the present-day ISM are unclear because there are no determinations for As in stars with metallicities approaching that of the Sun. (The most metal-rich star in the samples shown in Roederer (2012) and Roederer et al.~(2014b) has [Fe/H]~$\approx$~$-$0.8.)

As discussed in Section 3.1, the oscillator strength for the As~{\sc ii} transition at 1263.8~\AA{} has not been experimentally verified. Various theoretical determinations of the $f$-value range from 0.18 to 0.35 (e.g., Cardelli et al.~1993; Ganas 2000), and have a mean not very much different from the value of 0.259 adopted by Morton (2000) and used in this investigation. The oscillator strength would need to be a factor-of-three larger than the Morton (2000) value to account for the 0.47 dex discrepancy between the initial interstellar abundance of As (implied by the least-squares fit) and the solar system abundance. The situation is somewhat more complicated for P because different investigators have used a variety of P~{\sc ii} lines (e.g., $\lambda1152$, $\lambda1301$, and $\lambda1532$) to derive interstellar P abundances. The oscillator strength of the $\lambda1152$ transition is the most secure as a result of the beam-foil measurements performed by Federman et al.~(2007). While those experimental results were published after most studies of interstellar P~{\sc ii} abundances, the Federman et al.~(2007) $f$-value of $0.272\pm0.029$ for P~{\sc ii}~$\lambda1152$ essentially confirmed the value of 0.245 listed by Morton (2003), which was based on earlier experimental and theoretical results. Since J09 adopted the Morton (2003) $f$-values, the P abundances he uses in his analysis (at least those derived from the $\lambda1152$ line) should be secure. (Other P~{\sc ii} transitions such as $\lambda1301$ show larger discrepancies in their theoretical $f$-values. It remains to be seen whether future experiments can resolve such discrepancies.)

The [$X$/H]$_1$ values (plotted in the lower panel of Figure 21) are representative of the depletions seen in relatively cold, diffuse clouds, where substantial grain growth has already occurred. The trend with condensation temperature, which is evident in the figure, is reminiscent of the cold-cloud depletion pattern seen toward $\zeta$ Oph (e.g., Savage \& Sembach 1996). (This is not at all surprising, of course, since $\zeta$ Oph was used to define the depletion scale at $F_*=1$.) Most elements with $T_C<800$~K show relatively mild depletions at $F_*=1$ ($-$0.3 dex on average), while elements with higher condensation temperatures tend to show progressively stronger depletions (culminating in a depletion of $-$3.1 dex for Ti). However, just like at $F_*=0$, there are certain elements (i.e., As, P, Si, and Mg) that exhibit depletions at $F_*=1$ that are weaker than expected given the trend seen for the other elements. The depletion slopes for Si and Mg, like those for As and P, do not appear to be unusual (although there is a considerable spread in the values of the slopes for elements with condensation temperatures near that of Fe; Figure 22). Rather, the absolute depletions of As, P, Si, and Mg seem to be offset from those of other elements at all values of $F_*$. This seems to suggest that these elements are either prevented from being incorporated into the initial grains that condense (in the outer atmospheres of late-type stars, for example), or are incorporated primarily into grain mantles that are subsequently destroyed or disrupted by shocks after the grains are deposited into the ISM.\footnote{Jones (2000) argues that sputtering due to ion-grain collisions in supernova-generated shock waves will preferentially remove Si and Mg atoms over the heavier Fe atoms. If most grains deposited into the ISM are processed by shocks, then this phenomenon would seem to explain why only about 40\% of the available Si and Mg atoms are locked up in grains in low depletion sight lines, while nearly 90\% of the Fe atoms remain in the dust phase at $F_*=0$ (see Figure 21).} In either case, the fact that the depletion slopes seem to be normal suggests that these elements participate as expected in the mantling process that presumably takes place within interstellar clouds.

Joseph (1988) suggested that phosphorus may be chemically blocked from depleting in the outer atmospheres of late-type stars because it can form the stable molecule PN. In making this suggestion, Joseph (1988) cited the work of Gail \& Sedlmayr (1986), who had argued that nitrogen would be blocked from depleting because it forms N$_2$, a highly stable molecule with saturated valences that result in a high activation energy barrier for further gas-phase reactions. Since N and P have the same number of valence electrons, the idea was that PN would perform the same function for P as N$_2$ does for N. Upon entering the ISM, the PN molecules would be subject to the interstellar radiation field and would dissociate, allowing the P atoms to deplete along with the other refractory elements in interstellar clouds. We now have evidence that the depletion behavior of As in the ISM is similar to that of P. Both elements exhibit less depletion than expected at all values of $F_*$, but otherwise seem to behave normally, showing progressively stronger depletions as $F_*$ increases. Since As is in the same periodic group as P and N, perhaps it too is blocked from being incorporated into those initial dust grains that condense in the extended atmospheres of evolved stars. (While AsN, the As-bearing analog of N$_2$ and PN, has never been observed in interstellar or circumstellar environments, its predicted abundance would likely be below current detection limits given the low cosmic abundance of As.)

The chemistry of phosphorus in circumstellar envelopes remains uncertain despite recent progress on both the observational and theoretical fronts. A number of P-bearing molecules, including PN, HCP, CP, PO, and PH$_3$, have been detected in the circumstellar shells around carbon-rich and oxygen-rich evolved stars (e.g., Tenenbaum et al.~2007; Milam et al.~2008; Tenenbaum \& Ziurys 2008). However, there is still some debate regarding which molecules are the dominant reservoirs of P in these environments. Models that assume thermochemical equilibrium typically predict that HCP will be the dominant P-bearing species in C-rich envelopes, while PH$_3$, PS, or PO will dominate in O-rich environments (e.g., Ag{\'u}ndez et al.~2007; Milam et al.~2008). However, such models have struggled to explain the high abundance of PN observed in O-rich circumstellar envelopes (e.g., Milam et al.~2008; De Beck et al.~2013). De Beck et al.~(2013) find that the abundance of PN in the circumstellar shell of the O-rich AGB star IK Tau (PN/H$_2$~$\approx$~$3\times10^{-7}$) is comparable to that of PO, and conclude that these two species are the main gas-phase reservoirs of P in the circumstellar envelopes of O-rich stars. Gobrecht et al.~(2016) modelled the dust formation process in IK Tau, paying particular attention to the effects of non-equilibrium chemistry induced by periodic shocks related to stellar pulsations. They found that PN is formed very efficiently in the post-shock gas and maintains its high abundance throughout the dust formation region. Since their predicted abundance for PN matches the observed abundance (within a factor of two), and since the observed abundance is comparable to the solar P abundance, their results seem to validate the idea that P does not participate in the dust formation process because it is locked up in stable gas-phase molecules. (However, even if this idea turns out to be correct, and assuming it applies to As as well, it would not explain the apparent supersolar abundances of As and P in low depletion sight lines, a problem for which there is no obvious solution at present.)

While the [$X$/H]$_0$ and [$X$/H]$_1$ values are sensitive to the adopted reference abundances and to the oscillator strengths used to derive column densities from the interstellar absorption lines, the $A_X$ parameters do not depend on these factors. They reflect the changes in depletion between $F_*=0$ and $F_*=1$, and so depend only on the \emph{relative} abundances obtained for different sight lines. From a physical perspective, the depletion slopes are related to the rates at which different elements are incorporated into grain mantles under changing environmental conditions in interstellar clouds. As discussed in more detail in J09, the consumption rate for a given element depends on both the depletion slope and the gas-phase abundance of the element at a particular value of $F_*$ (see also Jenkins 2013). From Figure 22, it is clear that the $A_X$ parameters are correlated to some extent with the condensation temperatures of the elements. Elements with higher condensation temperatures tend to have steeper slopes, and therefore tend to be incorporated more readily into the grains, than elements with lower values of $T_C$. However, while this is generally true, there are a handful of elements that appear to exhibit unusual slopes compared to elements with similar condensation temperatures.

Both B and Cl exhibit much steeper depletion slopes than would be expected for elements with $T_C\sim900$~K. We have already discussed a likely explanation for this in the case of Cl (namely, that the J09 analysis neglected to include Cl~{\sc i} column densities when deriving gas-phase Cl abundances). However, the unusually steep slope of the B depletion trend remains unexplained. The B~{\sc ii} line at 1362.5~\AA{} is detected in a variety of different sight lines (R11), and the depletion slope for this element appears to be well constrained (see Appendix B). It is particularly unusual that the slope for B is so much steeper than that for Ga, which is chemically similar to B yet has a somewhat higher condensation temperature (Lodders 2003). It could be that the condensation temperature for B is larger than that calculated by Lodders (2003), but then the initial depletion of B would appear to be at odds with the trend seen for many other elements in the upper panel of Figure 21. (The nominal slopes derived for S and Pb also appear to be too steep in Figure 22. However, the slopes for these elements are not very well constrained and as a result have large associated uncertainties.)

At the other extreme, the depletion slope for Cd appears to be much shallower than expected compared to other elements with $T_C\sim700$~K. Indeed, the value we obtain for the slope ($A_{\mathrm{Cd}}=-0.028\pm0.221$) is consistent with zero, suggesting that Cd follows N in exhibiting no differential depletion in interstellar clouds. This conclusion is consistent with that of Sofia et al.~(1999), who found no changes in Cd depletion with increasing molecular fraction. Many trace elements condense by forming solid solutions with host phases composed of more abundant elements. According to Lodders (2003), the host phases for Cd condensation are enstatite (MgSiO$_3$) and troilite (FeS), with Cd replacing Mg and Fe in these compounds. Enstatite, along with forsterite (Mg$_2$SiO$_4$), is one of the host phases for Zn, and troilite is responsible for removing S from the gas phase (Lodders 2003). Since both S and Zn seem to show differential depletion (J09), it is not clear why Cd would not participate in this process.\footnote{In actuality, there are still considerable uncertainties regarding the depletion behavior of S in the diffuse ISM, primarily because the only available S~{\sc ii} lines are in most cases moderately, if not heavily, saturated, meaning that there are relatively few reliable abundance determinations and the measurements that do exist may be biased toward lower gas-phase abundances (see the discussion in J09).}

Another unusual result is that Kr exhibits a \emph{nonzero} depletion slope, contrary to expectations based on most earlier studies (e.g., Cardelli \& Meyer 1997; Cartledge et al.~2008) that found no changes in the mean gas-phase Kr abundance when the abundances were plotted against the molecular hydrogen fractions or the average line-of-sight hydrogen densities. We find the same negative value for $A_{\mathrm{Kr}}$ as J09 found. However, by re-examining some of the Kr measurements from the literature and by adding new measurements, we have reduced the uncertainty associated with that value by $\sim$40\%, strengthening the case that Kr participates in the depletion process occurring in interstellar clouds. Lodders (2003) argues that heavy noble gases like Kr can be sequestered by the formation of clathrate hydrates (e.g., Kr$\cdot$6H$_2$O), and that 50\% of the available Kr will be sequestered in this way at a temperature of 52~K. While this process may account for the change in Kr depletion between $F_*=0$ and $F_*=1$, it probably cannot explain the significant offset between the interstellar abundances and the solar system abundance, which even at $F_*=0$ is 0.23 dex. In low depletion sight lines, where $F_*$ approaches zero, the gas temperatures are likely not cold enough for significant Kr condensation in clathrate compounds.

By examining the depletion behaviors of rarely-studied elements, and evaluating the results alongside those for more abundant elements, we gain a deeper understanding of the processes that govern the gas-phase abundances of the elements in diffuse clouds. While many questions remain concerning the dust condensation and dust destruction processes, the results presented here, when combined with those in J09, should help to constrain future modelling efforts that seek to understand these processes. In particular, the depletion slopes, which are related to the rates at which the elements are incorporated into grain mantles, can serve as a guide to the changes that occur in the composition of the grains between low and high depletion sight lines. While the rare elements considered in this investigation contribute little to the total mass of interstellar dust grains, the depletion patterns they exhibit can still provide constraints on dust compositions. Typically, for trace elements to condense into solid compounds, certain host minerals (composed of major elements) must already be present in the grains (Lodders 2003). Thus, a detailed accounting of the depletions of many different elements can be useful for determining the most likely grain compositions (e.g., Savage \& Sembach 1996; Draine 2004). The initial depletions, represented by the [$X$/H]$_0$ values, should be of particular interest to those who model the dust formation processes in AGB stars and Type II supernovae (e.g., Sarangi \& Cherchneff 2015; Gobrecht et al.~2016).

\subsection{Implications for $s$- and $r$-Process Nucleosynthesis in the Present Epoch}
Having evaluated the depletion characteristics of Ga, Ge, As, Kr, Cd, Sn, and Pb, and having considered how the results compare to those of other more abundant elements, we are in a position to address claims regarding enhancements and deficiencies in the abundances of $n$-capture elements in the ISM. Our investigation was premised on the suggestion (e.g., Walker et al.~2009) that heavy elements synthesized primarily by massive stars are underabundant in the present-day ISM relative to the solar system, while those produced mainly by low- and intermediate-mass stars are not deficient, and may even show supersolar abundances (e.g., Sofia et al.~1999). Walker et al.~(2009) based their suggestion on the relatively limited data available at the time concerning the interstellar abundances of $n$-capture elements. While Ge, Kr, Rb, Cd, and Sn had been studied in a number of different sight lines, much less was known about the abundances of interstellar Ga, As, and Pb. This was also before J09 published his study, which considered the depletions of many different elements (including Ge and Kr) within a unified framework. Having extended that framework in this investigation to include Ga, As, Cd, Sn, and Pb, we are better positioned to critically evaluate the original suggestion by Walker et al.~(2009) of a deficit in the contribution from massive stars to the production of $n$-capture elements.

First, we should establish the dominant production routes for the elements of interest. Using an updated Galactic chemical evolution (GCE) model, which considers $s$-process yields from AGB stars of low and intermediate mass ($M=1.3$$-$$8M_{\sun}$), Bisterzo et al.~(2014) predict the main $s$ contributions to the solar abundances of the elements from Kr to Bi. For the elements relevant to our investigation, the predicted main $s$ contributions are 13.9\% for Kr, 18.1\% for Rb, 46.1\% for Cd, 52.5\% for Sn, and 87.2\% for Pb. This prediction for Pb includes, in addition to the main $s$-process, a major contribution to the synthesis of $^{208}$Pb by the so-called strong $s$-process in low-mass stars of low metallicity (Gallino et al.~1998). While Bisterzo et al.~(2014) do not give $s$-process predictions for Ga, Ge, or As, an earlier study (not based on GCE) does (Bisterzo et al.~2011). Following Arlandini et al.~(1999), Bisterzo et al.~(2011) derive main $s$ contributions to the solar abundances by averaging the yields of AGB models with $M=1.5$ and $3M_{\sun}$ at $Z=0.5Z_{\sun}$. They find main $s$ contributions of 4.4\% for Ga, 7.1\% for Ge, and 6.2\% for As. The remaining fractions for all of these elements are presumably contributed by massive stars through a combination of the weak $s$-process and the $r$-process.

If it were true that heavy elements produced primarily by massive stars were underabundant in the ISM relative to the solar system, then we would expect this to apply to the elements Kr, Rb, Ga, Ge, and As, which have massive star contributions in the range 82\% to 96\%. We would not expect the elements Cd, Sn, or Pb to be as affected since massive stars contribute much less to their production. From the depletion results presented in Section 4.2, and discussed above (Section 5.1), it is clear that Kr is underabundant in the ISM at all values of $F_*$, a conclusion consistent with the long-established deficit in the interstellar Kr abundance (e.g., Cardelli \& Meyer 1997; Cartledge et al.~2008). (Naturally, this conclusion regarding Kr rests on the assumption that the solar Kr abundance, which relies on theoretical $s$-process production rates, is accurate.) While the trapping of Kr atoms within clathrate compounds may explain the slight increase in the depletion of Kr with $F_*$, it probably cannot account for the overall deficit in the interstellar abundance since at least some sight lines should be characterized by temperatures that are too high for significant Kr condensation. Walker et al.~(2009) found evidence that the interstellar abundance of Rb is also lower than expected. That conclusion was based on their finding of subsolar Rb/K ratios since the dominant ionization stage of Rb is not observed and hence the total gas-phase Rb abundance is not known directly.

For the other three elements produced primarily by massive stars (Ga, Ge, and As) evidence of unusual abundance deficiencies relative to the solar system is less forthcoming. Both Ga and Ge are depleted in the ISM, even in low depletion sight lines; at $F_*=0$, Ga is depleted by 0.43 dex and Ge by 0.22 dex (Table 8). However, these initial depletions seem to follow the general trend in [$X$/H]$_0$ values exhibited by many other elements (i.e., Cu, Mn, Cr, Fe, Ni, and Ti; see Figure 21), a trend which may in fact correspond to a condensation sequence. Arsenic shows no sign of being deficient in the ISM, and may instead be enhanced in its abundance relative to the solar system. Recall that both As and P exhibit much less depletion at all values of $F_*$ compared to elements with similar condensation temperatures, and that, at $F_*=0$, the gas-phase As and P abundances appear to be supersolar (Section 5.1 and Figure 21). (These conclusions necessarily assume that the least-squares fits for As and P are truly representative and that the $f$-values of the relevant transitions are accurate.) Since Ga and Ge seem to exhibit normal depletion patterns, and As seems to be enhanced rather than deficient, we conclude that a simple dichotomy between the production of heavy elements by low-mass and high-mass stars cannot account for the unexpectedly low interstellar abundances of Kr and Rb.

While a comparison between the absolute gas-phase abundances of $n$-capture elements and their expected depletions can yield insight into the possible effects of $s$- and $r$-process nucleosynthesis, another way to explore those effects is to search for variations in the abundances among different lines of sight that are unrelated to differences in depletion. Having delineated the general trends due to the changes in depletion with $F_*$, we can search for any significant scatter in the abundances superimposed onto those trends. Such scatter could be an indication of nucleosynthetic enrichment and/or inefficient mixing in the ISM, for example. To investigate these possibilities, we take the least-squares linear fits (described in Section 4.2) as a basis, and calculate the root mean square (rms) deviations about those fits for all of the elements in our survey. Following J09, we define a residual equal to [$X$/H]$_{\mathrm{obs}}$~$-$~[$X$/H]$_{\mathrm{fit}}$, where [$X$/H]$_{\mathrm{obs}}$ is the observed depletion of element $X$ and [$X$/H]$_{\mathrm{fit}}$ is the depletion expected based on the linear fit defined by Equation (1). The errors in the residuals are calculated by adding in quadrature the errors in [$X$/H]$_{\mathrm{obs}}$ and [$X$/H]$_{\mathrm{fit}}$, where the latter account for errors in the best-fit coefficients $A_X$ and $B_X$ and the sight-line depletion factors $F_*$ (see Appendix B of J09).

The rms deviations in the depletion residuals with respect to the least-squares linear fits are presented in Table 10 for the elements O, Ga, Ge, As, Kr, Cd, Sn, and Pb. (Similar information is presented in Appendix B for the element B.) We also list separately in Table 10 the mean errors in the observed depletions [$X$/H]$_{\mathrm{obs}}$, in the expected depletions [$X$/H]$_{\mathrm{fit}}$, and in the residuals themselves, so that one can easily judge whether the uncertainties in the residuals are dominated by observational errors, uncertain fit parameters, or some combination of both. As an example, we can see that for the elements O and Kr, the fit parameters are well determined and the errors in the residuals are dominated by the observational uncertainties associated with the column density measurements. Only in the case of Pb is the mean error in the expected depletions larger than that in the observed depletions. This reflects the fact that the fit parameters for Pb are poorly constrained, a consequence of the Pb depletion measurements being clustered around only relatively high values of $F_*$.

For most of the elements examined here, we find that the scatter in the abundances with respect to the depletion trends is nearly equal to, if not somewhat less than, the mean error in the residuals, when one accounts for both the observational errors and the uncertainties in the fit parameters (Table 10). This is not the case for Sn, however, where the rms deviations about the least-squares fit are approximately 0.18 dex, while the mean error in the residuals is only 0.12 dex. This result is consistent with our having found an extremely low probability of obtaining a worse fit for Sn (as indicated in the last column of Table 8). While the scatter in the Pb abundances (which is approximately 0.21 dex) is even greater than the scatter for Sn, the mean error in the residuals for Pb is also much greater (due to the poorly constrained fit and the relatively large observational uncertainties). None of these conclusions regarding the scatter in the depletion data would be significantly altered were the survival analysis fits (described in Section 4.3) adopted as the basis for the residuals rather than the least-squares fits.

Another way to examine the scatter in the measured depletions with respect to the depletion trends is to search for significant deviations in the residuals for individual lines of sight. If we divide the depletion residuals by their respective errors, we can identify sight lines whose measured depletions deviate from expected values by some specified factor. Appyling this procedure to the O depletion data, for example, we find that in only 5 out of 91 sight lines ($\sim$5\%) do the measured depletions deviate from the expected ones by 2$\sigma$ or more (using the least-squares fit as the basis). Similar results are obtained for many of the $n$-capture elements. We find that 2$\sigma$ or greater deviations occur in only 3 out of the 43 sight lines with Ga depletion measurements, 4 out of the 66 sight lines with Ge depletions, and 1 out of the 12 sight lines with depletion measurements for Pb. For As, Kr, and Cd, there are no sight lines that deviate from the depletion trends by 2$\sigma$ or more. In the case of Sn, however, we find that 9 out of 37 sight lines ($\sim$24\%) exhibit depletions that deviate from expectations (based on the least-squares fit) by more than 2$\sigma$. While differences in ionization along different lines of sight could potentially lead to scatter in the abundances (if the ionization potential of the ion is significantly greater than that of H~{\sc i}), the ionization potential of Sn~{\sc ii} (14.63 eV) is among the lowest of the $n$-capture species examined here.

It thus remains an intriguing possibility that the significant scatter in the gas-phase Sn abundances has a nucleosynthetic origin. More than half of the solar system abundance of Sn is contributed by the main $s$-process occurring in low- and intermediate-mass AGB stars (Bisterzo et al.~2011, 2014). The long evolutionary lifetimes of such stars mean that their contributions tend to increase with time relative to those from massive stars. As a result, some evolution in the interstellar abundance of Sn could conceivably have taken place in the 4.6 Gyr since the formation of the solar system (Sofia et al.~1999). If the enrichment process is relatively localized and/or if the timescale for enrichment is significantly shorter than the timescale on which the interstellar gas becomes well mixed, then the variations we find in the gas-phase Sn abundances might be expected. However, if $s$-process enrichment combined with inefficient mixing is the cause of the observed variations for Sn, then similar variations might be expected for Cd and Pb, both of which, like Sn, owe much of their production to the main $s$-process. There are significantly fewer interstellar detections of Cd~{\sc ii} and Pb~{\sc ii} than there are of Sn~{\sc ii}. However, the gas-phase Cd abundances are remarkably consistent from one line of sight to another. The Pb abundances seem to show a high degree of scatter (despite the fact that most sight lines with Pb~{\sc ii} detections have similar values of $F_*$), but that scatter is matched by large uncertainties in the depletion residuals (Table 10).

Support for our finding that the local Galactic ISM shows evidence of enrichment due to $s$-process nucleosynthesis is provided by abundance determinations of $n$-capture elements in Galactic planetary nebulae (Sterling \& Dinerstein 2008; Sterling et al.~2015, 2016). Planetary nebulae are particularly interesting in this context as they can be used to study the products of the final stages of AGB star evolution. Many planetary nebulae are found to be enriched in $n$-capture elements produced by the main $s$-process during the thermally pulsing AGB stage of their progenitor stars (Sterling \& Dinerstein 2008; Sterling et al.~2015). Sterling et al.~(2015) find that nearly 40\% of planetary nebulae with determined Se and/or Kr abundances are $s$-process enriched, with the degree of enrichment varying significantly from one object to the next. Other heavy elements found to be enhanced in planetary nebulae include Rb, Cd, and Xe (e.g., Sterling et al.~2016). While Sn has yet to be detected in such objects, its abundance in the envelopes of low-mass stars at the end of the AGB phase is predicted to be enhanced considerably (e.g., Cristallo et al.~2015). The substantial $s$-process enhancements that have already been detected in planetary nebulae indicate that enrichment of the ISM in the products of $s$-process nucleosynthesis must be occurring at some level.

\section{SUMMARY AND CONCLUSIONS}
In this investigation, we have examined the gas-phase abundances and depletion behaviors of the $n$-capture elements Ga, Ge, As, Kr, Cd, Sn, and Pb, which are the only elements heavier than Zn that have been detected (in their dominant ionization stage) in multiple interstellar sight lines. Our study was motivated by previous findings that were suggestive of unusual deficiencies and potential enhancements in the interstellar abundances of elements produced through $s$- and $r$-process nucleosynthesis. We carried out our survey by analyzing the interstellar absorption profiles for a sample of 69 sight lines with high- and/or medium-resolution STIS archival spectra covering the species of interest. Column densities (or upper limits to the column densities) of Ga~{\sc ii}, Ge~{\sc ii}, Kr~{\sc i}, As~{\sc ii}, Cd~{\sc ii}, Sn~{\sc ii}, and Pb~{\sc ii} were derived for our sample through detailed profile synthesis fits, often using the O~{\sc i}~$\lambda1355$ line as a template for the line-of-sight component structure. We note that our survey has produced the first reported detections of the Pb~{\sc ii}~$\lambda1203$ transition in individual sight lines (after the discovery of this feature in a composite spectrum by Heidarian et al.~2015). Column density measurements for an additional 59 sight lines with abundance determinations in the literature, from STIS and GHRS observations, were included in our analysis (as described in Appendix A).

Having collected as much information as possible on the gas-phase abundances of $n$-capture elements, either through our own analysis of archival STIS spectra or by adopting measurements from the literature, we determined depletion parameters for each element following the methodology of J09, who developed a unified framework for describing the depletion behaviors of 17 different elements (including O, Ge, and Kr). Our work extends the analysis of J09 to B, Ga, As, Cd, Sn, and Pb, and provides updated results for O, Ge, and Kr. The two key parameters needed to characterize the depletion trends for different elements are the initial depletions [$X$/H]$_0$ and the depletion slopes $A_X$. Values of these parameters, derived through least-squares linear fits, are provided in Table 8 for O, Ga, Ge, As, Kr, Cd, Sn, and Pb (and in Appendix B for B). For the hard-to-detect elements As and Pb, there are more nondetections than actual measurements, and the detections tend to be clustered at relatively high values of the sight-line depletion factor $F_*$. This leads to depletion parameters that are poorly constrained. To address this situation, we performed a survival analysis on the gas-phase abundance data for each element, the results of which are presented in Table 9. While the survival analysis regression fits consider both the actual measurements and any upper limits, they do not account for errors in the independent and dependent variables like the least-squares fits do. Nevertheless, the survival analysis regressions suggest that the depletion slopes for As and Pb may not be as steep as indicated by the least-squares fits. More precise measurements are needed, particularly in low depletion sight lines, to better constrain the depletion slopes in these cases.

The derived depletion parameters for the elements studied in this investigation were compared with those for elements considered by J09 and trends were sought between the ensemble of depletion results and the condensation temperatures of the elements. These efforts have provided us with a better understanding of the depletion behaviors of elements with low-to-moderate condensation temperatures, specifically those with $T_C$ between 600~K and 1100~K, where the transition from relatively mild to relatively strong depletions is seen to occur. By examining the depletion properties of many different elements at once, we are better able to identify peculiarities in specific cases. For the $n$-capture elements that were the focus of our investigation, such peculiarities could have implications not only for the dust-grain depletion processes but also for the nucleosynthetic processes responsible for the production of heavy elements in the current epoch of Galactic evolution. We summarize our main findings for specific elements below:

\begin{enumerate}
\item The two most readily detectable $n$-capture elements, Ga and Ge, exhibit normal depletion patterns, meaning that their initial depletions and depletion slopes are in line with expectations based on the depletion results for many other elements with a range of condensation temperatures. Indeed, there seems to be a fairly well-defined trend of increasing initial depletion with increasing $T_C$ for the elements Ge, Ga, Cu, Mn, Cr, Fe, Ni, and Ti. This could represent a condensation sequence resulting from dust formation in the outer atmospheres of late-type stars or supernovae. (Note, however, that B, Cl, As, P, Si, and Mg do not seem to follow this trend.)
\item Arsenic appears to exhibit less depletion than expected at all values of $F_*$, a trait it shares with the chemically-similar element P. Moreover, the gas-phase abundances of As and P appear to be supersolar in low-depletion sight lines (if the slope indicated by the least-squares fit for As is accurate). While As and P may be chemically blocked from depleting in the circumstellar envelopes of evolved stars, because these elements can form stable molecules with N, such a phenomenon would not explain their enhanced abundances at $F_*=0$. In the case of As, it would be helpful if the oscillator strength of the As~{\sc ii} transition at 1263.8 \AA{} were experimentally verified.
\item Our analysis of the depletion trend for Kr strengthens the finding of J09 that this noble gas participates in the collective depletion behavior exhibited by most other elements. We find the same negative value for the slope parameter $A_{\mathrm{Kr}}$ as J09 did, yet our analysis has reduced the error in this quantity by $\sim$40\%. The small increase in Kr depletion with $F_*$ may be due to the sequestering of Kr atoms within clathrate hydrates. However, it seems unlikely that such a process could account for the significant offset between the interstellar Kr abundances and the solar abundance since the temperature required for Kr condensation is well below that which characterizes an average interstellar sight line.
\item Among the elements not previously examined by J09, only Cd shows no evidence of differential depletion, making it only the second element (after N) to exhibit that trait. Since the host minerals necessary for Cd condensation appear to be the same as those responsible for removing S and Zn from the gas phase, and since both of the latter elements do seem to exhibit differential depletion, it is not clear why Cd would not also participate in that process. It may be that the interstellar measurements are not precise enough to detect the small changes in Cd depletion that would generally be expected.
\item The interstellar gas-phase abundances of Sn and Pb show a large amount of scatter at any given value of $F_*$. In the case of Pb, this may simply reflect the relatively large observational uncertainties. For Sn, however, the scatter appears to be real and may be evidence of intrinsic variations in the abundance of Sn from one location to another. Such variations could conceivably result from $s$-process enrichment by low- and intermediate-mass AGB stars if the enrichment timescale is sufficiently short compared to the timescale on which the ISM becomes well mixed.
\end{enumerate}

Significant progress has been made in understanding the interstellar abundances of some heavy elements. For the most part, that progress was made possible by the ability of \emph{HST} to acquire high-resolution UV spectra at relatively high signal to noise for targets probing the local part of the Milky Way Galaxy. Further advances, however, will likely require use of a next-generation UV space telescope with a significantly larger aperture compared to \emph{HST}. A large aperture UV space telescope would yield more detections of rare species in low density sight lines (which could help to better define the depletion behaviors of As and Pb, for example). Such an instrument might also yield the first detections of as-yet-undetected elements (particularly those with low intrinsic abundances but low condensation temperatures, like Xe and Hg, or those with somewhat higher abundances but high condensation temperatures, like Sr and Ba). Finally, a large aperture space telescope with UV and optical capabilities would enable a broader study of the abundances of $n$-capture elements in the ISM of nearby galaxies and high-redshift absorption systems.

\acknowledgments
This research has made use of the SIMBAD database operated at CDS, Strasbourg, France. A.M.R.~acknowledges support from the Kenilworth Fund of the New York Community Trust. D.L.L.~thanks the Robert A.~Welch Foundation of Houston, TX, for support through grant F-634. Additional support for this work was provided by the Space Telescope Science Institute through grant HST-AR-12123.001-A. Observations were obtained from the Mikulski Archive for Space Telescopes (MAST). STScI is operated by the Association of Universities for Research in Astronomy, Inc., under NASA contract NAS5-26555.

\facility{\emph{HST} (STIS)}

\software{ISMOD (see Sheffer et al.~2008), STSDAS}

\appendix
\section{COMPILATION OF COLUMN DENSITY MEASUREMENTS FROM THE LITERATURE}
In constructing our primary sample of interstellar sight lines from data available in the STIS archive, we focused on directions showing evidence of absorption from the hard-to-detect species As~{\sc ii}, Cd~{\sc ii}, Sn~{\sc ii}, and Pb~{\sc ii}. We also included sight lines that had been previously analyzed by R11 since much was already known about the component structures in those directions and it was straightforward to calculate upper limits to the column densities of any species that were not detected. However, so as not to ignore potentially useful measurements from the literature, we compiled a list of previously published column densities for species of interest to our investigation along sight lines not included in our primary sample, focusing only on measurements derived from GHRS or STIS observations. This compilation is presented in Table 11. All column densities reported in the table have been corrected to be consistent with the set of $f$-values adopted in this investigation (Table 3). We note particularly large corrections in the following cases. The Ge~{\sc ii} column densities from Welty et al.~(1999), Cartledge et al.~(2006), and Welty (2007) had to be increased by 0.15 dex to be consistent with the new experimental $f$-value for Ge~{\sc ii}~$\lambda1237$ reported by Heidarian et al.~(2017). The Pb~{\sc ii} column density toward 1 Sco from Welty et al.~(1995) had to be increased by 0.43 dex to be consistent with the experimental $f$-value for Pb~{\sc ii}~$\lambda1433$ provided by Heidarian et al.~(2015). The As~{\sc ii} and Pb~{\sc ii} column densities toward $\zeta$ Oph were calculated from the published equivalent widths (Cardelli et al.~1993; Cardelli 1994), and our adopted $f$-values, under the assumption that the absorption is optically thin.

\begin{figure}
\centering
\includegraphics[width=0.5\textwidth]{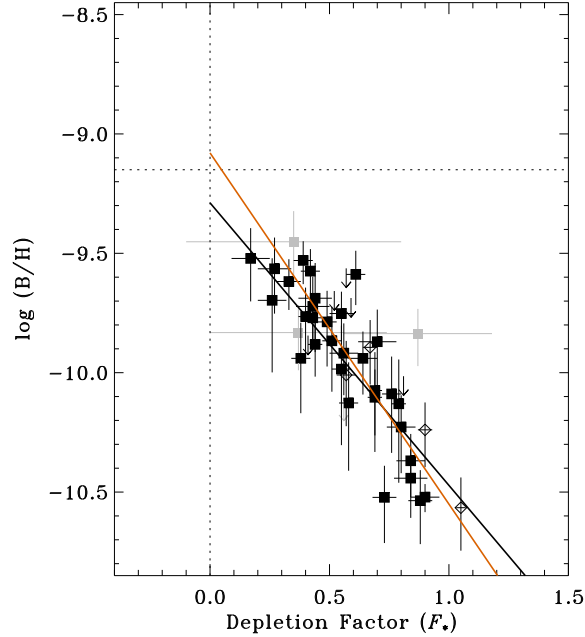}
\caption{Gas-phase B abundances as a function of the sight line depletion factor ($F_*$) from J09. Solid symbols (and upper limits) represent abundances derived by R11, with two additional determinations given in Appendix B. Open symbols are used for other results obtained from the literature. Grey symbols denote sight lines where $\sigma(F_*)\ge0.30$. The solid orange line shows the linear fit based on the methodology of J09. The solid black line shows the fit based on a survival analysis. Parameters for both fits are given in Appendix B. The horizontal dotted line gives the adopted solar system abundance from Lodders (2003).}
\end{figure}

\section{APPLICATION OF THE JENKINS (2009) METHODOLOGY TO BORON}
Since the boron abundance study of R11 was published after the depletion study of J09, the latter paper did not consider B when deriving depletion parameters for the elements. Thus, in order to include B in our discussion of element depletions in Section 5.1, we derive the relevant parameters here. In R11, B~{\sc ii} column densities were obtained for 47 sight lines with high- and/or medium-resolution STIS spectra covering the B~{\sc ii} resonance line at 1362.5 \AA{}. Through the course of this investigation, we discovered two additional detections of the B~{\sc ii}~$\lambda1362$ feature. Using a profile synthesis approach (similar to that described in Section 3.2), we find log~$N$(B~{\sc ii})~=~$11.65^{+0.08}_{-0.09}$ toward HD~147889 and log~$N$(B~{\sc ii})~=~$11.30^{+0.09}_{-0.12}$ toward HDE~232522 (adopting the same $f$-value for the B~{\sc ii} transition as in R11). We also redetermined the B~{\sc ii} column density toward HD~99890 (using high-resolution STIS spectra not available previously) finding log~$N$(B~{\sc ii})~=~$11.44^{+0.04}_{-0.04}$. If we include the five sight lines with B~{\sc ii} measurements from GHRS observations (see R11 and references therein), then we have a total of 54 sight lines with significant detections of the B~{\sc ii} absorption feature. Of these, 38 have the necessary data on $N$(H$_{\mathrm{tot}}$) and have $F_*$ values from J09. We plot the B abundances for these directions against the sight line depletion factors in Figure 23. A least-squares linear fit (adopting the same formalism as described in Section 4.2) yields the following parameters: $A_{\mathrm{B}}$~=~$-1.471\pm0.110$, $B_{\mathrm{B}}$~=~$-0.801\pm0.047$, $z_{\mathrm{B}}$~=~0.592, [B/H]$_0$~=~$+0.071\pm0.080$, [B/H]$_1$~=~$-1.401\pm0.065$. The $\chi^2$ statistic for the fit is 44.3 with 36 degrees of freedom, yielding a probability of obtaining a worse fit of 0.162. For consistency, we also performed a survival analysis on the B abundance data (following the same procedure as described in Section 4.3). The survival analysis regression fit yields a slope of $-1.183\pm0.123$ and an intercept of $-0.139\pm0.079$ (for 38 detections and six upper limits). The scatter in the depletion residuals about the least-squares fit for B is 0.178 dex, while the mean error in the residuals is 0.184 dex. (For the observed and expected depletions, separately, the mean errors are 0.125 dex and 0.116 dex.) These quantities for B may be compared to those listed in Tables 8, 9, and 10 for the other elements analyzed in this investigation.

\clearpage

\startlongtable
\begin{deluxetable}{lccccccccc}
\tablecolumns{10}
\tablewidth{0pt}
\tabletypesize{\footnotesize}
\tablecaption{Stellar Data}
\tablehead{ \colhead{Star} & \colhead{Name} & \colhead{Sp.~Type} & \colhead{Ref.} & \colhead{$V$} & \colhead{$E(\bv)$\tablenotemark{a}} & \colhead{$l$} & \colhead{$b$} & \colhead{$d$\tablenotemark{a}} & \colhead{$z$} \\
\colhead{} & \colhead{} & \colhead{} & \colhead{} & \colhead{(mag)} & \colhead{(mag)} & \colhead{(deg)} & \colhead{(deg)} & \colhead{(kpc)} & \colhead{(kpc)} }
\startdata
HD~1383     & \ldots         & B1 II     & 1 & 7.63 & 0.37 & 119.02 &  $-$0.89 & 2.9 & $-$0.05 \\
HD~13268    & \ldots         & O8 Vnn    & 1 & 8.18 & 0.23 & 133.96 &  $-$4.99 & 2.1 & $-$0.18 \\
HD~13745    & V354 Per       & O9.7 IIn  & 1 & 7.90 & 0.34 & 134.58 &  $-$4.96 & 3.2 & $-$0.28 \\
HD~14434    & \ldots         & O5.5 Vnfp & 1 & 8.49 & 0.38 & 135.08 &  $-$3.82 & 3.5 & $-$0.23 \\
HD~22951    & 40 Per         & B0.5 V    & 1 & 4.98 & 0.19 & 158.92 & $-$16.70 & $0.32\pm0.02$ & $-$0.05 \\
HD~23180    & $o$ Per        & B1 III    & 1 & 3.83 & 0.22 & 160.36 & $-$17.74 & $0.34\pm0.09$ & $-$0.18 \\
HD~24190    & \ldots         & B2 Vn     & 1 & 7.45 & 0.23 & 160.39 & $-$15.18 & $0.27\pm0.05$ & $-$0.28 \\
HD~24398    & $\zeta$ Per    & B1 Ib     & 1 & 2.85 & 0.27 & 162.29 & $-$16.69 & $0.23\pm0.01$ & $-$0.23 \\
\enddata
\tablenotetext{a}{Values of $E(\bv)$ and $d$ are mainly from J09. Distances with error bars are from $\geq$$4\sigma$ \emph{Hipparcos} results (van Leeuwen 2007). Other exceptions are noted.}
\tablenotetext{b}{These values of $E(\bv)$ and $d$ were derived following the set of procedures outlined in Appendix B of B08, using intrinsic colors from Wegner (1994) and absolute visual magnitudes from B08.}
\tablenotetext{c}{These values of $E(\bv)$ and $d$ were obtained directly from Table 1 of B08.}
\tablenotetext{d}{The apparent $V$-band magnitude of this star is dominated by the cool component, which has $V=5.71$ (based on a transformation of the \emph{Tycho} magnitudes for this component; ESA 1997). The hot component has $V=9.24$ and is the one used to calculate the distance of 1.4~kpc.}
\tablerefs{(1) J09; (2) Houk \& Cowley 1975; (3) B08; (4) Houk \& Smith-Moore 1988; (5) Sota et al.~2014; (6) Hill 1970; (7) Simonson 1968; (8) Venn et al.~1998.}
\end{deluxetable}

\clearpage

\startlongtable
\begin{deluxetable}{lccccc}
\tablecolumns{6}
\tablewidth{0pt}
\tabletypesize{\footnotesize}
\tablecaption{STIS Data Sets}
\tablehead{ \colhead{Star} & \colhead{Data Set} & \colhead{Exp.~Time} & \colhead{Cen.~Wave.} & \colhead{Grating} & \colhead{Slit} \\
\colhead{} & \colhead{} & \colhead{(s)} & \colhead{(\AA{})} & \colhead{} & \colhead{(arcsec)} }
\startdata
HD~1383      & o5c07c     & 1440   & 1271      & E140H   & $0.2\times0.2$ \\
             & obkr7c     & 1200   & 1271      & E140H   & $0.2\times0.2$ \\
             & o6lz01     & 1200   & 1425      & E140M   & $0.2\times0.2$ \\
HD~13268     & o63506     & 1440   & 1425      & E140M   & $0.2\times0.2$ \\
HD~13745     & o6lz05     & 1200   & 1425      & E140M   & $0.2\times0.2$ \\
HD~14434     & o63508     & 1440   & 1425      & E140M   & $0.2\times0.2$ \\
HD~22951     & o64805--09 & 36133  & 1416      & E140H   & $0.2\times0.05$ \\
             & o6lv01     & 2773   & 1978      & E230M   & $0.2\times0.05$ \\
\enddata
\tablenotetext{a}{In addition to the STIS data sets listed here for X~Per (HD~24534), we also examined the GHRS data set z2do0107t and the \emph{FUSE} data set p1930201.}
\end{deluxetable}

\clearpage

\begin{deluxetable}{lccc}
\tablecolumns{4}
\tablewidth{0pt}
\tabletypesize{\footnotesize}
\tablecaption{Atomic Data}
\tablehead{ \colhead{Species} & \colhead{$\lambda_0$} & \colhead{$f$} & \colhead{Ref.} \\
\colhead{} & \colhead{(\AA{})} & \colhead{} & \colhead{} }
\startdata
O~{\sc i}   & 1355.598 & 0.00000116 & 1 \\
Ga~{\sc ii} & 1414.402 & 1.77\phn\phn\phn\phn\phn\phn & 1 \\
Ge~{\sc ii} & 1237.059 & 0.872\phn\phn\phn\phn\phn & 2 \\
            & 1602.486 & 0.144\phn\phn\phn\phn\phn & 3 \\
As~{\sc ii} & 1263.770 & 0.259\phn\phn\phn\phn\phn & 3 \\
Kr~{\sc i}  & 1164.867 & 0.188\phn\phn\phn\phn\phn & 3 \\
            & 1235.838 & 0.204\phn\phn\phn\phn\phn & 3 \\
Cd~{\sc ii} & 2145.070 & 0.498\phn\phn\phn\phn\phn & 3 \\
            & 2265.715 & 0.247\phn\phn\phn\phn\phn & 3 \\
Sn~{\sc ii} & 1400.440 & 1.03\phn\phn\phn\phn\phn\phn & 3 \\
Pb~{\sc ii} & 1203.616 & 0.75\phn\phn\phn\phn\phn\phn & 4 \\
            & 1433.906 & 0.321\phn\phn\phn\phn\phn & 4 \\
\enddata
\tablerefs{(1) Morton 2003; (2) Heidarian et al.~2017; (3) Morton 2000; (4) Heidarian et al.~2015.}
\end{deluxetable}

\begin{deluxetable}{lccccccccccc}
\tablecolumns{12}
\tablewidth{0pt}
\tabletypesize{\footnotesize}
\tablecaption{Component Structure}
\tablehead{\colhead{Star} & \multicolumn{3}{c}{O~{\sc i}~$\lambda$1355} & \colhead{} & \multicolumn{3}{c}{Ga~{\sc ii}~$\lambda$1414} & \colhead{} & \multicolumn{3}{c}{Ge~{\sc ii}~$\lambda$1237} \\
\cline{2-4} \cline{6-8} \cline{10-12} \\
\colhead{} & \colhead{$v_{\mathrm{LSR}}$} & \colhead{$b$} & \colhead{$N$/$N_{\mathrm{tot}}$} & \colhead{} & \colhead{$v_{\mathrm{LSR}}$} & \colhead{$b$} & \colhead{$N$/$N_{\mathrm{tot}}$} & \colhead{} & \colhead{$v_{\mathrm{LSR}}$} & \colhead{$b$} & \colhead{$N$/$N_{\mathrm{tot}}$} \\
\colhead{} & \colhead{(km s$^{-1}$)} & \colhead{(km s$^{-1}$)} & \colhead{} & \colhead{} & \colhead{(km s$^{-1}$)} & \colhead{(km s$^{-1}$)} & \colhead{} & \colhead{} & \colhead{(km s$^{-1}$)} & \colhead{(km s$^{-1}$)} & \colhead{} }
\startdata
HD~14434    & $-$52.8  & 3.2    & 0.11   && $-$52.6  & 3.2  & 0.11   && $-$57.0  & 3.2    & 0.14 \\
            & $-$42.3  & 1.0    & 0.10   && $-$43.5  & 1.6  & 0.01   && $-$46.5  & 0.9    & 0.30 \\
            & \ldots   & \ldots & \ldots && $-$36.3  & 1.6  & 0.19   && \ldots   & \ldots & \ldots \\
            & $-$19.4  & 0.7    & 0.07   && $-$20.1  & 1.6  & 0.22   && $-$23.6  & 0.8    & 0.09 \\
            & $-$10.0  & 1.3    & 0.25   && $-$10.8  & 0.7  & 0.19   && $-$14.2  & 1.3    & 0.14 \\
            & $-$3.9   & 1.1    & 0.24   && $-$4.6   & 0.7  & 0.18   && $-$8.2   & 0.7    & 0.03 \\
            & +1.4     & 0.7    & 0.23   && +1.8     & 1.6  & 0.10   && $-$2.8   & 3.2    & 0.30 \\
\enddata
\tablecomments{Table 4 is published in its entirety in the machine-readable format. A portion is shown here for guidance regarding its form and content.}
\end{deluxetable}

\clearpage

\begin{longrotatetable}
\begin{deluxetable}{lcccccccc}
\tablecolumns{9}
\tablewidth{0pt}
\tabletypesize{\scriptsize}
\tablecaption{Column Densities}
\tablehead{ \colhead{Star} & \colhead{log $N$(O~{\sc i})} & \colhead{log $N$(Ga~{\sc ii})} & \colhead{log $N$(Ge~{\sc ii})} & \colhead{log $N$(As~{\sc ii})} & \colhead{log $N$(Kr~{\sc i})} & \colhead{log $N$(Cd~{\sc ii})} & \colhead{log $N$(Sn~{\sc ii})} & \colhead{log $N$(Pb~{\sc ii})} }
\startdata
HD~1383     & $18.15^{+0.02}_{-0.02}$\tablenotemark{a}   & $12.07^{+0.05}_{-0.05}$   & $12.74^{+0.02}_{-0.02}$\tablenotemark{a}   & $<12.10$   & $12.52^{+0.08}_{-0.09}$\tablenotemark{a}   & \ldots   & $11.88^{+0.12}_{-0.17}$   & $<12.17$ \\
HD~13268    & $18.04^{+0.05}_{-0.06}$   & $11.58^{+0.10}_{-0.13}$   & $12.71^{+0.05}_{-0.06}$   & $<12.37$   & $<13.00$   & \ldots   & $<11.76$   & $<12.10$ \\
HD~13745    & $17.97^{+0.06}_{-0.06}$   & $11.76^{+0.06}_{-0.07}$   & $12.60^{+0.05}_{-0.06}$   & $<12.34$   & $<13.07$   & \ldots   & $<11.76$   & $<12.08$ \\
HD~14434    & $18.18^{+0.05}_{-0.06}$   & $11.94^{+0.06}_{-0.07}$   & $12.88^{+0.07}_{-0.08}$   & $<12.45$   & $<13.32$   & \ldots   & $<11.75$   & $<12.13$ \\
HD~22951    & $17.75^{+0.01}_{-0.01}$   & $11.20^{+0.02}_{-0.03}$   & \ldots   & \ldots   & \ldots   & \ldots   & $11.05^{+0.03}_{-0.03}$   & $<11.00$ \\
HD~23180    & $17.75^{+0.01}_{-0.01}$   & $11.10^{+0.04}_{-0.04}$   & \ldots   & \ldots   & \ldots   & \ldots   & $11.14^{+0.02}_{-0.02}$   & $11.27^{+0.08}_{-0.10}$ \\
HD~24190    & $17.86^{+0.03}_{-0.03}$   & $11.34^{+0.05}_{-0.06}$   & $12.31^{+0.03}_{-0.03}$   & $11.55^{+0.11}_{-0.14}$   & $12.27^{+0.03}_{-0.03}$   & \ldots   & $<11.13$   & $11.33^{+0.16}_{-0.24}$\tablenotemark{b} \\
HD~24398    & $17.78^{+0.01}_{-0.01}$   & $11.23^{+0.02}_{-0.02}$   & \ldots   & \ldots   & \ldots   & \ldots   & $11.16^{+0.03}_{-0.04}$   & $11.21^{+0.08}_{-0.10}$ \\
\enddata
\tablenotetext{a}{Weighted mean of the column densities obtained from high-resolution (E140H) and medium-resolution (E140M) data.}
\tablenotetext{b}{Column density obtained from the Pb~{\sc ii}~$\lambda1203$ line.}
\tablenotetext{c}{Weighted mean of the column densities obtained from the Ge~{\sc ii}~$\lambda1237$ line (from GHRS data) and the Ge~{\sc ii}~$\lambda1602$ line (from STIS data).}
\tablenotetext{d}{Column density obtained from the Kr~{\sc i}~$\lambda1164$ line from \emph{FUSE} data.}
\tablenotetext{e}{Weighted mean of the column densities obtained from the Cd~{\sc ii}~$\lambda2145$ and $\lambda2265$ lines.}
\tablenotetext{f}{The expected position of the As~{\sc ii}~$\lambda1263$ line overlaps with a high-velocity component of Si~{\sc ii}*~$\lambda1264$, precluding a detection of As~{\sc ii} in this direction.}
\end{deluxetable}
\end{longrotatetable}

\clearpage

\begin{deluxetable}{lccccccc}
\tablecolumns{8}
\tablewidth{0pt}
\tabletypesize{\scriptsize}
\tablecaption{Comparison with Previous Studies: Ge~{\sc ii} and Kr~{\sc i}}
\tablehead{ \colhead{Star} & \multicolumn{3}{c}{log $N$(Ge~{\sc ii})} & \colhead{} & \multicolumn{3}{c}{log $N$(Kr~{\sc i})} \\
\cline{2-4} \cline{6-8} \\
\colhead{} & \colhead{This Work} & \colhead{Previous Result} & \colhead{Ref.} & \colhead{} & \colhead{This Work} & \colhead{Previous Result} & \colhead{Ref.} }
\startdata
HD~1383 & $12.74^{+0.02}_{-0.02}$ & $12.60\pm0.09$\tablenotemark{a} & 1 && $12.52^{+0.08}_{-0.09}$ & \ldots \\
HD~13268 & $12.71^{+0.05}_{-0.06}$ & $12.59\pm0.07$\tablenotemark{a} & 1 && $<13.00$ & \ldots \\
HD~24190 & $12.31^{+0.03}_{-0.03}$ & \ldots & && $12.27^{+0.03}_{-0.03}$ & $12.35\pm0.06$ & 2 \\
HD~27778 & $12.23^{+0.03}_{-0.03}$ & $12.29\pm0.04$\tablenotemark{a} & 1 && $12.28^{+0.04}_{-0.04}$ & $12.37\pm0.05$ & 3 \\
HD~37021 & $12.51^{+0.05}_{-0.05}$ & $12.47\pm0.04$\tablenotemark{a} & 1 && $12.64^{+0.03}_{-0.04}$ & $12.63\pm0.04$ & 3 \\
HD~37061 & $12.55^{+0.05}_{-0.05}$ & $12.66\pm0.05$\tablenotemark{a} & 1 && $12.72^{+0.01}_{-0.01}$ & $12.72\pm0.03$ & 3 \\
HD~43818 & $12.80^{+0.06}_{-0.07}$ & $12.73\pm0.06$\tablenotemark{a} & 1 && $12.76^{+0.12}_{-0.17}$ & $12.69\pm0.09$ & 4 \\
HD~52266 & $12.50^{+0.03}_{-0.04}$ & $12.58\pm0.04$\tablenotemark{a} & 1 && $12.16^{+0.06}_{-0.08}$ & \ldots \\
\enddata
\tablenotetext{a}{Revised using the new experimental $f$-value for Ge~{\sc ii}~$\lambda1237$ from Heidarian et al.~(2017).}
\tablerefs{(1) Cartledge et al. 2006; (2) Cartledge et al. 2008; (3) Cartledge et al. 2003; (4) Cartledge et al. 2004.}
\end{deluxetable}

\clearpage

\begin{longrotatetable}
\begin{deluxetable}{lcccccccccc}
\tablecolumns{11}
\tablewidth{0pt}
\tabletypesize{\scriptsize}
\tablecaption{Elemental Abundances}
\tablehead{ \colhead{Star} & \colhead{log $N$(H$_{\mathrm{tot}}$)\tablenotemark{a}} & \colhead{$F_*$\tablenotemark{a}} & \colhead{log (O/H)} & \colhead{log (Ga/H)} & \colhead{log (Ge/H)} & \colhead{log (As/H)} & \colhead{log (Kr/H)} & \colhead{log (Cd/H)} & \colhead{log (Sn/H)} & \colhead{log (Pb/H)} }
\startdata
HD~1383     & $21.50^{+0.08}_{-0.09}$ & $0.61\pm0.04$ & $-3.35^{+0.08}_{-0.10}$ & $-9.44^{+0.09}_{-0.11}$ & $-8.77^{+0.08}_{-0.10}$ & $<-9.41$ & $-8.99^{+0.10}_{-0.14}$ & \ldots & $-9.62^{+0.14}_{-0.21}$ & $<-9.33$ \\
HD~13268    & $21.42^{+0.07}_{-0.09}$ & $0.51\pm0.04$ & $-3.37^{+0.09}_{-0.11}$ & $-9.83^{+0.12}_{-0.17}$ & $-8.71^{+0.09}_{-0.11}$ & $<-9.05$ & $<-8.42$ & \ldots & $<-9.66$ & $<-9.32$ \\
HD~13745    & $21.37^{+0.08}_{-0.10}$ & $0.43\pm0.07$ & $-3.40^{+0.09}_{-0.12}$ & $-9.61^{+0.10}_{-0.12}$ & $-8.77^{+0.09}_{-0.12}$ & $<-9.04$ & $<-8.31$ & \ldots & $<-9.62$ & $<-9.30$ \\
HD~14434    & $21.47^{+0.07}_{-0.09}$ & $0.52\pm0.04$ & $-3.29^{+0.09}_{-0.11}$ & $-9.53^{+0.09}_{-0.12}$ & $-8.58^{+0.10}_{-0.12}$ & $<-9.02$ & $<-8.14$ & \ldots & $<-9.72$ & $<-9.33$ \\
HD~22951    & $21.22^{+0.08}_{-0.10}$ & $0.73\pm0.05$ & $-3.48^{+0.08}_{-0.10}$ & $-10.02^{+0.09}_{-0.11}$ & \ldots & \ldots & \ldots & \ldots & $-10.17^{+0.09}_{-0.11}$ & $<-10.23$ \\
HD~23180    & $21.20^{+0.08}_{-0.10}$ & $0.84\pm0.06$ & $-3.45^{+0.08}_{-0.10}$ & $-10.10^{+0.08}_{-0.11}$ & \ldots & \ldots & \ldots & \ldots & $-10.06^{+0.08}_{-0.10}$ & $-9.93^{+0.11}_{-0.15}$ \\
HD~24190    & $21.30^{+0.05}_{-0.06}$ & $0.63\pm0.24$ & $-3.44^{+0.06}_{-0.06}$ & $-9.96^{+0.07}_{-0.08}$ & $-8.99^{+0.06}_{-0.06}$ & $-9.75^{+0.12}_{-0.16}$ & $-9.03^{+0.06}_{-0.06}$ & \ldots & $<-10.17$ & $-9.97^{+0.16}_{-0.26}$ \\
HD~24398    & $21.19^{+0.06}_{-0.08}$ & $0.88\pm0.05$ & $-3.41^{+0.06}_{-0.08}$ & $-9.96^{+0.07}_{-0.08}$ & \ldots & \ldots & \ldots & \ldots & $-10.04^{+0.07}_{-0.08}$ & $-9.98^{+0.10}_{-0.13}$ \\
\enddata
\tablenotetext{a}{Total hydrogen column densities were calculated from the values of $N$(H~{\sc i}) and $N$(H$_2$) listed in J09, unless otherwise noted. The line-of-sight depletion factors $F_*$ are the observed values from J09.}
\tablenotetext{b}{The large correction for stellar Ly$\alpha$ contamination applied to $N$(H~{\sc i}) for this star by J09 may not be necessary. Here, we adopt the uncorrected value of $N$(H~{\sc i}) from Cartledge et al.~(2004). This sight line is not used in the depletion analysis.}
\tablenotetext{c}{Measurement of $N$(H$_2$) is not available. However, the weak or absent absorption from Cl~{\sc i}~$\lambda1347$ indicates that there is very little H$_2$ in the line of sight, and we adopt $N$(H~{\sc i}) (from J09) as representative of $N$(H$_{\mathrm{tot}}$). The sight line is not included in the depletion analysis.}
\tablenotetext{d}{This star is not listed in J09; the total hydrogen column density was calculated using $N$(H~{\sc i}) from Diplas \& Savage (1994) and $N$(H$_2$) from Sheffer et al.~(2007).}
\tablenotetext{e}{The value of $N$(H~{\sc i}) for this star is listed incorrectly in J09. Here, we adopt the correct value from Diplas \& Savage (1994), and provide an updated value for the sight-line depletion factor $F_*$ (E.~Jenkins 2012, private communication).}
\end{deluxetable}
\end{longrotatetable}

\clearpage

\begin{deluxetable}{lccccccccc}
\tablecolumns{10}
\tablewidth{0pt}
\tabletypesize{\footnotesize}
\tablecaption{Element Depletion Parameters}
\tablehead{ \colhead{Elem.} & \colhead{log ($X$/H)$_{\sun}$\tablenotemark{a}} & \colhead{$A_X$} & \colhead{$B_X$} & \colhead{$z_X$} & \colhead{[$X$/H]$_0$} & \colhead{[$X$/H]$_1$} & \colhead{$\chi^2$} & \colhead{$\nu$} & \colhead{Prob. Worse Fit} }
\startdata
O   &  \phn$-3.24\pm0.05$ & $-0.284\pm0.042$  & $-0.153\pm0.051$  & 0.609  & $+0.020\pm0.057$  & $-0.264\pm0.053$  & 71.9    & 89    & \phantom{$<$}0.907 \\
Ga  &  \phn$-8.83\pm0.06$ & $-0.834\pm0.064$  & $-0.936\pm0.062$  & 0.607  & $-0.430\pm0.073$  & $-1.264\pm0.067$  & 60.3    & 41    & \phantom{$<$}0.027 \\
Ge  &  \phn$-8.30\pm0.05$ & $-0.526\pm0.051$  & $-0.539\pm0.051$  & 0.609  & $-0.219\pm0.060$  & $-0.745\pm0.055$  & 61.5    & 64    & \phantom{$<$}0.565 \\
As  &  \phn$-9.60\pm0.05$ & $-0.873\pm0.213$  & $-0.280\pm0.069$  & 0.856  & $+0.467\pm0.195$  & $-0.405\pm0.076$  & \phn6.8 & \phn9 & \phantom{$<$}0.654 \\
Kr  &  \phn$-8.64\pm0.08$ & $-0.166\pm0.059$  & $-0.342\pm0.081$  & 0.663  & $-0.232\pm0.090$  & $-0.398\pm0.083$  & 32.9    & 57    & \phantom{$<$}0.996 \\
Cd  &     $-10.19\pm0.03$ & $-0.028\pm0.221$  & $-0.108\pm0.050$  & 0.839  & $-0.085\pm0.192$  & $-0.113\pm0.062$  & \phn7.1 & \phn7 & \phantom{$<$}0.416 \\
Sn  &  \phn$-9.81\pm0.04$ & $-0.517\pm0.070$  & $-0.148\pm0.044$  & 0.691  & $+0.209\pm0.065$  & $-0.308\pm0.049$  & 89.6    & 35    & $<$0.001 \\
Pb  &  \phn$-9.87\pm0.04$ & $-1.077\pm0.396$  & $-0.179\pm0.064$  & 0.834  & $+0.720\pm0.337$  & $-0.358\pm0.092$  & 19.4    & 10    & \phantom{$<$}0.035 \\
\enddata
\tablenotetext{a}{Recommended solar system abundances from Lodders (2003).}
\end{deluxetable}

\begin{deluxetable}{lcccc}
\tablecolumns{5}
\tablewidth{0pt}
\tabletypesize{\footnotesize}
\tablecaption{Survival Analysis Coefficients}
\tablehead{ \colhead{Elem.} & \colhead{Slope} & \colhead{Intercept} & \colhead{$N_{\mathrm{det}}$\tablenotemark{a}} & \colhead{$N_{\mathrm{upp}}$\tablenotemark{b}} }
\startdata
O   & $-0.235\pm0.031$  & $-0.010\pm0.054$  & 91    & \phn0 \\
Ga  & $-0.691\pm0.085$  & $-0.510\pm0.079$  & 43    & \phn0 \\
Ge  & $-0.371\pm0.081$  & $-0.309\pm0.067$  & 66    & \phn0 \\
As  & $-0.479\pm0.193$  & $+0.035\pm0.164$  & 11    & 36 \\
Kr  & $-0.126\pm0.054$  & $-0.270\pm0.088$  & 59    & 12 \\
Cd  & $-0.257\pm0.181$  & $+0.084\pm0.153$  & \phn9 & \phn2 \\
Sn  & $-0.212\pm0.107$  & $-0.050\pm0.083$  & 37    & 14 \\
Pb  & $-0.221\pm0.185$  & $-0.046\pm0.125$  & 12    & 29 \\
\enddata
\tablenotetext{a}{Number of sight lines with detections.}
\tablenotetext{b}{Number of sight lines with upper limits only.}
\end{deluxetable}

\clearpage

\begin{deluxetable}{lcccc}
\tablecolumns{5}
\tablewidth{0pt}
\tabletypesize{\footnotesize}
\tablecaption{Scatter and Mean Error in Depletion Residuals}
\tablehead{ \colhead{Elem.} & \colhead{Scatter\tablenotemark{a}} & \multicolumn{3}{c}{Mean Error\tablenotemark{b}} \\
\cline{3-5} \\
\colhead{} & \colhead{(dex)} & \colhead{Obs.} & \colhead{Fit} & \colhead{Res.} }
\startdata
O    & 0.071    & 0.085    & 0.028    & 0.093 \\
Ga   & 0.138    & 0.085    & 0.080    & 0.126 \\
Ge   & 0.103    & 0.084    & 0.050    & 0.105 \\
As   & 0.111    & 0.135    & 0.095    & 0.169 \\
Kr   & 0.073    & 0.101    & 0.027    & 0.106 \\
Cd   & 0.089    & 0.112    & 0.058    & 0.127 \\
Sn   & 0.176    & 0.106    & 0.048    & 0.121 \\
Pb   & 0.206    & 0.130    & 0.154    & 0.209 \\
\enddata
\tablenotetext{a}{Root mean square deviations in the depletion residuals with respect to the least-squares linear fits described in Section 4.2.}
\tablenotetext{b}{We list separately the mean errors in the observed depletions [$X$/H]$_{\mathrm{obs}}$, in the expected depletions [$X$/H]$_{\mathrm{fit}}$, and in the residuals [$X$/H]$_{\mathrm{obs}}$~$-$~[$X$/H]$_{\mathrm{fit}}$.}
\end{deluxetable}

\clearpage

\begin{longrotatetable}
\begin{deluxetable}{lcccccccccccc}
\setlength{\tabcolsep}{0.04in}
\tablecolumns{13}
\tablewidth{0pt}
\tabletypesize{\scriptsize}
\tablecaption{Column Densities Obtained from the Literature}
\tablehead{ \colhead{Star} & \colhead{Name} & \colhead{log $N$(H$_{\mathrm{tot}}$)\tablenotemark{a}} & \colhead{$F_*$\tablenotemark{a}} & \colhead{log $N$(O~{\sc i})} & \colhead{log $N$(Ga~{\sc ii})} & \colhead{log $N$(Ge~{\sc ii})} & \colhead{log $N$(As~{\sc ii})} & \colhead{log $N$(Kr~{\sc i})} & \colhead{log $N$(Cd~{\sc ii})} & \colhead{log $N$(Sn~{\sc ii})} & \colhead{log $N$(Pb~{\sc ii})} & \colhead{References} }
\startdata
HD~5394  & $\gamma$ Cas    & $20.16^{+0.08}_{-0.10}$ & $0.52\pm0.04$ & $16.76\pm0.04$ & \ldots & \ldots & \ldots & \ldots & \ldots & \ldots & \ldots  & 1 \\
HD~12323 &                 & $21.29^{+0.07}_{-0.09}$ & $0.52\pm0.04$ & $18.02\pm0.06$ & \ldots & $12.30\pm0.06$ & \ldots & \ldots & \ldots & \ldots & \ldots & 2, 3 \\
HD~23478 &                 & $21.05^{+0.09}_{-0.12}$ & $-$$0.00\pm0.50$\phantom{$-$} & $17.79\pm0.09$ & \ldots & \ldots & \ldots & $12.23\pm0.05$ & \ldots & \ldots & \ldots & 4 \\
HD~24398 & $\zeta$ Per     & $21.19^{+0.06}_{-0.08}$ & $0.88\pm0.05$ & $17.71\pm0.05$\tablenotemark{b} & \ldots & \ldots & \ldots & $12.17\pm0.03$\tablenotemark{b} & \ldots & $11.08\pm0.15$\tablenotemark{b} & \ldots  & 1, 5, 6 \\
HD~24760 & $\epsilon$ Per  & $20.50^{+0.07}_{-0.09}$ & $0.68\pm0.04$ & $17.03\pm0.04$ & \ldots & \ldots & \ldots & $11.46\pm0.07$ & \ldots & \ldots & \ldots  & 1, 5 \\
HD~24912 & $\xi$ Per       & $21.29^{+0.06}_{-0.07}$ & $0.83\pm0.02$ & $17.81\pm0.05$ & \ldots & \ldots & \ldots & \ldots & \ldots & $11.09\pm0.04$ & \ldots  & 1, 6 \\
HD~35149 & 23 Ori          & $20.63^{+0.11}_{-0.15}$ & $0.54\pm0.04$ & $17.32\pm0.10$ & $11.20\pm0.03$ & $12.04\pm0.04$ & \ldots  & \ldots & \ldots & $11.00\pm0.18$ & \ldots  & 7, 6 \\
HD~36486 & $\delta$ Ori A  & $20.19^{+0.04}_{-0.04}$ & $0.54\pm0.02$ & $16.68\pm0.05$ & \ldots & \ldots & \ldots & \ldots & \ldots & \ldots & \ldots  & 1 \\
\enddata
\tablenotetext{a}{Total hydrogen column densities were calculated from the values of $N$(H~{\sc i}) and $N$(H$_2$) listed in J09, unless otherwise noted. The line-of-sight depletion factors $F_*$ are the observed values from J09.}
\tablenotetext{b}{The O~{\sc i} and Sn~{\sc ii} column densities listed here for $\zeta$~Per are from GHRS observations (Meyer et al.~1998; Sofia et al.~1999) and are not used in the depletion analysis. We use the values we obtain from STIS observations instead (see Table 5). The GHRS result for Kr~{\sc i} listed here (from Cardelli \& Meyer 1997) is used in the analysis, however, because the STIS observations do not cover the Kr~{\sc i}~$\lambda1235$ line.}
\tablenotetext{c}{This sight line has an O~{\sc i} column density from \emph{Copernicus} observations (Meyer et al.~1998), which is not used in the depletion analysis. We restrict our analysis of O abundances to sight lines with column density measurements from STIS or GHRS observations.}
\tablenotetext{d}{The large correction for stellar Ly$\alpha$ contamination applied to $N$(H~{\sc i}) for this star by J09 may not be necessary. Here, we adopt the uncorrected value of $N$(H~{\sc i}) from Cartledge et al.~(2004). This sight line is not used in the depletion analysis.}
\tablerefs{(1) Meyer et al.~1998; (2) Cartledge et al.~2004; (3) Cartledge et al.~2006; (4) Cartledge et al.~2008; (5) Cardelli \& Meyer 1997; (6) Sofia et al.~1999; (7) Welty et al.~1999; (8) Cartledge et al.~2003; (9) Hobbs et al.~1993; (10) Andr{\'e} et al.~2003; (11) Welty et al.~1995; (12) Federman et al.~2003; (13) Savage et al.~1992; (14) Cardelli et al.~1993; (15) Cardelli 1994; (16) Snow et al.~1996; (17) Cartledge et al.~2001; (18) Welty 2007.}
\end{deluxetable}
\end{longrotatetable}

\end{document}